\documentclass[11pt]{article}
\usepackage[utf8]{inputenc}

\usepackage{stmaryrd}
\usepackage{varioref}
\usepackage{amsmath,amsfonts,amsthm,mathrsfs}
\usepackage{mathabx}
\usepackage[normalem]{ulem}
\usepackage{multirow,color,graphics}
\usepackage{amsmath}
\usepackage{amsfonts}
\usepackage{amssymb}
\usepackage{jheppub}
\usepackage{enumerate}
\usepackage{fancyvrb}
\usepackage{verbatim}
\usepackage{wrapfig}
\usepackage{appendix}
\usepackage{amstext}
\usepackage{amssymb}
\usepackage[bbgreekl]{mathbbol}
\usepackage{bbm}
\usepackage{graphicx}
\usepackage{color}
\usepackage{varioref}
\usepackage{multirow,graphics}
\usepackage{epstopdf}

\usepackage{mmacells}

\mmaSet{
  morefv={gobble=2},
  linklocaluri=mma/symbol/definition:#1,
  morecellgraphics={yoffset=1.9ex}
}

\newcommand{\nn}{\nonumber}

\DeclareSymbolFontAlphabet{\mathbbm}{bbold}
\DeclareSymbolFontAlphabet{\mathbb}{AMSb}

\renewcommand{\comment}[1]{}

\numberwithin{equation}{section}

\def\[{\left[}
\def\]{\right]}
\def\({\left(}
\def\){\right)}
\def\<{\left<}
\def\>{\right>}
\def\d{\partial}

    \newcommand{\beq}{\begin{equation}}
    \newcommand{\eeq}{\end{equation}}
    \newcommand\beqa{\begin{eqnarray}}
    \newcommand\eeqa{\end{eqnarray}}
\newcommand\bea{\begin{array}}
\newcommand\eea{\end{array}}

\newcommand{\bs}{{\bf s}}

\newcommand{\la}[1]{\label{#1}}
\newcommand{\eq}[1]{(\ref{#1})}

    \def\bT{{\bf T}}

\definecolor{cadmiumgreen}{rgb}{0.0, 0.42, 0.24}

\makeatletter
     \@ifundefined{usebibtex}{} {}
\makeatother

     \def\bT{{\bf T}}

     \def\Tr{\text{Tr}}
     \def\l{\lambda}

\newcommand{\bra}[1]{\langle #1 |}
\newcommand{\ket}[1]{| #1 \rangle}
\newcommand{\cN}{\mathcal{N}}

\renewcommand{\<}{\langle} 
\renewcommand{\>}{\rangle} 
\renewcommand{\sl}{\mathfrak{sl}}
\newcommand{\su}{\mathfrak{su}}

\newcommand{\lH}{\mathcal{H}}

\newcommand{\lN}{\mathcal{N}}
\newcommand{\lR}{\mathcal{R}}
\newcommand{\lT}{\mathcal{T}}
\newcommand{\lS}{\mathcal{S}}
\newcommand{\bB}{\textbf{B}}
\newcommand{\bC}{\textbf{C}}
\newcommand{\bbT}{\mathbb{T}}
\newcommand{\bbS}{\mathbb{S}}
\newcommand{\bS}{\textbf{S}}
\newcommand{\svx}{{\mathsf{x}}}

\newcommand{\svy}{{\mathsf{y}}}

\newcommand{\gl}{\mathfrak{gl}}

\newcommand{\cD}{{\cal D}}
\newcommand{\cW}{{\cal W}}

\newcommand{\bl}{\left(\hspace{-1.85mm}\left(\,}
\newcommand{\br}{\,\right)\hspace{-1.85mm}\right)}
\newcommand{\detl}{\left\llbracket\,}
\newcommand{\detr}{\,\right\rrbracket}

\newcommand{\mE}{\mathbb E}

\title{
Determinant Form of Correlators 
in High Rank Integrable Spin Chains via 
Separation of Variables
}

\emailAdd{nikgromov$\bullet$gmail.com}
\emailAdd{fedor.levkovich$\bullet$gmail.com} 
\emailAdd{pryan$\bullet$maths.tcd.ie} 

\author[a,b]{Nikolay Gromov,}
\author[c,1]{Fedor Levkovich-Maslyuk\note{Also at Institute for Information Transmission Problems, Moscow 127994, Russia},}
\author[e]{Paul Ryan}

\affiliation[a]{
Mathematics Department, King's College London,
The Strand, London WC2R 2LS, UK
}

\affiliation[b]{St.Petersburg INP, Gatchina, 188 300, St.Petersburg,
Russia}
\affiliation[c]{
Institut de Physique Th\'{e}orique, Universit\'{e} Paris Saclay, CEA, CNRS, F-91191 Gif-sur-Yvette, France
}
\affiliation[e]{School of Mathematics \& Hamilton Mathematics Institute, Trinity College Dublin, College Green, Dublin 2, Ireland
}

\abstract{In this paper we take further steps towards developing the separation of variables program for integrable spin chains with $\gl(N)$  symmetry. By finding, for the first time, the matrix elements of the SoV measure explicitly we were able to compute correlation functions and wave function overlaps in a simple determinant form. In particular, we show how an overlap between on-shell and off-shell algebraic Bethe states can be written as a determinant. Another result, particularly useful for  AdS/CFT applications, is an overlap between two Bethe states with different twists, which also takes a determinant form in our approach. Our results also extend our previous works in collaboration with A.~Cavaglia and D.~Volin to general values of the spin, including the SoV construction in the higher-rank non-compact case for the first time.
}

\begin{document}
\maketitle
\section{Introduction}

More and more frequently integrability has been found to make its appearance in modern theoretical and mathematical physics in a wide range of problems, sometimes very unexpectedly (see the ongoing seminar series \cite{LIJC}). To a large extent this is because these models hide beautiful and very rich mathematical structures. They are universal enough to accommodate the complicated combinatorics of Feynman diagrams in 4D and 3D gauge theories and at the same time describe the motion of classically extended objects in curved spaces making the AdS/CFT correspondence almost manifest in many cases \cite{Beisert:2010jr}.

One of the main features of integrable models is separability of variables: the possibility of choosing a rather non-trivial coordinate system where the dynamics of the system simplifies dramatically and often can be solved exactly. At the quantum level this frequently implies the existence of a separation of variables (SoV) basis in the Hilbert space in which the wave function factorises into simple universal blocks. Ultimately, this factorisation should allow one to compute non-trivial expectation values of various observables with qualitatively much less computational effort than solving the same problem in a direct brute-force way.

SoV methods for quantum spin chains were pioneered by Sklyanin in \cite{Sklyanin:1984sb, Sklyanin:1987ih,Sklyanin:1991ss,SklyaninFBA,Sklyanin:1995bm}.
The generalisation to higher rank was also initiated by Sklyanin \cite{Sklyanin:1992sm} and later extended by Smirnov \cite{Smirnov2001}, but its explicit realization for models such as the Heisenberg XXX spin chain took more time and required new tricks to be developed.
One of the crucial steps was done in \cite{Gromov:2016itr}
where the basis factorizing the stationary wave functions of the spin chain was built explicitly and it was also understood into which blocks the wave function factorises. These findings started a new wave of results in the subject \cite{Liashyk:2018qfc,Maillet:2018bim,Ryan:2018fyo,Maillet:2018czd,Maillet:2018rto,Maillet:2019nsy,Maillet:2019hdq,Maillet:2019ayx,Ryan:2020rfk} which remains very active. Since a large class of integrable models can be related in one way or another to a generalisation or deformation of the Heisenberg spin chain -- even a complicated and powerful model such as planar ${\cal N}=4$ SYM which is essentially a version of the XXX spin chain with $PSU(2,2|4)$ symmetry \cite{Beisert:2005fw} -- this makes the problem of understanding how separation of variables works in models with high rank symmetry pivotal for progress in a number of directions.

In order to compute non-trivial expectation values between two stationary spin chain states one needs an additional ingredient -- the measure or, equivalently, the scalar product in the SoV basis. For $\gl(2)$ (rank 1) models it was found in numerous examples, but for higher rank cases it has remained unknown for a long time. Only recently it was obtained for spin chains at any rank, in a series of papers
\cite{Cavaglia:2019pow} and \cite{Gromov:2019wmz}.  The result for the scalar product was shown to take a very compact determinant form, and is also in perfect agreement with the general structure anticipated earlier from a semiclassical picture \cite{SmirnovClassM,SmirnovQuantM}.  This finding also encouraged development of an alternative approach to obtaining the measure via recursion relations for its elements in \cite{Maillet:2020ykb}, which to a large extent was shown to be equivalent. 
In this paper we extend the results of \cite{Cavaglia:2019pow,Gromov:2019wmz}  to non-compact spin chains with spin $\bs$, and also show how to use the measure to compute some very non-trivial overlaps and expectation values.

The SoV program has a strong motivation from the perspective of the AdS/CFT correspondence. At the moment it is well understood how to compute, using integrability, the exact non-perturbative spectrum of anomalous dimensions in planar ${\cal N}=4$ super Yang-Mills theory in 4D \cite{Gromov:2013pga}, and there are more examples of non-trivial QFTs which can be studied. At the same time much less is known about correlation functions -- for very long operators there is the powerful hexagon approach \cite{Basso:2015zoa}, which, however, fails at a certain loop order. 
There are strong indications that 
in order to access truly non-perturbative correlation functions one should apply SoV methods~\cite{Cavaglia:2018lxi,Giombi:2018qox,Giombi:2018hsx,Cavaglia:2020hdb}
and in this paper we show how very similar objects can be efficiently computed in the SoV framework.  Related overlaps were studied for spin chains in \cite{Belliard:2012pr,Belliard:2012is,Slavnov:2015qoa,Hutsalyuk:2016yii}, and determinant representations for them are not known beyond the rank-2 $\sl(3)$ case. Here we derive a (different) determinant form for the overlaps at any rank of the symmetry group. 

The paper is organised as follows. In Section \ref{sec2} we discuss some generalities of the Heisenberg spin chain which will be used throughout the main text and explain how to implement the SoV procedure for $\sl(2)$ spin chains. In Section \ref{sec:SoV} we extend the previous $\sl(2)$ results to the case of $\sl(3)$ case and discuss the details and subtleties which appear beyond the rank $1$ case in depth with the main focus being the construction of the SoV bases and the corresponding wave functions. In Section \ref{sec:measureint} we explain the functional approach to scalar products and demonstrate that this formalism matches what is expected from the operatorial construction of the wave functions in the SoV bases. In Section \ref{sec:measureint} we explain how to extract the measure in the SoV bases directly from the integral formalism and give an explicit formula for its matrix elements. In Section \ref{sec:secslN} we extend the previous construction of SoV states and wave functions to the $\sl(N)$ case and explain the technical aspects omitted from previous sections. In Section \ref{detrep} we apply our new techniques to the computation of various observables such as overlaps and correlation functions and show that they take a simple determinant form. Afterwards, in the Outlook section, we present various interesting avenues for future research. We also include various appendices to supplement the main text.

\section{Warm up: Notations and ${\mathfrak {sl}}(2)$ Example}\label{sec2}
In this section we collect our main definitions which we will be using in the remainder of the text.

\subsection{Heisenberg spin chain generalities}\la{sec:heis}
We begin by introducing the Heisenberg spin chain. Throughout the text we will assume that the spin chain is built out of spins in some highest-weight (HW) representation of $\mathfrak{sl}(N)$ with the same spin\footnote{Actually it is $-\bs$ which corresponds to the physical notion of spin -- the defining representation of $\sl(2)$ corresponds to $\bs=-\frac{1}{2}$.} ${\bf s}$ which we define below in terms of the highest-weight of the representation. Let ${\mathbb E}_{a,b}$ be the generators of $\mathfrak{sl}(N)$, satisfying
\beq\la{comL}
[{\mathbb E}_{ab},{\mathbb E}_{cd}]=\delta_{cb}{\mathbb E}_{ad}-\delta_{ad}{\mathbb E}_{cb}\;.
\eeq
We use the HW representation where ${\mathbb E}_{ab}|0\rangle = 0$ for $b>a$, ${\mathbb E}_{11}|0\rangle = -\bs$
and ${\mathbb E}_{bb}|0\rangle = +\bs$ for $b>1$, which we refer to as the spin $\bs$ representations (explicit form of the generators can be found in Appendix~\ref{oscrep}). This class of representations is a generalisation of the symmetric powers of the defining representation which are labelled by Young diagrams with a single row. Such finite-dimensional representations are obtained by setting $\bs=0,-\frac{1}{2},-1,\dots$. Highest-weight representations of $\sl(N)$ can be constructed on the space of polynomials in some number of variables. In general it is $\frac{N}{2}(N-1)$ \cite{Derkachov:2008aq} but for the class of representations we consider it reduces to $N-1$. Hence, our spin chain of length $L$ has $L(N-1)$ degrees of freedom. 

In terms of the Lax operator
\beq
\label{LaxGen}
{\cal L}_{a,b}(u)=u\, \delta_{ab}+i\, {\mathbb E}_{b,a}\;,
\eeq
and a constant {\it twist} matrix $\Lambda_{ab}$ we define the monodromy matrix
\beq\la{Tdefab}
{\bf T}_{cb}(u)=\sum_{b_i}{\cal L}^{(1)}_{c b_1}(u-\theta_1)
{\cal L}^{(2)}_{b_1b_2}(u-\theta_2)\dots
{\cal L}^{(L)}_{b_{L-1}b_{L}}(u-\theta_L)\Lambda_{b_L b}\;.
\eeq
The transfer matrix is then the following operator acting on the spin chain
\beq\la{Tdef}
{\mathbb T}(u)=\sum_c{\bf T}_{cc}(u)\;.
\eeq
The key property leading to the integrability of the model is that $[{\mathbb T}(u),{\mathbb T}(v)]=0$ and so the coefficients of the operator $\bbT(u)$ in the $u$ expansion are integrals of motion. However, ${\mathbb T}(u)$, which is the transfer-matrix in fundamental representation\footnote{of the auxiliary space}, is a polynomial of degree $L$ and does not contain the complete set of mutually commuting operators. To complete the set of the mutually commuting integrals of motion we have to additionally introduce the transfer matrices in all anti-symmetric representations, which we denote as 
${\mathbb T}_{a,1}$. We define\footnote{We define antisymmetrization as $A_{[i_1\dots i_k]}=\frac{1}{k!}\sum_{\sigma\in S_k} (-1)^\sigma A_{i_{\sigma(1)}\dots i_{\sigma(k)}}$.} 
\begin{equation}\la{La1}
    \mathcal{L}^{a,1}_{\bar{b},\bar{c}}={\cal L}^{b_1}_{\;\;\;\;[c_1}\(
u+i\frac{a-1}{2}\){\cal L}^{b_2}_{\;\;\;\;c_2}\(
u+i\frac{a-3}{2}\)
\dots
{\cal L}^{b_a}_{\;\;\;\;c_a]}\(
u-i\frac{a-1}{2}\)
\end{equation}
where $\bar{b}:=\{b_1,b_2,\dots,b_a\}$ and similarly for $\bar{c}$\footnote{The r.h.s. of \eq{La1} is antisymmetric in both $b_1,\dots,b_a$ and $c_1,\dots,c_a$.} and in the same way define the twist matrix in the anti-symmetric representation $\Lambda^{a,1}$
\begin{equation}
    {\Lambda}^{a,1}_{\bar{b},\bar{c}}={\Lambda}^{b_1}_{\;\;\;\;[c_1}{\Lambda}^{b_2}_{\;\;\;\;c_2}
\dots
{\Lambda}^{b_a}_{\;\;\;\;c_a]}\,.
\end{equation}
We then define the transfer matrix in the antisymmetric representation ${\mathbb T}_{ a,1}$ by using the corresponding ${\cal L}^{ a, 1}$
and $\Lambda^{ a, 1}$ in place of the the original fundamental representation building blocks in \eq{Tdefab}
\begin{equation}
    \bbT_{a,1}(u)=\sum_{\bar{b},\bar{b}_i}{\cal L}^{a,1(1)}_{\bar{b} \bar{b}_1}(u-\theta_1)
{\cal L}^{a,1(2)}_{\bar{b}_1 \bar{b}_2}(u-\theta_2)\dots
{\cal L}^{a,1(L)}_{\bar{b}_{L-1}\bar{b}_{L}}(u-\theta_L)\Lambda^{a,1}_{\bar{b}_L \bar{b}}\;.
\end{equation}
Defined in this way ${\mathbb T}_{a,1}(u)$ is a polynomial of degree $a\times L$. However, for $a>1$ one usually finds that ${\mathbb T}_{ a,1}(u)$ contains trivial factors, see Appendix \ref{oscrep}. In particular ${\mathbb T}_{N, 1}(u)$ is just proportional to the unit operator and is called quantum determinant. After removing trivial factors, each $\bbT_{a,1}$ for $a=1,\dots,N-1$ can be reduced to a polynomial of degree $L$ and hence the total number of non-trivial conserved charges is $L(N-1)$ -- precisely matching the number of degrees of freedom of the system and implying complete integrability.  

\paragraph{Twist matrix.}
The purpose of twisting is to remove degeneracies in the spectrum of the integrals of motion. For this a sufficient condition is that the twist matrix $\Lambda$ has pairwise distinct eigenvalues $\lambda_1,\dots,\lambda_N$. In principle the twist matrix could be any diagonalisable matrix with these eigenvalues, but for the purpose of this paper we assume a particular form of it and take
\beq\la{twistM}
\Lambda_{ij} =(-1)^{j-1}\chi_j \delta_{i1}+\delta_{i,j+1}
\eeq
where $\chi_j$ denote the elementary symmetric polynomials in $\lambda$ 
\begin{equation}\label{comptwist}
    \prod_{j=1}^n (t+\lambda_j)=\sum_{j=0}^n \chi_j\, t^{n-j}
\end{equation}
and we further constrain the eigenvalues with $\lambda_1\dots \lambda_n=1$.
One can perform a similarity transformation, simultaneously in the physical and auxiliary spaces, to bring the twist to diagonal form and the matrix which diagonalises $\Lambda$ is given by a simple Vandermonde-type matrix
\beq
S_{ij}=\lambda_i^{N-j+1}\;.
\eeq However, for the purpose of separation of variables \eqref{twistM} is the most convenient as we will see (see also \cite{Ryan:2018fyo,Ryan:2020rfk}).

\paragraph{Wave functions.}
We will frequently refer to the eivenvectors of the transfer matrices simply as wave functions, and denote them as $|\Psi\rangle$ for the right eigenvectors and $\langle\Psi|$ for the left eigenvectors. We denote the corresponding eigenvalues as $T_{a,1}(u)$
\beq
{\mathbb T}_{a,1}|\Psi\rangle =
T_{a,1}|\Psi\rangle\;\;,\;\;
\langle\Psi|{\mathbb T}_{a,1} =
\langle\Psi|T_{a,1}
\;.
\eeq

\paragraph{Spectrum and Q-functions.}
The expressions for the transfer matrix eigenvalues can be conveniently written in terms of the so-called Q-functions or Q-polynomials.
We define the Q-functions to be ``twisted" 
polynomials 
\beq
\label{qgen}
Q_{i_1,\dots,i_m}(u) = \(\lambda_{i_1}\dots \lambda_{i_m}\)^{i u}\prod_{k=1}^{M_{i_1,\dots,i_m}}
(u-u^{i_1,\dots,i_m}_k)\;.
\eeq
The numbers $u^{i_1,\dots,i_m}_k$ are the Bethe roots and they can be found, for example, from the Bethe ansatz (BA) equations which follow the following pattern
\beqa\la{BAEgeneral}
\frac{Q_\theta(u^1_k-i \bs)}
{Q_\theta(u_k^1+i \bs)}
&=&\quad\quad\quad\quad\quad\quad-\frac{Q_1(u^1_k+i)}{Q_1(u^1_k-i)}
\frac{Q_{12}(u^1_k-\tfrac{i}{2})}{Q_{12}(u^1_k+\tfrac{i}{2})}\;\;,\;\;k=1,\dots,M_{1}\\
\nn 1&=&-
\frac{Q_{1}(u^{12}_k-\tfrac{i}{2})}{Q_{1}(u^{12}_k+\tfrac{i}{2})}
\frac{Q_{12}(u^{12}_k+i)}{Q_{12}(u^{12}_k-i)}
\frac{Q_{123}(u^{123}_k-\tfrac{i}{2})}{Q_{123}(u^{123}_k+\tfrac{i}{2})}
\;\;,\;\;k=1,\dots,M_{12}\\
\nn &&\vdots
\eeqa
and we have introduced the Baxter polynomial $Q_\theta(u)=\prod_{\alpha=1}^L (u-\theta_\alpha)$. 
As the above BA equations originate from a nesting procedure, they contain some arbitrariness. 
Namely, at the $m^{\rm th}$ step of nesting one can choose one of $N-m$ ``vacua". This arbitrariness 
results in the existence of a large number of equivalent BAs related by the duality relations
\beq\la{DualityB}
Q_{I,i}(u+\tfrac i2)Q_{I,j}(u-\tfrac i2)-Q_{I,j}(u+\tfrac i2)Q_{I,i}(u-\tfrac i2)\;\propto\; Q_I(u) Q_{I,i,j}(u)\;
\eeq
where $I$ is a multi-index containing $1$.
We also assume the boundary condition $Q_{12\dots N}=1$ in the recursion relation \eq{DualityB}. 
By evaluating \eq{DualityB} 
at $u= u_{I,i}+\tfrac{i}{2}$ and $u= u_{I,i}-\tfrac{i}{2}$ and then dividing 
the results by each other we obtain the general form of the BA for auxiliary Bethe roots\footnote{In some degenerate cases the set of \eq{BAEgeneral} may not produce one solution for each state. In this case one could either consider the full set \eq{BAEgeneral2} or use the Baxter equation instead.}
\beq\la{BAEgeneral2}
1=-
\left.\frac{Q_{I}^-}{Q_{I}^+}
\frac{Q_{I,i}^{++}}{Q_{I,i}^{--}}
\frac{Q_{I,i,j}^-}{Q_{I,i,j}^+}\right|_{u=u_k^{I,i}}\;\;,\;\;k=1,\dots,M_{I,i}\;.
\eeq
Above we introduced the standard notation
\beq
f^\pm = f(u\pm\tfrac i2)\;\;,\;\;
f^{\pm\pm} = f(u\pm i)\;\;,\;\;
f^{[a]} = f(u+\tfrac {ia}{2})\;.
\eeq
Having the Q-functions defined one can express all eigenvalues $T_{a,1}(u)$ in terms of Q's.
In particular
\beq\la{T11sln}
T_{1,1}(u) = {Q^{[-2\bs]}_{\theta}}
\frac{
Q_{1}^{--}
}{Q_{1}}
+
Q_{\theta}^{[+2\bs]}
\frac{
Q_1^{++}
}{Q_1}
\frac{
Q_{12}^{-}
}{Q^{+}_{12}}
+\dots+
Q_{\theta}^{[+2\bs]}
\frac{
Q_{1\dots,{i-1}}^{[+i]}
}{Q^{[-2+i]}_{1\dots,{i-1}}}
\frac{
Q_{1 \dots i}^{[-3+i]}
}{Q^{[-1+i]}_{1 \dots i}}+\dots\;.
\eeq
The above expression is indeed a polynomial
of degree $L$
when the BA equations \eq{BAEgeneral} and \eq{BAEgeneral2} are satisfied.
General expressions for $T_{a,1}(u)$ can be found in appendix~\ref{app:slNbax}.

\subsection{Warm up example -- $\sl(2)$ case}\la{sec:sl2}
To give a simple example of the known construction we described above and to set the stage for the more complicated and original $\sl(3)$ and $\sl(N)$ cases in this section we consider the simplest $\sl(2)$ case, very well known in the existing literature, see for example~\cite{Derkachov:2001yn,Derkachov:2002tf}.

\paragraph{Representation.}
We are considering general non-compact HW representations, where each site carries the spin $\bs$ representation. The $\sl(2)$  raising and lowering operators are given respectively by
\begin{equation}
    {\mathbb E}_{12}=\partial_x,\quad {\mathbb E}_{21}=-x^2\partial_x -2 \bs\ x
\end{equation}
and the Cartan generators are 
\begin{equation}
    {\mathbb E}_{11}=-x \partial_x-\bs,\quad  {\mathbb E}_{22}=x \partial_x+\bs\;.
\end{equation}
The representation space is then the space $\mathbb{C}[x]$ of polynomials in $x$ which is spanned by monomials $x^n$, $n\geq 0$. For generic $\bs$ this space is irreducible and infinite-dimensional. However, for special values of $\bs$, in particular when $\bs\in\{0,-\frac{1}{2},-1,\dots\}$ the representation becomes reducible with a finite-dimensional irreducible part. 
It is obvious that the raising operator annihilates the state given by a constant and so the highest-weight state is simply given by the polynomial $1$. 

\paragraph{Scalar product.}
We define the scalar product $\langle\cdot|\cdot\rangle$ on our Hilbert space by introducing an orthonormal basis $e_n$, $n=0,1,2,\dots$ and imposing $\langle e_n| e_m\rangle = \delta_{nm}$. Naively, one would take $e_n=x^n$ as a normalised basis, however, this will not result in correct conjugation properties for the generators. Instead one can define $e_n=c_n x^n$ and require the matrix of the operator ${\mathbb E}_{12}$ to be minus transposed\footnote{It may be tempting to instead require that the operators are related by transposition instead of minus transposition. However this is not compatible with the normalisation of our generators - our choice corresponds to conjugating the usual defining representation generators with a simple diagonal matrix.} of the matrix of the operator ${\mathbb E}_{21}$, i.e. ${\mathbb E}_{21}=-{\mathbb E}_{12}^T$, where transposition is defined in the usual way $\langle \Psi_1|O \Psi_2\rangle = \langle O^T\Psi_1| \Psi_2\rangle$. Together with the requirement $e_0=1$ this fixes $c_n$ and we find\footnote{With this definition our representation is equivalent to the well known harmonic oscillator construction.}
\beq
e_n=x^n\sqrt{\frac{\Gamma (n+2 s)}{\Gamma (n+1) \Gamma (2
   s)}}\;.
\eeq

At this point we should mention that our scalar product does not involve any complex conjugation and the scalar product is linear in both arguments.
To promote it to Hermitian conjugation we need to impose that $\bs$ is real. Furthermore, in order for Hermitian conjugation to lift to the spin chain Hilbert space one should make certain choices on the reality of the parameters of the model such as inhomogeneities $\theta_\alpha$ and twists $\lambda_i$. Instead we will view our scalar product as simply defining the action of a dual vector on a vector.

\paragraph{Bethe ansatz and Baxter equation.} 
There is only one non-trivial transfer matrix in anti-symmetric representations, the fundamental representation, whose eigenvalue is given by
\beq
\label{Tsl2Q}
T_{1,1}(u) = {Q^{[-2\bs]}_{\theta}}
\frac{
Q_{1}^{--}
}{Q_{1}}
+
Q_{\theta}^{[+2\bs]}
\frac{
Q_1^{++}
}{Q_1}
\;.
\eeq
The above expression becomes polynomial when the BA equations are satisfied
\beq\la{BAEgeneralSL2}
\frac{Q_\theta(u^1_k-i \bs)}
{Q_\theta(u_k^1+i \bs)}
=-\frac{Q_1(u^1_k+i)}{Q_1(u^1_k-i)}
\;\;,\;\;k=1,\dots,M_{1}\;.
\eeq

\paragraph{Twist and the ground state.}
For the $\sl(2)$ case the expression \eqref{twistM} is simply given by
\begin{equation}\label{MCT2}
    \Lambda=\left(\begin{array}{cc}
       \lambda_1+\lambda_1^{-1}  & -1 \\
        1 & 0
    \end{array}\right)\;.
\end{equation}
For diagonal twist the ground state state, corresponding to $Q_1(u)=\lambda_1^{i u}$, would be just a constant polynomial. However, since our twist is non-trivial the constant polynomial gets transformed into the following expression
\begin{equation}\la{sl2omega}
 \ket{\Omega}=\prod_{\alpha=1}^L\lambda_1^{ i\theta_\alpha-\bs}{\(1+\frac{1}{\lambda_1}x_\alpha\)^{-2\bs}}\;\;,\;\;
  \bra{\Omega}=\prod_{\alpha=1}^L\lambda_1^{ i\theta_\alpha-\bs}{\(1+\frac{1}{\lambda_1}x_\alpha\)^{-2\bs}}
 \;.
\end{equation}
The normalisation here is chosen for later convenience as we will see soon. Whereas these states are clearly not polynomial, they can be expanded into an infinite series. The Hilbert space should be understood as a completed space of polynomials, where such analytic functions which are regular at the origin are included.
The scalar products of such states are computed as a limit of a scalar product of the truncated series, which additionally imposes (for $\bs>0$) that the convergence radius of the series should be $\ge 1$.
In our particular case convergence is guaranteed for $|\lambda_1|>1$ which we assume to be satisfied\footnote{Typically our results are analytic in the twist so one may be able to go to other regimes by a careful analytic continuation.}. Then the overlap of $\langle\Omega|$ with $|\Omega\rangle$ is
\beq\la{OO}
\bra{\Omega}\Omega\rangle= \left(1-\frac{1}{\lambda _1^2}\right)^{-2
   \bs}\times \prod_{\alpha=1}^L\lambda_1^{ 2i\theta_\alpha-2\bs}\;.
\eeq

\paragraph{Excited states.}
The excited states (those with non-trivial Bethe roots) can be obtained by consecutive action of the ${\bf B}(u)={\bf T}_{12}(u)$ operator
\beq\la{Bom}
|\Psi\rangle\; \propto \prod_{k=1}^{M_1}{\bf B}(u^1_k)|\Omega\rangle\;.
\eeq
In the case of ${\mathfrak{sl}}(2)$ the left eigenvectors of the transfer matrix can be built in the same way\footnote{In the earlier literature operator ${\bf C}(u)=T_{21}(u)$ was used to create the left eigenstates. The reason for this was that in the case of diagonal twist ${\bf B}(u)$ would annihilate the left ground state. One of course can start from our current construction and diagonalise the twist by a global rotation, then, however, one will get a ${\bf B}^{\rm good}(u)$ operator, which is a linear combination of all $4$ matrix elements ${\bf T}_{ab}$, like described in~\cite{Gromov:2016itr,Sklyanin:1989cg}.}
\beq
\langle \Psi| \;\propto\;
\langle\Omega |\prod_{k=1}^{M_1}{\bf B}(u^1_k)\;.
\eeq
\paragraph{SoV basis.}
Another advantage of the non-diagonal twist \eqref{twistM} is that ${\bf B}(u)$ is diagonalisable. 
Furthermore, an important property of our twist, which we will use below, is that the ${\bf B}(u)$ operator does not actually depend on $\lambda$ -- indeed if we denote the untwisted\footnote{corresponding to the case where the twist matrix is the identity operator} monodromy matrix elements by $\lT_{ij}$ then the twisted monodromy matrix is given by 
\begin{equation}
    \left(\begin{array}{cc}
      \bT_{11}   &  \bT_{12}\\
       \bT_{21}  & \bT_{22}
    \end{array}\right)=  \left(\begin{array}{cc}
      \lT_{11}   &  \lT_{12}\\
       \lT_{21}  & \lT_{22}
    \end{array}\right)\left(\begin{array}{cc}
       \lambda_1+\lambda_1^{-1}  & -1 \\
        1 & 0
    \end{array}\right)
\end{equation}
and hence we see that ${\bf T}_{12}=-\lT_{11}$. In fact this is how this twist was initially introduced in \cite{Ryan:2018fyo}.
The eigenvectors of ${\bf B}(u)$ form left and right SoV bases, which we denote it as $\langle \svx|$ and $|\svx \rangle$. The ${\bf B}(u)$ operator has a very simple spectrum
\beq\la{BUX}
{\bf B}(u)|\svx \rangle = -\prod_{\alpha=1}^L(u-\theta_\alpha -i n_\alpha-i \bs)|\svx \rangle\equiv 
 -\prod_{\alpha=1}^L(u-\svx_\alpha)|\svx \rangle,\quad n_\alpha=0,1,2,\dots\;
\eeq
so that $\svx_\alpha$ take values
\beq
\label{xvalsl2}
    \svx_\alpha=\theta_\alpha+i\bs+in_\alpha \; , \ \ \ n_\alpha=0,1,2,\dots \; .
\eeq
The SoV states can be uniquely labelled by the non-negative integers $n_\alpha$ so one can also denote $|\svx \rangle=|n_1,\dots, n_L\rangle$.
The SoV states $|\svx \rangle$ are the homogeneous polynomials of degree $\sum_\alpha n_\alpha$ -- we will also refer to this number as SoV charge. 
There is only one SoV charge zero state, which we call the SoV vacuum and denote as $|0\rangle$. 
Finally, we define ${\bf b}(u)= -{\bf B}(u)$ in order to make this operator a monic polynomial. 

\paragraph{Normalisation of the wavefunctions and SoV states.}
The reason we mainly use the non-diagonal twist \eqref{twistM} in this paper is to make the SoV basis simple. In particular the SoV states are simply polynomials and the SoV vacuum is a constant. To fix the normalisation we define
\beq
\label{00vacsl2}
|0\rangle = 1\;\;,\;\;\langle 0 | = 1\;.
\eeq
The transfer matrix ground states \eq{sl2omega} are already normalised so that
\beq
\langle \Omega |0\rangle = \langle 0 |\Omega\rangle = \prod_{\alpha=1}^L\lambda_1^{i\svx_0^\alpha}\;\;,\;\;\svx^\alpha_0\equiv \theta_\alpha+i\bs\;.
\eeq
We fix the normalisation of the excited states by  
\beq\la{Psib}
|\Psi\rangle = (-1)^{M_1 L}\prod_{k=1}^{M_1}{\bf b}(u^1_k)|\Omega\rangle
\;\;,\;\;
\langle \Psi|= \langle \Omega |(-1)^{M_1 L}\prod_{k=1}^{M_1}{\bf b}(u^1_k)\;.
\eeq
It remains to fix the normalisation of the excited SoV states $\langle \svx|$
and $|\svx \rangle$. We notice that as a consequence of \eq{Psib}
and \eq{BUX} we have
\beq
\langle \svx|\Psi\rangle =\langle \svx|\Omega\rangle \prod_{\alpha=1}^L\prod_{k=1}^{M_1}(\svx_\alpha - u^1_k)
\eeq
and similar for the $\langle \Psi|\svx \rangle$.
We fix the remaining scale of the SoV states by requiring 
\beq\la{lxover}
\langle \svx|\Omega\rangle = 
\langle \Omega|\svx \rangle 
= \prod_{\alpha=1}^L\lambda_1^{i \svx^\alpha}\;.
\eeq
Even though this normalisation of the SoV basis does not look very natural, it actually makes the SoV states independent of $\lambda_1$ as was initially shown in \cite{Ryan:2018fyo}.
For example for the case $L=1$ the SoV states have to be proportional to $x_1^{n_\alpha}$, fixing the coefficient according to \eq{lxover} we get 
\beq
\langle \svx |_{L=1} = (-x_1)^{n_{\alpha}}\;,
\eeq
which indeed does not depend on the twist.
For the most general proof of this see \cite{Ryan:2018fyo} and some details also in in Appendix~\ref{antipodeapp}.

Finally, in this normalisation we get
\beq
\label{sovwfsl2}
\langle \svx|\Psi\rangle =
\langle \Psi|\svx \rangle =\prod_{\alpha=1}^L Q_1(\svx^\alpha)\;,
\eeq
which indeed shows that in the SoV basis the wave functions factorize into a product of Q-functions.
For the general $\mathfrak{sl}(n)$ case we will be using analogous conventions for normalisations.

\paragraph{Measure and scalar products in SoV basis.}
Since the normalisation of the SoV states is completely fixed,
their overlap could be a non-trivial number. 
As they are left and right eigenstates of the same operator ${\bf B}(u)$
 they are orthogonal. The overlap $\langle \svx| \svx \rangle\equiv M_\svx^{-1}$ is nontrivial and is given by 
 
\beq
\label{mucompact}
    M_{n_1,\dots,n_L}=\frac{\Delta(\{\svx^\alpha\})}{\Delta(\{\theta_\alpha\})}\prod_{\alpha=1}^L\frac{r_{\alpha,n_\alpha}}{r_{\alpha,0}}
\eeq
where we defined\footnote{the factor $-\frac{1}{2\pi}$ in \eq{resm} is chosen for future convenience in section \ref{sec:expmeasure}}
\beq\la{resm}
r_{\alpha,n}=-\frac{1}{2\pi}\prod_{\beta=1}^L
(n+1-i \theta _{\alpha}+i\theta_{\beta})_{2 \bs-1}\;
\eeq
and $(f)_s$ is the Pochhammer function\footnote{For general values it is defined as $(f)_s=\frac{\Gamma (f+s)}{\Gamma (f)}$, in addition for a  particular case
when the arguments are integers and their sum is zero we define $(-n)_n=(-1)^n \Gamma (n+1)$. This is how it is defined in {\it Mathematica} in particular. }
while $\Delta$ is the Vandermonde determinant,
\beq
    \Delta(z_1,\dots,z_n)=\prod_{i<j}(z_i-z_j) \;.
\eeq
This result will be derived in section \ref{sec:intsl2} using an integral representation of the scalar product. At the same time, we expect it to match the overlaps $\bra{\svx}\svx\rangle$ and we we have checked this directly for all states for $L=1$ and also for states with SoV charge $\leq 2$ for $L=2$. For example, by explicitly diagonalizing the ${\bf B}(u)$ operator for the $L=2$ case and considering the states with $n_1=1,n_2=0$, we find that the right eigenstate (i.e. the  $\ket{\svx}$ state) reads
\beq
    \ket{1,0}=c_1 x_1\;,
\eeq
while the left one  (i.e. the  $\bra{\svx}$ state) is
\beq
    \bra{1,0}=c_2\(x_1+\frac{2i\bs}{\theta_{12}}x_2\) \;,
\eeq
where $\theta_{12}\equiv \theta_1-\theta_2$. The normalisations $c_1,c_2$ are fixed by requiring \eq{lxover} which gives $c_1=-1, \ c_2=\frac{i\theta_{12}}{2\bs-i\theta_{12}}$ . Finally, computing the inverse of their  overlap we find
\beq
    M_{1,0} = \frac{1}{\bra{1,0}1,0\rangle} = \frac{2\bs(\theta_{12}+2i\bs)}{\theta_{12}}\;,
\eeq
in full agreement with \eq{mucompact}.

Knowing the measure \eq{mucompact}, we can write the resolution of identity as
\beq
\label{sl2comp}
    \sum_\svx M_\svx\ket{\svx}\bra{\svx}={ 1} \;,
\eeq
which is the key completeness relation crucial for the computation of various scalar products as we will see below. As an example, we can write the overlap of left and right transfer matrix eigenstates $\bra{\Psi_A}$ and $\ket{\Psi_B}$ corresponding to Q-functions $Q_1^A$ and $Q_1^B$ as
\beqa
\label{PsiABsl2}
    \bra{\Psi_A}\Psi_B\rangle
    &=&\sum_\svx
M_\svx
\langle \Psi_A |\svx \rangle 
\langle \svx | \Psi_B\rangle
\\ \nn
&=&
\sum_\svx M_\svx\(
\prod_{\alpha=1}^L Q_1^A(\svx^\alpha)\)\(
\prod_{\alpha=1}^L Q_1^B(\svx^\alpha)\) \;, 
\eeqa
where we used the SoV wavefunctions \eq{sovwfsl2}.

\paragraph{Overlaps of off-shell states.}
Another representation for the measure \eq{mucompact} is
\beq
M_{n_1,\dots,n_L}=\frac{d(n_1,\dots,n_L)}{d(0,\dots,0)}\;\;,\;\;d(n_1,\dots,n_L)=\det_{\alpha,\beta}\frac{\left(\theta _{\alpha }+i n_{\alpha }\right){}^{\beta -1} \left(n_{\alpha
   }+1\right)_{2 \bs-1}}{\prod\limits_{\gamma\neq\alpha}\left(
   -n_{\alpha }+i \theta _{\alpha
   }-i \theta _\gamma\right)_{1-2 \bs}}\;,
\eeq
which is equivalent to \eq{mucompact}. The fact that it can be written as a determinant is quite significant, as this implies that some overlaps can be also written as determinants as well.
For example, let us demonstrate that the overlap of any two states
$\langle \Phi |$ and $| \Theta\rangle$
, which satisfy the separability condition i.e.
\beq
\langle \Phi | \svx \rangle = \prod_{\alpha=1}^L F(\svx^\alpha)\;\;,\;\;
\langle \svx | \Theta\rangle = \prod_{\alpha=1}^L G(\svx^\alpha)\;
\eeq
can be written in the form of determinant.
Indeed
\beq
\langle \Phi  | \Theta\rangle
=
\sum_\svx
M_\svx
\langle \Phi |\svx \rangle 
\langle \svx | \Theta\rangle
=
\frac{1}{d_0}
\det_{\alpha,\beta}
\sum_{n_\alpha=0}^\infty
\frac{F(\svx^\alpha)G(\svx^\alpha)\left(\theta _{\alpha }+i n_{\alpha }\right){}^{\beta -1} \left(n_{\alpha
   }+1\right)_{2 \bs-1}}{\prod\limits_{\gamma\neq\alpha}\left(
   -n_{\alpha }+i \theta _{\alpha
   }-i \theta _\gamma\right)_{1-2 \bs}}\;,
\eeq
where $d_0\equiv d(0,\dots,0)$.
Examples of such states are off-shell algebraic Bethe states
with two different twists
\beq
|\Theta\rangle =(-1)^{K_1L} \prod_i^{K_1} {\bf b}(v_i)|\Omega_{\lambda_1}\rangle\;\;,\;\;
\langle\Phi| = (-1)^{K_2L}\langle \Omega_{\tilde\lambda_1}|\prod_i^{K_2} {\bf b}(w_i)
\eeq
with
$G(\svx^\alpha)=\lambda_1^{i\svx_\alpha}\prod_{i=1}^{K_1}(\svx^\alpha-v_i)$
and 
$F(\svx^\alpha)=\tilde\lambda_1^{i\svx_\alpha}\prod_{i=1}^{K_2}(\svx^\alpha-w_i)$.
In particular, for the simplest case $K_1=K_2=0$ and $L=1$ we get
\beq
\langle \Omega_{\tilde\lambda_1}|
\Omega_{\lambda_1}\rangle = \frac{1}{d_0}\prod_{\alpha=1}^L(\lambda_1\tilde\lambda_1)^{ i\svx_0^\alpha}\sum_{n=0}^\infty
\frac{(n+1)_{2 s-1}}{
(\lambda_1\tilde\lambda_1)^{n}
}=\(1-\frac{1}{\lambda_1\tilde\lambda_1}\)^{-2\bs}\prod_{\alpha=1}^L(\lambda_1\tilde\lambda_1)^{ i\svx_0^\alpha}\;,
\eeq
which correctly extends the relation \eq{OO}.

The main goal of this paper is to show how to generalise these types of 
results to the $\mathfrak{sl}(N)$ case.
In the next section we will extend our consideration to $\mathfrak{sl}(3)$, mostly following the same steps as in this section.

 \section{$\sl(3)$ spin chain}\la{sec:SoV}
 In this section we describe our general construction for the case of $\sl(3)$. For brevity we will leave the proof of certain technical details until we describe the general $\sl(N)$ case. 
 The main purpose of this section is to demonstrate the main ideas and our techniques. In comparison to the $\sl(2)$ case there will be new ingredients involved, such as the ${\bf C}(u)$ operator. Also, unlike in the $\sl(2)$ case the SoV measure is unavoidably non-diagonal\footnote{Different SoV bases which lead to a diagonal measure were constructed in \cite{Maillet:2020ykb} but to our knowledge these
 bases do not diagonalise any well-defined operators such as $\bB$ and $\bC$, nor can the measure be efficiently extracted from the Baxter equation.} for the general case, but it nevertheless leads to simple determinant expressions for various correlators and overlaps as we will see. 
 \subsection{Representation}
 As we explained in the previous section in this paper we consider HW representations on the space of polynomials. 
 More specifically we consider the representation of spin ${\bf s}$, 
 which we define in terms of the operators acting on a space of polynomials
  $\mathbb{C}[x,y]$ in two variables as follows:\\
 {\bf Raising operators}
 \begin{equation}\label{sl3raise}
     {\mathbb E}_{12}=\partial_x,\quad {\mathbb E}_{13}=\partial_y,\quad {\mathbb E}_{23}=x\partial_y
 \end{equation}
{\bf Lowering operators}
 \begin{align}\label{sl3lower}
     {\mathbb E}_{21} & =-x^2 \partial_x-x y \partial_y-2\bs\, x \\
     {\mathbb E}_{31} & =-y^2\partial_y-y x\partial_x-2\bs\, y\\
     {\mathbb E}_{32} & =+y\partial_x 
     \end{align}
{\bf Cartan generators} 
 \begin{equation}\label{sl3cartan}
     \begin{split}
         {\mathbb H}_1 & =-2x\partial_x-y\partial_y-2\bs \\
         {\mathbb H}_2 & =-y \partial_y+x\partial_x \;.
     \end{split}
 \end{equation}
 It is also convenient to repackage the $\sl(3)$ Cartan generators into $\gl(3)$ Cartan generators
 \begin{align}
     {\mathbb E}_{11}& =\frac{2}{3}{\mathbb H}_1+\frac{1}{3}{\mathbb H}_2+\frac{\bs}{3} \\
     {\mathbb E}_{22}& =\frac{1}{3}{\mathbb H}_1-\frac{1}{3}{\mathbb H}_2+\frac{\bs}{3} \\
     {\mathbb E}_{33}& =-\frac{1}{3}{\mathbb H}_1-\frac{2}{3}{\mathbb H}_2+\frac{\bs}{3} \;.
 \end{align}
In this way the generators satisfy  the commutation relations \eq{comL}.
The HW state is simply a constant polynomial $|0\rangle = 1$ and the diagonal generators ${\mathbb E}_{aa}$ have the eigenvalues $\{-\bs,+\bs,+\bs\}$ on the HW state.
The eigenstates of the Cartan generators are homogeneous polynomials in $x$ and $y$ and
the lowering generators increase the degree by $1$.
 
\paragraph{Lax operators} 
For $\sl(3)$ there are two non-trivial Lax operators $\mathcal{L}^{a,1}$ in anti-symmetric representations. Denoting $\mathcal{L}^{1,1}$ as simply $\mathcal{L}$, it is an easy calculation to show directly from the definition of $\mathcal{L}^{a,1}$ \eqref{La1} that 
\begin{equation}
    \mathcal{L}^{2,1}(u) = \left(u+i\,\bs-\frac{i}{2}\right) \left[-\mathcal{L}\left(-u-\frac{i}{2}\right)\right]^t\,,
\end{equation}
where $t$ denotes the transpose of $\mathcal{L}$ written as a $3\times 3$ matrix. 

\subsubsection{Scalar product}
Like in the $\mathfrak{sl}(2)$ case we define the scalar product by introducing an 
orthonormal basis
\beq\la{basis}
e_{n,k}\equiv 
x^n y^k  \sqrt{\frac{\Gamma\left(n+k+2\bs\right)}{ \Gamma(n+1)\Gamma(k+1) \Gamma(2\bs)}}
\eeq
and define the bracket $\langle \cdot|\cdot\rangle$ by 
\beq\la{biform}
\langle e_{nk}, e_{n'k'}\rangle\equiv \delta_{nn'}
\delta_{kk'}\;.
\eeq
As any polynomial can be expressed as a finite linear combination of $e_{nk}$, this defines the scalar product on the space of all polynomials of $x$ and $y$. It also defines the scalar product between a polynomial and any function analytic at the origin. In order to have the scalar product between two analytic functions finite one should impose some constraints on the convergence radius. More precisely we need to require the limit of the scalar products between two truncated expansions to have a finite limit.

Like in the case of $\mathfrak{sl}(2)$, the factor of gamma functions in \eq{biform} is needed to ensure that the generators ${\mathbb E}_{ab}$ are either self-conjugate or anti-self-conjugate to ${\mathbb E}_{ba}$. This requirement fixes \eq{basis} completely (up to an overall real factor).
Similar to the $\sl(2)$ case we find the following conjugation properties of the generators ${\mathbb E}^T_{12}=-{\mathbb E}_{21}$, ${\mathbb E}^T_{13}=-{\mathbb E}_{31}$ and ${\mathbb E}^T_{23}=+{\mathbb E}_{32}$. Finally, since the Cartan generators act diagonally we also have ${\mathbb E}^T_{aa}=+{\mathbb E}_{aa}$ for $a=1,2,3$.

\subsection{Transfer matrix and Integrability}\la{sec:sl3int}
Having the representation defined we follow the general steps outlined in section \ref{sec:heis}. In this section we explicitly write some of the expressions from section \ref{sec:heis} and give a few more details specific to the ${\mathfrak{sl}}(3)$ case.

For the case $N=3$ the twist matrix \eq{twistM} becomes
\beq\la{twistform}
\Lambda = \left(
\begin{array}{ccc}
 \lambda _1+\lambda _2+\lambda_3 & -\frac{1}{\lambda_1}-\frac{1}{\lambda _2}-\frac{1}{\lambda _3} & 1 \\
 1 & 0 & 0 \\
 0 & 1 & 0 \\
\end{array}
\right)\;.
\eeq
As before it can be brought to diagonal form and has eigenvalues $\lambda_1,\;\lambda_2$ and $\lambda_3\equiv\frac{1}{\lambda_1\lambda_2}$.

The transfer matrix \eq{Tdefab} is a differential operator in $2L$
variables $x_\alpha, y_\alpha,\;\alpha=1,\dots,L$. The complete set of conserved quantities is contained in the two non-trivial transfer matrices in fundamental $\bbT_{1,1}(u)$ and anti-fundamental $\bbT_{2,1}(u)$ representations. The transfer matrix $\bbT_{3,1}(u)$ correspnding to the totally antisymmetric representatio does not contain any new conserved quantities, but is a non-trivial function of $u$
\beq
 {\mathbb T}_{3,1}(u)=Q_\theta(u-i \bs+i)  Q_\theta(u+i \bs-i)Q_\theta(u+i \bs)\;{\mathbb I}\;,
\eeq 
where ${\mathbb I}$ denotes the identity operator.

\paragraph{Bethe Ansatz and the Transfer Matrix Eigenvalues.}
The set of Bethe Ansatz equations (BAE), relevant for our discussion,
is the following
\beqa
\frac{Q_\theta(u_k-i \bs)}
{Q_\theta(u_k+i \bs)}
&=&-\frac{Q_1(u^1_k+i)}{Q_1(u^1_k-i)}
\frac{Q_{12}(u^1_k-\tfrac{i}{2})}{Q_{12}(u^1_k+\tfrac{i}{2})}\;\;,\;\;k=1,\dots,M_{1}\\
1&=&-
\frac{Q_{12}(u^{12}_k+i)}{Q_{12}(u^{12}_k-i)}
\frac{Q_{1}(u^{12}_k-\tfrac{i}{2})}{Q_{1}(u^{12}_k+\tfrac{i}{2})}\;\;,\;\;k=1,\dots,M_{12}\\
1&=&-
\frac{Q_{13}(u^{13}_k+i)}{Q_{12}(u^{13}_k-i)}
\frac{Q_{1}(u^{13}_k-\tfrac{i}{2})}{Q_{1}(u^{13}_k+\tfrac{i}{2})}\;\;,\;\;k=1,\dots,M_{13}
\eeqa
where the twisted Baxter polynomials are $Q_1=\lambda_1^{i u}\prod_{i=1}^{M_1}(u-u^{1}_k)$, $Q_{12}=(\lambda_1\lambda_2)^{i u}\prod_{i=1}^{M_{12}}(u-u^{12}_k)$
and  $Q_{13}=(\lambda_1\lambda_3)^{i u}\prod_{i=1}^{M_{13}}(u-u^{13}_k)$ (from \eq{DualityB} one should have $M_1=M_{12}+M_{13}$).
Note that for the purposes of finding the spectrum the first two equations are usually sufficient. However, for the SoV construction we describe below one also needs to find $Q_{13}$, which is a dualised Baxter polynomial, corresponding to an alternative nesting path in the nested BAE terminology.

Once the Q-functions are known, the eigenvalues of the transfer matrices are given by simple expressions (e.g. \eq{T11sln}). It will be convenient to introduce the notation
\beqa
\label{tau1Q}
\tau_1&=&
Q_\theta(u-i\bs)
\frac{Q_1^{--}}{Q_1}
+
Q_\theta(u+i\bs)
\frac{Q_1^{++}}{Q_1}
\frac{Q_{12}^-}{Q_{12}^+}
+
Q_\theta(u+i\bs)
\frac{Q_{12}^{[+3]}}{
Q_{12}^{+}
}\;,\\
\label{tau2Q}
\tau_2&=&
Q_\theta(u-i\bs)
\frac{Q_{12}^{[-3]}}{Q_{12}^{-}}
+
Q_\theta(u-i\bs)
\frac{Q_1^{--}}{Q_1}
\frac{Q_{12}^+}{Q_{12}^-}
+
Q_\theta(u+i\bs)
\frac{Q_{1}^{++}}{
Q_{1}
}\;.
\eeqa
In terms of these functions the eigenvalues of the transfer matrices become
\beqa\la{PSi}
{\mathbb T}_{1,1}(u)|\Psi\rangle
&=&\tau_1(u)
|\Psi\rangle\;,\\
\label{Ttau3}
{\mathbb T}_{2,1}(u)|\Psi\rangle
&=&
Q_\theta(u+i\bs-\tfrac i2)\;
\tau_2(u+\tfrac{i}{2})
|\Psi\rangle\;.
\eeqa

\paragraph{Baxter TQ-relations.}
As there is certain confusion in the literature about the completeness of the Bethe ansatz\footnote{Recently completelness has been proven for supersymmetric spin chains in the defining representation \cite{Chernyak:2020lgw}.}, one may like to have an alternative way to define the Baxter polynomials which can be done by means of the Baxter TQ-relations.

The Baxter polynomials $Q_1,\;Q_{12},\;Q_{13}$ and simultaneously $\tau_1$ and $\tau_2$ can be determined by requiring
their polynomiality (up to the twist factor $\lambda_1^{i u}$ in $Q_1$ and $(\lambda_1\lambda_a)^{i u}$ in $Q_{1a}$ for $a=2,3$) and that they satisfy the following finite-difference equations
\beq
\label{Bax1sl3}
   {Q_\theta^{[+2\bs+1]}Q_\theta^{[+2\bs-1]}}Q_1^{[+3]}
   - 
    \tau_{2}^{+}{Q_\theta^{[+2\bs-1]}}Q_1^{+}
    +
     {Q_\theta^{[-2\bs+1]}}\tau_1^{-}Q_1^{-}
      -
{Q_\theta^{[-2\bs+1]}}Q_\theta^{[-2\bs-1]}Q_1^{[-3]}=0
\eeq
and
\beq
\label{Bax12s}
Q_\theta^{[-2s]}Q_{1\alpha}^{[-3]}-\tau_2Q_{1\alpha}^- + \tau_1Q_{1\alpha}^+ - Q_\theta^{[+2s]}Q_{1\alpha}^{[+3]}=0\;\;,\;\;\alpha=2,3\;.
\eeq
This set of requirements is a way to define the Bethe roots, alternative to the Bethe ansatz, by first finding the Baxter polynomials satisfying the above equations and the analyticity requirements.

\paragraph{Ground state wave function.}
The ground state of the transfer matrix -- the state corresponding to the trivial Baxter polynomials $M_1=M_{12}=M_{13}=0$ is particularly simple.
If we were considering a diagonal twist, it would simply be a constant polynomial. As our twist is non-diagonal, but diagonalisable, the ground state can be obtained as a result of rotation of the constant function with a $\mathsf{GL}(3)$ group element and is a non-trivial function like in the case of $\mathfrak{sl}(2)$ -- \eq{sl2omega}.
Instead of diagonalising the twist and rotating the ground state it is simpler to construct $\bbT_{ 1,1}(u)$ explicitly for the length $L=1$ case first. Then requiring
\beqa
\bbT_{ 1,1}(0)|\Omega_{L=1}\rangle
=\tau_1(0)|\Omega_{L=1}\rangle\;\;,\;\;
\bbT_{ 2,1}(0)|\Omega_{L=1}\rangle
&=&
Q_\theta(i\bs-\tfrac i2)\;
\tau_2(\tfrac{i}{2})
|\Omega_{L=1}\rangle\;,
\eeqa
we obtain two first order PDE's on $|\Omega_{L=1}\rangle$, fixing it uniquely up to a constant factor to
\beq\la{rvac}
|\Omega_{L=1}\rangle = 
\left(1+\frac{x}{\lambda _1}+\frac{ y}{\lambda _1^2}\right)^{-2 \bs}\times \lambda_1^{2i(\theta_1+i\bs)}\;.
\eeq
Similarly, one can find the left ground state\footnote{i.e. eigenvector of the transposed transfer matrix w.r.t. the quadratic form \eq{biform}.}
\beq\la{lvac}
\langle \Omega_{L=1}| = 
\left(1+x \left(\lambda _2+\lambda_3\right)-\frac{y}{\lambda _1}\right)^{-2 \bs}\times \[\lambda_3^{-i(\theta_1+i\bs)+1/2}\lambda_2^{-i(\theta_1+i\bs)-1/2}-(\lambda_2\leftrightarrow \lambda_3)\]\;.
\eeq
Here the nontrivial overall normalisation is chosen so as to simplify the main results later on. 
These functions are analytic near the origin and can be expanded into a series in $x$ and $y$.

Note that their scalar product, obtained as a limit of the product of the truncated series expansions, is a nontrivial number, which we denote as ${\cal N}_1$\footnote{The series is convergent for large enough $|\lambda_1|$ (and $\lambda_2$ fixed).  }
\beqa\la{vacvac}
&&\langle \Omega_{L=1}|
\Omega_{L=1}\rangle = {\cal N}_1\;\;,\;\;\\ \nn
&&
{\cal N}_1=
\left(1-\lambda^2 _2\lambda_3-\lambda_2\lambda_3^2+\lambda_2^3\lambda_3^3\right)^{-2 \bs}   \lambda_1^{2i(\theta_1+i\bs)}\[\lambda_3^{-i(\theta_1+i\bs)+1/2}\lambda_2^{-i(\theta_1+i\bs)-1/2}-(\lambda_2\leftrightarrow \lambda_3)\]
   \;. 
\eeqa

For general $L$ the ground state is simply a tensor product of $L$ copies of $| \Omega_{L=1}\rangle$
(or $\langle \Omega_{L=1}|$, for the left eigenvector)\footnote{With the replacement $\theta_1\to\theta_\alpha$ in the overall normalisation}.

Having the ground state explicitly will allow us to build the excited states by means of the creation operator ${\bf B}(u)$~\cite{Gromov:2016itr,Sklyanin:1992sm}, as we describe in the next section.

\subsection{Wave functions and SoV}\la{sec:wfs}
In this section we explain how to construct excited states of the transfer matrices by action of a creation operator ${\bf B}(u)$ on the ground state, in analogy with the $\mathfrak{sl}(2)$ case \eq{Bom}. The ${\bf B}(u)$ operator was first proposed in the context of SoV by Sklyanin in the seminal paper~\cite{Sklyanin:1992sm}. It was much later in~\cite{Gromov:2016itr} when it was realised that the the same operator can be used to diagonalise the transfer matrix and explicitly build its eigenstates. We will review this construction in this section and explain how it leads to the separation of variables.

Another operator, which was recently shown in~\cite{Gromov:2019wmz} to also play a key role in the SoV construction, is the ${\bf C}(u)$ operator which will be used to produce the dual SoV basis in the next section.
Both of these operators have a similar structure in terms of the 
monodromy matrix elements\footnote{For the finite dimensional case and with diagonal twist matrix the ${\bf B}(u)$-operator defined in~\cite{Sklyanin:1992sm} is nilpotent and cannot be used for construction of the SoV basis. In \cite{Gromov:2016itr} it was shown that there is a family of operators ${\bf B}^{\rm good}(u)$ which can create eigenstates of the transfer matrix and at the same time are diagonalisable.}
\beqa\la{befB}
{\bf B}(u)&=&{\bf T}_{23}(u)
{\bf T}_{12}(u-i){\bf T}_{23}(u)-
{\bf T}_{23}(u){\bf T}_{22}(u-i){\bf T}_{13}(u)\nn\\
&+&{\bf T}_{13}(u){\bf T}_{11}(u-i){\bf T}_{23}(u)-{\bf T}_{13}(u){\bf T}_{21}(u-i)
{\bf T}_{13}(u)\;,
\eeqa
and
\beqa\la{befC}
{\bf C}(u)&=&{\bf T}_{23}(u){\bf T}_{12}(u){\bf T}_{23}(u+i)-
{\bf T}_{23}(u){\bf T}_{22}(u){\bf T}_{13}(u+i)\nn\\
&+&{\bf T}_{13}(u){\bf T}_{11}(u){\bf T}_{23}(u+i)-{\bf T}_{13}(u){\bf T}_{21}(u)
{\bf T}_{13}(u+i)\;.
\eeqa
A simple observation, which one can immediately make from the form of the ${\bf B}$ and ${\bf C}$ operators, is that due to the particular choice of the twist matrix \eq{twistform} both of them do not depend on the twist eigenvalues $\lambda_a$ which can be checked by a direct calculation similarly to the $\mathfrak{sl}(2)$ case~\cite{Ryan:2018fyo}.

\paragraph{Creating excited states with ${
\bf B}(u)$.}
The key formula,
\beq\la{BBP}
|\Psi\rangle \;\propto \; \prod_{k=1}^{M_1}{\bf B}(u^1_k) |\Omega\rangle
\eeq
where $|\Psi\rangle$ is the transfer matrix eigenvector with eigenvalues as in \eq{PSi} and $u^1_k$ are the (momentum carrying) Bethe roots corresponding to this state, first found in \cite{Gromov:2016itr} for the fundamental representation $\bs=-1/2$ is valid for general $\bs$. Note that ${\bf B}(u)$ is built out of ${\bf T}_{ab}(u)$ and thus is a differential operator. 
Hence \eq{BBP} implies that once the momentum-carrying Bethe roots $u_k$ are found (from the TQ-relations or from the Bethe ansatz equations), one can immediately build the corresponding eigenvector in terms of partial derivatives of the ground state in full analogy with the $\mathfrak{sl}(2)$ algebraic Bethe ansatz construction. This is a huge simplification in comparison with the old nested Bethe ansatz construction~\cite{Kulish:1983rd}, which involves all the auxiliary roots and is a hybrid between the algebraic and coordinate Bethe ansatz construction for $\mathfrak{sl}(2)$.

In order to fix the normalisation of $|\Psi\rangle$ it is convenient to extract a trivial scalar factor\footnote{The presence of this trivial factor stems from the fact that we consider a special class of representations of $\sl(3)$ on $2$ variables instead of the $3$ needed for generic representations, see \cite{Ryan:2018fyo,Ryan:2020rfk}.} from the ${\bf B}(u)$ operator 
\beq\la{Bb}
{\bf B}(u)=
-Q_\theta(u + i \bs - i)
{\bf b}(u)\;.
\eeq
The remaining operator ${\bf b}(u)$
is a polynomial of degree $2L$. After that we define
\beq\la{bbP}
|\Psi\rangle =  \prod_{k=1}^{M_1}{\bf b}(u_k) |\Omega\rangle
\eeq
exactly like in the $\sl(2)$ case.

\paragraph{Eigenvalues of ${\bf B}(u)$.} Another key observation of
\cite{Gromov:2016itr} which generalises to our case is that the eigenvalues of the operator 
${\bf B}(u)$  are very simple. Because ${\bf B}(u)$ and thus ${\bf b}(u)$ \eq{Bb} commute with themselves
for different $u$ their eigenvalues are also polynomials
\beq\la{bu}
\langle \svx|{\bf b}(u)=
\prod_{\alpha=1}^L(u-\svx_{\alpha,1})(u-\svx_{\alpha,2})\;\langle \svx|
\eeq
where 
\beq\la{ntox}
\svx_{\alpha,a} = \theta_\alpha + i\bs+ i n_{\alpha, a}\;\;,\;\;a=1,2
\eeq
and $n_{\alpha a}$ are integers such that
\beq
0 \leq n_{\alpha. 2} \leq n_{\alpha, 1}\;.\la{neq}
\eeq
For convenience we also introduce
\beq
\svx_{\alpha,0}\equiv \theta_\alpha+i\bs\;.
\eeq
The spectrum of ${\bf b}$ is non-degenerate so the eigenstates $\langle \svx|$
are well defined. As we will explain below $n$'s could take any integer values as long as the constraint \eq{neq} is satisfied\footnote{In order to find the eigenvectors one needs to know how various operators acts on the right states. As ${\bf B}(u)$
or  ${\bf C}(u)$ are built out of elements of monodromy matrix, which is built out of generators of $\frak{sl}(3)$ we can easily transpose those operators by flipping signs of some of the generators accordingly. As a result the action of these operators to the right is also by a partial derivatives. Equation \eq{bu} is simply a set of PDE on the function $\langle \svx|$. We will give some explicit examples below in section~\ref{sec:l2sl3}.}.

\paragraph{Separation of variables.}
Combining \eq{BBP}, \eq{Bb} and \eq{bu} we obtain
\begin{equation}
    \Psi(x)\equiv\langle\svx|\Psi\rangle=\langle\svx|\Omega\rangle\prod_{\alpha=1}^L\prod_{k=1}^{M_1} (u_k-\theta_\alpha-i\bs-i n_{\alpha,1})
    (u_k-\theta_\alpha-i\bs-i n_{\alpha,2})
\;.
\end{equation}
As we still have not defined the normalisation of the SoV states $\langle \svx |$ we fix it by requiring
\beq\la{xnrm}
\langle\svx|\Omega\rangle =\prod_{\alpha=1}^L\lambda_1^{i \svx_{\alpha,1}+i \svx_{\alpha,2}} \;,
\eeq
which then results in\footnote{We recall that we define the Baxter polynomials with the twist factors as $Q_1(u)=\lambda_1^{i u} \prod_{k=1}^{M_1}(u-u_k)$.}
\begin{equation}\la{SOVx}
    \Psi(\svx)\equiv
    \langle\svx|\Psi\rangle=\prod_{\alpha=1}^L Q_1(\svx_{\alpha,1})
    Q_1(\svx_{\alpha,2})\;
\;,
\end{equation}
which is the very essence of separation of variables as our wave function is now explicitly a product of one-dimensional factors.

Let us again emphasise the importance of the normalisation \eq{xnrm}. It was shown in \cite{Ryan:2018fyo} that this normalisation ensures that there is no dependence on the twist eigenvalues $\lambda_1,\lambda_2$ in the SoV states $\langle \svx|$. We will demonstrate this later on an explicit example for $L=2$.

\paragraph{SoV charge operator.}
While ${\bf B}(u)$ and $\bC(u)$ commute with themselves for
different arguments $u$ they do not commute with each other and it is 
this property which leads to a non-diagonal measure for $\sl(3)$. However, as we see from the definitions \eq{befB} and \eq{befC} they only differ by shifts in $u$
and thus commute at large $u$.
Whereas the leading coefficient is simply a constant the subleading coefficients in ${\bf B}(u)$ and ${\bf C}(u)$
contain the same non-trivial operator ${\bf N}$ which thus commutes with both ${\bf B}(u)$ and ${\bf C}(u)$ at any $u$. We refer to this non-trivial operator as SoV charge~\cite{Gromov:2019wmz} and in proper normalisation it is given by
\beq
{\bf b}(u)=u^{2L}+u^{2L-1}\[i{\bf N}+2\sum_{\alpha=1}^L (\theta_\alpha+i\bs)\]+{\cal O}(u^{2L-2})\;.
\eeq\label{SoVcharge}
The SoV charge operator ${\bf N}$ defined in this way satisfies
\beq
[{\bf N},{\bf B}(u)]=0\;\;,\;\;
[{\bf N},{\bf C}(u)]=0\;.
\eeq
When acting on the left SoV state $\langle \svx|$
it gives a non-negative integer number
\beq\la{Nn}
\langle \svx|{\bf N}=\sum_{\alpha,a} n_{\alpha,a}\langle \svx|
\eeq
and so counts the excitations above the SoV vacuum. It is straightforward to find the explicit form of the operator ${\bf N}$ straight from its definition
\beq\la{Nexplicit}
{\bf N}={\bf N}^T=
\sum_{\alpha=1}^L
\(x_\alpha \partial_{x_\alpha}+
2y_\alpha \partial_{y_\alpha}
\)\;.
\eeq
We can deduce from this that the SoV states have to be homogeneous polynomials in $x_\alpha,\; y_\alpha$, with $x_\alpha$ contributing one unit of SoV charge whereas each $y_\alpha$ adds two units.

\paragraph{Eigenstates of ${\bf B}(u)$ operator.}
As it was shown above the SoV states are polynomials. As the ground state $\langle 0|$ (i.e. the left eigenstate of 
${\bf B}(u)$ with all $n_{\alpha,a}=0$) according to \eq{Nn} has SoV charge $0$ it must be  a constant function.
Furthermore, its normalisation is fixed by \eq{xnrm}, which implies
\beq
\langle 0| = 1\;.
\eeq

All the other SoV states $\langle \svx |$
can be obtained by consecutive action of transfer matrices on the SoV ground state $\langle 0 |$.
One way to see this is by observing that
the eigenvalue of ${\mathbb T}_{2,1}(u)$
at a special values of the spectral parameter $u=\theta_\alpha+i\bs-\tfrac{i}{2}$ simplifies to (see \eq{PSi})
\beq
{\mathbb T}_{2,1}(\theta_\alpha+i\bs-\tfrac{i}{2})|\Psi\rangle
=
Q_\theta(\theta_\alpha+2i\bs-i)
Q_\theta(\theta_\alpha+2i\bs)
\frac{Q_1(\theta_\alpha+i\bs+i)}{Q_1(\theta_\alpha+i\bs)}
|\Psi\rangle\;.
\eeq
Similar relations hold true for the transfer matrices operators in higher representations.
Denoting by ${\mathbb T}_{a,s}$ the transfer matrix corresponding to the rectangular $a\times s$ Young diagram and introducing the SoV creation operator ${\bf A}_{\alpha,s}$ defined by
\beq
{\bf A}_{\alpha,s}\equiv 
\frac{{\mathbb T}_{2,s}(\theta_\alpha+i\bs+\tfrac{is}{2}-i)}{
\prod\limits_{k=1}^s Q_\theta(\theta_\alpha+2i\bs+ik-i)
Q_\theta(\theta_\alpha+2i\bs+ik-2i)
}\;.
\eeq
we have\footnote{A simple way to verify this relation is by using the Hirota identity, which states that the eigenvalues of the transfer matrices in rectangular representations $T_{a,s}$ satisfy
$T_{a,s}(u+\tfrac i2)T_{a,s}(u-\tfrac i2)=
T_{a+1,s}(u)T_{a-1,s}(u)+T_{a,s+1}(u)T_{a,s-1}(u)$\;.
Using the known eigenvalues $T_{a,0}=1,\;T_{0,s}=1,\;T_{4,s}=0$ and $T_{1,1}=T^{\bf 3},\;T_{2,1}=T^{\bar{\bf 3}},\;
T_{3,1}=T^{\bar{\bf 1}}$ one can find all $T_{a,s}(u)$ recursively using the Hirota identity.
Alternatively, one can use the Wronskian solution of the Hirota identity, which gives $T_{a,s}$ explicitly in terms of $3$ Q-functions.}
\beq
{\bf A}_{\alpha,s}|\Psi\rangle
=
\frac{Q_1(\theta_\alpha+i\bs+i s)}{Q_1(\theta_\alpha+i\bs)}
|\Psi\rangle\;.
\eeq
From this we obtain that
\beq\la{APsi}
\langle 0|\prod_{\alpha=1}^L\prod_{a=1}^2 {\bf A}_{\alpha,n_{\alpha,a}}
|\Psi\rangle
=
\langle 0|\Psi\rangle
\prod_{\alpha=1}^L
\prod_{a=1}^2
\frac{Q_1(\svx_{\alpha,a})}{Q_1(\svx_{\alpha,0})}
 = \langle \svx|\Psi\rangle
\eeq
where we used \eq{SOVx} to get the last equality.
So we conclude that the state 
$\langle 0|\prod_{\alpha,a} {\bf A}_{\alpha,n_{\alpha,a}}
$ has the same overlap as $\langle \svx|$ with {\it all} eigenstates of the transfer matrix $|\Psi\rangle$,
which of course means that they are equal\footnote{Note that it is manifest from this construction that the SoV states $\langle\svx|$ are rational functions of the spin $\bs$ since the transfer matricies are polynomial functions of the Lax operators \eqref{LaxGen} which themselves are polynomial functions of $\bs$. }:
\beq\la{sl3x}
\langle \svx|=
\langle 0|\prod_{\alpha=1}^L\prod_{a=1}^2 {\bf A}_{\alpha,n_{\alpha,a}}\;\;.
\eeq
As ${\bf A}_{\alpha,s}$
are obtained from the transfer matrix in the representation $2\times s$, which themselves
can be built out of $s$ copies of
${\cal L}^{2,1}_{a,b}(u)$ at each site,
we conclude that ${\bf A}_{\alpha,s}$ is an $s\times L$
order partial differential operator with polynomial coefficients, which makes the equation \eq{sl3x}
very convenient for the building of the SoV states.

\subsection{Dual SoV states}
In our construction the left eigenstates $\langle \Psi|$ are not related to the right eigenstates of the monodromy matrix in an obvious way.
Consequently, the basis which separates states $\langle \Psi|$ has to be built from scratch. 

In this section we build the right SoV basis $|\svy\rangle$ as an eigenbasis of the $\bC(u)$ operator.
Like the original $\bB(u)$ operator it commutes with itself $[\bC(u),\bC(v)]=0$.
We will again see that the spectrum of operator $\bC(u)$ is very simple. For example the right SoV ground state $|0\rangle$, is the only state with the SoV charge ${\bf N}=0$, which again must be a constant.
We fix its normalisation so that $|0\rangle=1$, or equivalently
\beq
\langle \Omega|0 \rangle
=
\prod_{\alpha=1}^L
{
\[\lambda_{3}^{-i\svy_{\alpha,0}+1/2}\lambda_{2}^{-i\svy_{\alpha,0}-1/2}-
\lambda_{2}^{-i\svy_{\alpha,0}+1/2}\lambda_{3}^{-i\svy_{\alpha,0}-1/2}\]
}\;
\eeq
where we defined for future convenience
\beq
\label{defy00}
\svy_{\alpha,0} = \theta_\alpha+i\bs\;.
\eeq

The eigenvectors of $\bC(u)$ can be constructed in a similar way to $\bB(u)$ -- using the transfer matrices in anti-symmetric representations as building blocks. 
Namely, we define the following combinations
\begin{equation}\la{Tstar}
    {\mathbb  T}^*_{\{m_1,m_2\}}(u)=\det_{1\leq j,k\leq m_1}{\mathbb  T}_{\mu'_j-j+k,1}\left(u+\frac{i}{2}\left(\mu'_1-\mu'_j-m_1+j+k-1\right)\right)\;\;,\;\;0\leq m_2 \leq m_1
\end{equation}
where $\mu_j'=2,\;j=1,\dots,m_2$
and is $1$ otherwise. The combination \eq{Tstar}
is reminiscent of the Cherednik-Bazhanov-Reshetikhin formula~\cite{Cherednik,Bazhanov:1989yk} for that transfer matrix in 
an irrep with a Young diagram $\mu$ (and $\mu'$  being the transposed of $\mu$,see Fig.\ref{fig:Youngtranspose}), which states
\begin{equation}
    {\mathbb T}_{\mu}(u)=\det_{1\leq j,k\leq \mu_1}{\mathbb  T}_{\mu'_j-j+k,1}\left(u-\frac{i}{2}\left(\mu'_1-\mu_1-\mu'_j+j+k-1\right)\right)\;.
\end{equation}
In \eq{Tstar} we had to replace $i$ by $-i$ in the shift of the argument. The reason for such replacement will be clear in section~\ref{sec:secslN}.

Like in the case with the eigenvalues of the operator
${\bf B}(u)$ we introduce the ``creation operators",
which are simply properly normalised combinations of the integrals of motion~\eq{Tstar}
\beq
{\bf D}_{\alpha,m_1,m_2}\equiv 
\frac{{\mathbb T}^*_{\{m_1,m_2\}}\left(\theta_\alpha+i\bs+i\frac{m_1-\mu'_1}{2}\right)}{
\prod\limits_{k=0}^{m_1} Q_\theta(\theta_\alpha+2i\bs+ik-i)
\prod\limits_{k=1}^{m_2-1} Q_\theta(\theta_\alpha+2i\bs+ik-i)
}\;,
\eeq
where like before $\mu'_1=2$ for $m_2>0$ and $1$ otherwise.
The ${\bf C}(u)$-operator eigenvectors are then given by
\beq\la{sl3y}
|\svy \rangle =
\prod_{\alpha=1}^L {\bf D}_{\alpha,m_{\alpha,1},m_{\alpha,2}}|0 \rangle\;\;,\;\;0\leq m_{\alpha,2} \leq m_{\alpha,1}\;
\eeq
with the eigenvalue
\begin{equation}
{\bf C}(u)|\svy \rangle =
   - Q_\theta(u+i\bs)\prod_{\alpha=1}^L(u-\svy_{1}^\alpha)(u-\svy_{2}^\alpha)|\svy \rangle\;
\end{equation}
where we introduced the notation\footnote{We see that in these conventions for $m_{\alpha,1}=m_{\alpha,2}=0$ we have $\svy_{\alpha,1}=\svy_{\alpha,0},\ \svy_{\alpha,2}=\svy_{\alpha,0}-i$ with $\svy_{\alpha,0}$ defined in \eq{defy00}.}
\beq\la{mtoy}
\svy_{\alpha,1} = \theta_\alpha+i\bs+i m_{\alpha,1}\;\;,\;\;\svy_{\alpha,2} = \theta_\alpha+i\bs+i m_{\alpha,2}-i\;\;.
\eeq

The SoV charge operator ${\bf N}$ \eq{Nexplicit}
appears in the subleading coefficient of the $1/u$ expansion of ${\bf C}(u)$ and so its eigenvalue is given by the sum of all  $m_{\alpha,a}$,
\beq
{\bf N}|\svy \rangle=\sum_{\alpha,a} m_{\alpha,a}|\svy \rangle\;.
\eeq

Finally, let us state the analogue of the relation \eq{SOVx} for the contraction of the eigenstate of the transfer matrix $\langle \Psi|$ and the eigenstate of the ${\bf C}(u)$-operator.
For that we notice that $\langle \Psi|$ also diagonalises ${\bf D}_{\alpha,m_1,m_2}$ with the following eigenvalue:
\beq
\langle \Psi|
{\bf D}_{\alpha,m_1,m_2}=
\frac{
Q_{12}(\svy_{\alpha,1} +\tfrac{i}{2})
Q_{13}(\svy_{\alpha,2} +\tfrac{i}{2})
-(Q_{13} \leftrightarrow Q_{12})
}{
Q_{12}(\svy_{\alpha,0} +\tfrac{i}{2})
Q_{13}(\svy_{\alpha,0} -\tfrac{i}{2})
-
(Q_{13} \leftrightarrow Q_{12})
}
\langle \Psi|\;.
\eeq
We normalise $\langle \Psi|$ so that
\beq
\langle \Psi|0\rangle = 
\prod_{\alpha=1}^L
\[{Q_{12}(\svy_{\alpha,0} +\tfrac{i}{2})
Q_{13}(\svy_{\alpha,0} -\tfrac{i}{2})
-
(Q_{13} \leftrightarrow Q_{12})}\]\;.
\eeq
After that we get a factorised expression for the wave function in the SoV basis
\beq
\langle \Psi|\svy \rangle
=
\prod_{\alpha=1}^L{\[
Q_{12}(\svy_{\alpha,1}+\tfrac{i}{2})
Q_{13}(\svy_{\alpha,2}+\tfrac{i}{2})
-(Q_{13} \leftrightarrow Q_{12})\]}
\;.
\eeq
In particular for the ground state we can read off the following normalisation of the right SoV states
\beq
\langle \Omega|\svy \rangle
=
\prod_{\alpha=1}^L
{
\[\lambda_{3}^{-i\svy_{\alpha,1}+1/2}\lambda_{2}^{-i\svy_{\alpha,2}+1/2}-
\lambda_{2}^{-i\svy_{\alpha,1}+1/2}\lambda_{3}^{-i\svy_{\alpha,2}+1/2}\]
}\;.
\eeq
Even though the above normalisation looks rather complicated, like in the case with $\langle \svx|$ it ensures that there is no $\lambda_a$ dependence in the $|\svy\rangle$ state either (for general proof of this see~\cite{Ryan:2018fyo}). We will see some explicit example below in section~\ref{sec:l2sl3}.

\subsection{Overlap of the SoV states}
In order to be able to use the factorised representation of the wave function one also needs to know the measure in the SoV basis.
We will see that unlike in the $\mathfrak{sl}(2)$ case 
the left and right SoV states are not orthogonal to each other. Nevertheless, we can write an analog of the $\sl(2)$ completeness relation \eq{sl2comp} as
\beq
    \sum_{\svx, \svy}M_{\svy,\svx}\ket{\svy}\bra{\svx}=1
\eeq
where $M_{\svy,\svx}$ is an infinite set of nontrivial coefficients that form the SoV measure, a key part of the whole construction. As a matrix it is the inverse of the infinite matrix of overlaps $\langle \svx | \svy \rangle$. Knowledge of the matrix $M_{\svy,\svx}$ 
in particular would allow the calculation of overlaps between two Bethe states $\Psi^A$ and $\Psi^B$
\beqa\la{oversov}
\langle \Psi^A|\Psi^B\rangle
&=&
\sum_{\svx, \svy}M_{\svy,\svx}\langle \Psi^A
\ket{\svy}\bra{\svx}
\Psi^B\rangle=\\
\nn&&\sum_{\svx, \svy}M_{\svy,\svx}
\prod_{\alpha=1}^L
{
\[Q^A_{12}(\svy_{\alpha,1}+\tfrac{i}{2})
Q^A_{13}(\svy_{\alpha,2}+\tfrac{i}{2})
-(Q^A_{13} \leftrightarrow Q^A_{12})\]}
{Q^B_1(\svx_{\alpha,1})
    Q^B_1(\svx_{\alpha,2})}\;.
\eeqa
The overlaps matrix has in fact a nice and simple structure.
First, due to the existence of the SoV charge operator it is block-diagonal. Second, each block is a triangular matrix for a particular ordering of the states $\langle \svx |$ and $| \svy \rangle$.
More precisely the left and right SoV states are in one-to-one
correspondence as both are labelled by a set of $2L$ integers constrained by $0\leq n_{\alpha,2}\leq n_{\alpha,1}$
and $0\leq m_{\alpha,2}\leq m_{\alpha,1}$. In section~\ref{intmeasure},
we show that the overlap matrix becomes upper triangular
when we order both SoV states lexicographically with words $(n_{1,1},n_{2,1},\dots,n_{L,1},n_{L,2},\dots,
n_{L,N-1})$ and same for $m$'s.
We will see that in general it is a rather sparse matrix with elements accumulating near the diagonal.  In section~\ref{intmeasure}
we derive the general form of this matrix giving an explicit relation for its matrix elements,
using an integral representation, generalising the results of \cite{Cavaglia:2019pow,Gromov:2019wmz}.

In the next section we report on some experimental observations coming from length two spin chains. 

\subsubsection{Length-two data}\la{sec:l2sl3}
We explicitly realised the above construction for a spin chain of length $L=2$. In particular we computed all SoV states 
$\langle \svx |$ and $| \svy \rangle$ up to the charge ${\bf N}=6$ analytically.

The eigenstates of ${\bf B}(u)$ are labelled by $4$ integer numbers
$n_{\alpha,a}$, such that $0\leq n_{\alpha,2}\leq n_{\alpha,1}$.
Correspondingly we denote the eigenstates
\beq
\langle n_{1,1},n_{1,2};n_{2,1},n_{2,2}|=\langle \svx |\;.
\eeq
The SoV vacuum in our normalisation is just $1$
\beqa
\langle 0,0;0,0|=1\;.
\eeqa
The first two excited states with SoV charge $1$ are
\beqa
\langle 0,0;1,0|&=&-{x_2}
\\
\langle 1,0;0,0|&=&
-\frac{2 i \bs x_2}{\left(\theta _{12}+2 i \bs\right)}-\frac{\theta _{12} x_1}{ \left(\theta _{12}+2 i \bs\right)}
\eeqa
where $\theta_{12}=\theta_1-\theta_2$.

At charge $2$ the states become more complicated
\beqa
\langle 0,0;1,1|&=&-\frac{i x_1 x_2}{2 i \bs-\theta _{12}}+\frac{(2 \bs+1) x_2^2}{2 \bs }+\frac{\theta _{12} y_2}{2 \bs \left(\theta _{12}-2 i
	\bs\right)}\\
\nn\langle 1,1;0,0|&=&
\frac{1}{
(\theta _{12}+2 i \bs)^2
} \left({x_1 x_2 \left(i \theta _{12} (4 \bs+1)+2 \bs\right)}+\frac{\theta _{12}^2 (2 \bs+1) x_1^2}{2 \bs }\right.\\
\nn&&\left.-{2 \bs (2 \bs+1)
	x_2^2}+\frac{\theta _{12}^2 y_1}{2 \bs }+{i \theta _{12} y_2}\right)\\
\nn\langle 0,0;2,0|&=&
{ x_2^2}\\
\nn\langle 1,0;1,0|&=&
\frac{1}{
 \theta_{12}+2i \bs
} \left({i (2 \bs+1) x_2^2}+{\left(\theta _{12}-i\right) x_1 x_2}\right)\\
\nn\langle 2,0;0,0|&=&
{i (2 \bs+1) x_2^2}+{\left(\theta _{12}-i\right) x_1 x_2}\;.
\eeqa
We see that if we assign to $x_\alpha$ a homogeneity weight $1$
and weight $2$ to $y_\alpha$, the eigenstates are homogeneous polynomials of weight equal to the SoV charge. 
We also note that the SoV basis does not have any dependence on the twist eigenvalues $\lambda_i$, as anticipated previously. 

In a similar way the eigenstates of ${\bf C}(u)$ are labelled by the $4$
integers $m_{\alpha,a}$. We denote
\beq
| \svy \rangle=
| m_{1,1},m_{1,2};m_{2,1},m_{2,2}\rangle\;.
\eeq
For the right SoV vacuum we have
\beqa
| 0,0;0,0\rangle=1\;.
\eeqa
At SoV charge $1$ we get
\beqa
| 0,0;1,0\rangle&=&
\frac{ 2 i \bs\, x_1-\theta _{12}  x_2}{\theta_{12}-2i \bs}
\\
| 1,0;0,0\rangle&=&
-x_1
\eeqa
At the level $2$ the states again become more complicated
\beqa
| 0,0;1,1\rangle&=&
\frac{\theta _{12}  y_2-2 i \bs\,  y_1}{\theta_{12}-2i \bs} 
\\
\nn
| 1,1;0,0\rangle&=&  y_1
\\
\nn| 0,0;2,0\rangle&=&
\frac{1 }{
	\theta_{12}-2i \bs}
\left(\frac{\theta _{12} \left(\theta _{12}-i\right) x_2^2}{\theta _{12}-i (2 \bs+1)}
+\frac{4 \theta _{12} \bs\,
	 x_1 x_2}{i \theta _{12}+2 \bs+1}
\right.\\
\nn&&\left.+\frac{2 \bs (2 \bs+1) x_1^2}{2 i \bs+i-\theta _{12}}+2 i
\bs\,  y_1-\theta _{12}  y_2\right)
\\
\nn| 1,0;1,0\rangle&=&
\frac{ -i (2 \bs+1)  x_1^2+\left(\theta _{12}+i\right)  x_1 x_2+i  y_1-i  y_2}{\theta_{12}-2i \bs}
\\
\nn| 2,0;0,0\rangle&=& x_1^2- y_1\;.
\eeqa
We notice once again that both left and right SoV states are homogeneous polynomials of degree equal to the SoV charge.

Let us give some examples of the overlaps for SoV charge $0$ and $1$. We get a very simple expression 
\beqa\la{cherge01}
\left.\langle \svx | \svy \rangle\right|_{{\bf N}\leq 1}=\left(
\begin{array}{ccc}
	1 & 0 & 0 \\
	0 & \frac{\theta _{12}}{2 \bs \left(\theta _{12}-2 i \bs\right)} & 0 \\
	0 & 0 & \frac{\theta _{12}}{2 \bs \left(\theta _{12}+2 i \bs\right)} \\
\end{array}
\right)\;.
\eeqa
For charge $2$ it is similarly simple, but we also get non-diagonal elements
\beqa\la{cherge2}
&&\left.\langle \svx | \svy \rangle\right|_{{\bf N}= 2}=\\
\nn&&{\tiny
\left(
\begin{array}{ccccc}
 \frac{-\theta _{12}^2}{4 \bs^2 \left(i \theta _{12}+2 \bs\right){}^2} & 0 & \frac{-\theta _{12}^2}{4 \bs^2 \left(i \theta _{12}+2 \bs\right){}^2} & \frac{1}{4 \bs^2 \left(i \theta _{12}+2 \bs\right){}^2} & 0 \\
 0 & \frac{-\theta _{12}^2}{4 \bs^2 \left(-i \theta _{12}+2 \bs\right){}^2} & 0 & \frac{1}{4 \bs^2 \left(-i \theta _{12}+2 \bs\right){}^2} & \frac{-\theta _{12}^2}{4 \bs^2 \left(-i \theta _{12}+2 \bs\right){}^2} \\
 0 & 0 & \frac{-\theta _{12}\left(\theta _{12}-i\right)}{\bs (2 \bs+1) \left(i \theta _{12}+2 \bs\right)
   \left(i \theta _{12}+2 \bs+1\right)} & 0 & 0 \\
 0 & 0 & 0 & \frac{\theta _{12}^2+1}{4 \bs^2 \left(\theta _{12}^2+4 \bs^2\right)} &
   0 \\
 0 & 0 & 0 & 0 & \frac{-\theta _{12} \left(\theta _{12}+i\right)}{\bs (2 \bs+1) \left(-i \theta _{12}+2
   \bs\right) \left(-i \theta _{12}+2 \bs+1\right)} \\
\end{array}
\right)
}\;.
\eeqa
\begin{figure}
    \centering
    \includegraphics[scale=0.5]{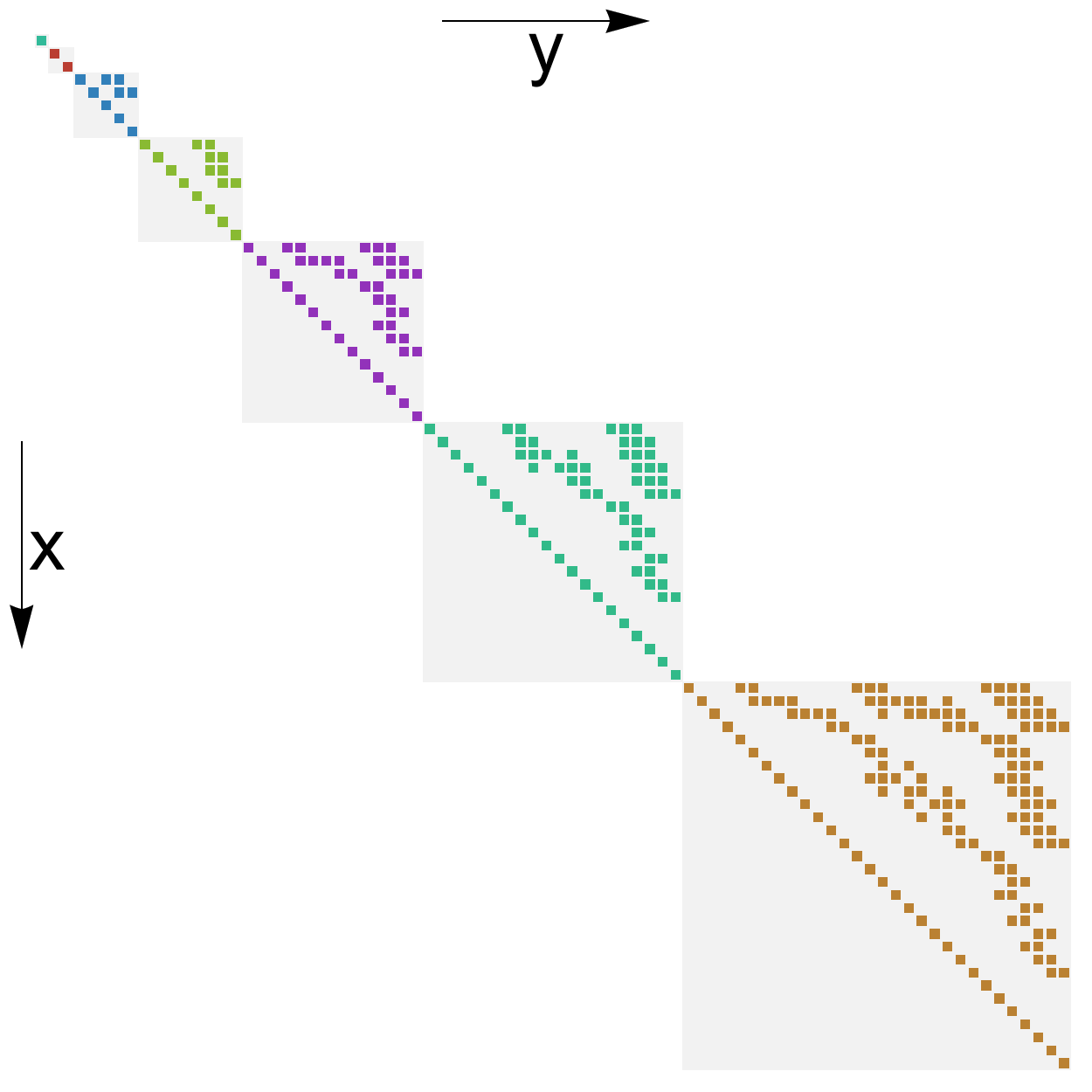}\quad\quad
        \includegraphics[scale=0.5]{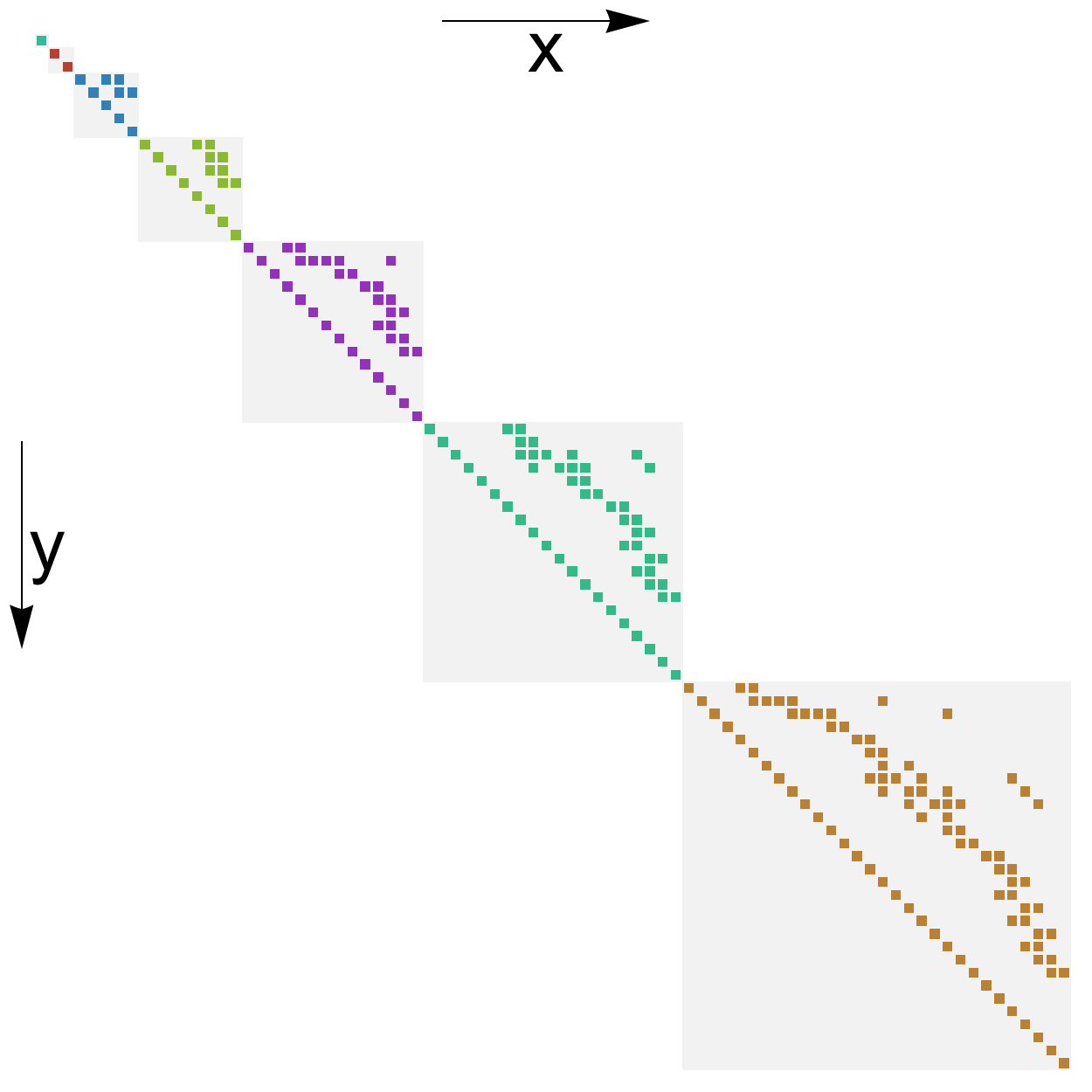}
    \caption{Non-zero elements of the matrix $\langle \svx|\svy \rangle$ (Left) and its inverse (Right) up to SoV charge $6$. Blocks indicate fixed SoV charges.}
    \label{fig:measurepic}
\end{figure}
In order to be able to compare with 
section~\ref{intmeasure},
where we found an analytic expression for the measure elements we need the inverse of the matrix~\eq{cherge2}
\beqa\la{measuredata}
&&\left.M_{\svy,\svx}\right|_{{\bf N}=2}=\\
\nn&&{\tiny
\left(
\begin{array}{ccccc}
 -\frac{4 \bs^2 \left(2 \bs+i \theta _{12}\right){}^2}{\theta _{12}^2} & 0 &
   \frac{\bs (2 \bs+1) \left(2 \bs+i \theta _{12}\right) \left(i \theta _{12}+2
   \bs+1\right)}{\theta _{12} \left(\theta _{12}-i\right)} & \frac{4 \bs^2
   \left(\theta _{12}^2+4 \bs^2\right)}{\theta _{12}^2 \left(\theta
   _{12}^2+1\right)} & 0 \\
 0 & -\frac{4 \bs^2 \left(2 \bs-i \theta _{12}\right){}^2}{\theta _{12}^2} &
   0 & \frac{4 s^2 \left(\theta _{12}^2+4 \bs^2\right)}{\theta _{12}^2
   \left(\theta _{12}^2+1\right)} & \frac{\bs (2 \bs+1) \left(2 \bs-i \theta
   _{12}\right) \left(-i \theta _{12}+2 \bs+1\right)}{\theta _{12}
   \left(\theta _{12}+i\right)} \\
 0 & 0 & -\frac{\bs (2 \bs+1) \left(2 \bs+i \theta _{12}\right) \left(i \theta
   _{12}+2 \bs+1\right)}{\theta _{12} \left(\theta _{12}-i\right)} & 0 & 0
   \\
 0 & 0 & 0 & \frac{4 \bs^2 \left(\theta _{12}^2+4 \bs^2\right)}{\theta
   _{12}^2+1} & 0 \\
 0 & 0 & 0 & 0 & -\frac{\bs (2 \bs+1) \left(2 \bs-i \theta _{12}\right)
   \left(-i \theta _{12}+2 s+1\right)}{\theta _{12} \left(\theta
   _{12}+i\right)} \\
\end{array}
\right)}\;.
\eeqa
The overlaps of the SoV states do not depend on twist eigenvalues $\lambda_a$.
This shows universality of these coefficients.
As it was advertised previously the matrix $M_{\svy,\svx}$ has upper triangular form, i.e. all elements with $m_{\alpha,1}>n_{\alpha,1}$ (for at least one $\alpha$) are zero,
furthermore if $m_{\alpha,1}=n_{\alpha,1}$ then one still gets zero if $m_{\alpha,2}>n_{\alpha,2}$. We computed explicitly the overlaps between the SoV states up to SoV charge $6$. On the figure \ref{fig:measurepic} we indicate with squares the non-zero elements.

In the section~\ref{sec:expmeasure} we will explain how to obtain the overlap coefficients bypassing explicit construction of the SoV states.

\section{Integral orthogonality relations}\la{sec:measureint}
In this section we will describe the method of~\cite{Cavaglia:2018lxi,Cavaglia:2019pow} for finding the SoV measure factor $M_{\svy,\svx}$ bypassing explicit calculation of the overlaps of the  SoV states and then inverting the matrix. We  derive the so-called orthogonality relation and then use it to 
find the matrix elements of the SoV measure explicitly in the next section.

The idea of the method of~\cite{Cavaglia:2018lxi,Cavaglia:2019pow} is the following: imagine we knew the measure, then we would be able to compute the scalar products between left and right eigenvectors of the transfer matrix in terms of the Q-functions corresponding to the states.
Due to the orthogonality of the eigenstates corresponding to different eigenvalues we then have that a combination
 of Q-functions, corresponding to any two different states, vanishes.
 At the same time the same combination where we only use the Q-functions of the same state should be non-zero. Firstly, without knowing about the SoV framework it may be even surprising that such combinations exist.
 At the same time if we find a combination of the Q-functions which can be interpreted as an SoV product with some state-independent measure, which has the above properties, it will most likely be unique and thus should produce the SoV measure up to an overall factor.
 
 In~\cite{Gromov:2019wmz} it was shown for a finite dimensional case how to build such combinations of Q-functions with the orthogonality properties satisfied. In~\cite{Gromov:2019wmz} the orthogonality relations were then interpreted as a system of linear equations for the measure matrix elements $M_{\svy,\svx}$ and from the counting of the equations it was argued that they fix the measure factors uniquely up to an overall factor.
In the infinite dimensional case, it is harder to make a totally rigorous argument as the system of equations becomes infinite. However, we will see that the dependence on the spin can be factorised and thus it is sufficient to prove this statement for a finite dimensional case only. Furthermore, the existence of the SoV charge implies that the measure factor is block diagonal with each block being of a finite size, which indeed helps to extend the previous proof to the general spin $\bs$ case.  
Furthermore, we explicitly verified our result for short lengths.

In this section we will generalize the results of~\cite{Cavaglia:2019pow} for $\sl(N)$ spin chains in the simplest infinite-dimensional representation (i.e. $\bs=1/2$ in our notations) to general values of $\bs$. We will first discuss the $\sl(2)$ case, and then move on to $\sl(3)$. Finally, in appendix \ref{app:slNbax} we give the generalization to any $\sl(N)$ \footnote{The notation we use in this paper differs by $i \to -i$ from \cite{Cavaglia:2019pow}. The notation we choose here is consistent with the more recent paper \cite{Gromov:2019wmz}. }.

\subsection{The $\sl(2)$ case}\la{sec:intsl2}

Before considering the non-trivial $\sl(3)$
case, we first re-derive the known $\sl(2)$ results in a way suitable for the generalisation in the next section.
We are following the derivation of~\cite{Cavaglia:2019pow}, which we generalise to the general $\bs>0$ case.

\subsubsection{Integral form of the scalar product}
In the $\sl(2)$ case the only nontrivial polynomial Q-function (with a twist factor) is $Q_1$ which satisfies the  Baxter equation which follows from \eq{Tsl2Q} or \eq{T11sln},
\beq
\label{baxsl2}
    Q_\theta^{[+2\bs]}Q_a^{++} - \tau_1 Q_a + Q_\theta^{[-2\bs]}Q_a^{--}=0\;.
\eeq
Here $\tau_1(u)\equiv T_{1,1}(u)$ is the eigenvalue of the transfer matrix with fundamental representation in the auxiliary space. Let us note that
for general $\bs$ only one of the two solutions of this equation will be regular and that is the one corresponding to $Q_1=\lambda_1^{i u}\prod_{k=1}^{M_1}(u-u_k^1)$.

The main idea of our approach is to introduce a scalar product on functions of one variable with respect to which the difference operator $\hat O$ defining the Baxter equation \eq{baxsl2},
\beq
        \hat O=Q_\theta^{[+2s]} \cD^{+2} - \tau_1+
    Q_\theta^{[-2s]} \cD^{-2} \ , \ \ \ \ \ \ \ \hat O Q_a=0
\eeq
will be     ``self-adjoint". Here $\cD$ is the shift operator, 
\beq
\label{ddef}
    \cD f(u)=f(u+i/2) \ .
\eeq
We write this scalar product as
\beq\la{scalargf}
    \bl g f\br_\alpha\equiv\frac{1}{2\pi i}\int_{-\infty}^{+\infty}du\;\mu_\alpha(u)g(u) f(u) \ 
\eeq
and the self-adjoint property is
\beq
\label{fgO}
    \bl f\hat O g\br_\alpha= 
   \bl g\hat {  O} f\br_\alpha
\eeq
where $f$ and $g$ are arbitrary twisted polynomials\footnote{with asymptotics such that the integral converges at infinity. We will specify the convergence condition later on.}.
This requirement constrains the integration measure $\mu_\alpha$.
In fact we will find several such measure factors and the index $\alpha$ labels different possible choices. Let us write more explicitly the l.h.s. of \eq{fgO},
\beq
\label{fgO2}
    \bl f\hat O g\br_\alpha=\int_{-\infty}^{+\infty}du\mu_\alpha f (Q_\theta^{[+2\bs]}g^{++} - \tau_1 g + Q_\theta^{[-2\bs]}g^{--})\;.
\eeq
For the $\bs=1/2$ case studied in \cite{Cavaglia:2019pow} it was sufficient to assume that $\mu_\alpha$ is $i$-periodic. 
Then we can shift the integration contour (assuming there are no poles, which could give an additional contribution) in each of the terms in \eq{fgO2} up or down by $i$ so as to remove the shifts of the argument in $g$. As a result we find precisely the same operator $\hat O$ acting now on $f$, thus proving the self-adjointness property \eq{fgO}. 

For generic $\bs$ the $i$-periodicity of $\mu_\alpha$ is obviously not sufficient. 
Assuming that we can shift the contour by $\pm i$, without getting any extra pole contributions we find that $\mu_\alpha$ has to satisfy
\beq
    \frac{\mu_\alpha^{++}}{\mu_\alpha}=\frac{Q_\theta^{[+2\bs]}}{Q_\theta^{[-2\bs+2]}}\;.
\eeq
The general solution to this first order difference equation is
\beq\la{defvarep}
    \mu_\alpha={\varepsilon}\times p_\alpha\;\;,\;\;
    \varepsilon=\prod_{\beta=1}^L\frac{\Gamma(\bs-i(u-\theta_\beta))}{\Gamma(1-\bs-i(u-\theta_\beta))}\;\;,\;\;p_\alpha^{++}=p_\alpha\;. 
\eeq
The factor $\varepsilon$ is chosen so that it is
analytic for all ${\rm Im}\,u>-\bs$ (assuming $\theta_\beta$'s are real),
it has poles at $u=\theta_\beta-i\bs-i n,\;n\ge 0$
and zeros at $u=\theta_\beta+i\bs-i n,\;n\ge 1$
and behaves at infinity as a power $\sim u^{1-2\bs}$.
It remains to determine the $i$-periodic factors $p_\alpha$.
 
The functions $p_\alpha$ have to be chosen such that 1) the integral is convergent 2) there are no extra poles contributing to the integral when we shift the contour.

For simplicity let's assume $\bs>0$ to ensure that poles of $\varepsilon$ are below the real axis.
Let's first look at the factor $p_\alpha\;\varepsilon\; Q_\theta^{[-2\bs]} g^{--}$
and we need to make sure there are no poles in the strip $0\leq {\rm Im}\,u\leq 1$. The only pole can come from the $p_\alpha$ factor, however, since there is always one zero $u=\theta_\beta+i\bs-i n$
for some $n\ge 0$ inside this strip, coming from $\varepsilon\times Q_\theta^{[-2\bs]}$ we still can allow for the $p_\alpha$
to have poles at $u=\theta_\beta+i\bs-i m,\;m\in{\mathbb Z}$.
Similarly for the term $\mu_\alpha Q_\theta^{[+2\bs ]} g^{++}$
there should not be poles at $-1\leq {\rm Im}\,u\leq 0$, this time $\varepsilon$ has a dangerous pole at $\theta_\beta-i\bs$, however, this one is luckily cancelled by the factor $Q_\theta^{[+2\bs ]}$;
and similarly to the previous term we still can allow for $p_\alpha$
to have poles at $u=\theta_\beta+i\bs-i m$.

Further constraints on $p_\alpha$ are coming from the convergence requirement. Assuming that both $f$ and $g$ behave as $\lambda_1^{i u}u^t$ at infinity, and assuming that $-\pi<{\rm arg}\;\lambda_1<0$ for definiteness, we see that in order for the integral to converge, $p_\alpha$
has to decay exponentially and faster than $\lambda_1^{i u}$ at $u\to+\infty$ and at the same time not
grow faster than $\lambda_1^{-i u}$ at $u\to-\infty$. Since, furthermore,
we are only allowed to have simple poles at $\theta_\beta+i\bs- im$ the most general $i$-periodic function with these properties should have the form
\beq
\sum_{\beta=1}^L \frac{C_\beta}{1-e^{2\pi(u-\theta_\beta-i\bs)}}\;.
\eeq
Thus we conclude there are $L$ linearly independent measures with the 
specified properties, which we denote as
\beq
\label{muj}
    \mu_\alpha=\frac{\varepsilon}{1-e^{2\pi (u-\theta_\alpha-i\bs)}} \ \ , \ \ \ \ \alpha=1,\dots,L\;.
\eeq
Note that for this choice of the basis, for any given $\alpha$ the poles of the measure $\mu_\alpha$ in the upper half plane are  at 
$u=+i\bs+in+\theta_\alpha$ with $n=0,1,\dots$. 
For the case $\bs=1/2$ the expression \eq{muj} reproduces the result of \cite{Cavaglia:2019pow}.

Finally, let us point out that for the case 
$0<{\rm arg}\;\lambda_1<\pi$ we would have to simply replace
the sign in the exponent in denominator of \eq{muj} to ensure the convergence.

\bigskip

\par\noindent
\textbf{Orthogonality. } Having the self-adjoint property \eq{fgO}, we can now use standard arguments from linear algebra in order to deduce orthogonality relations for Q-functions corresponding to different states\footnote{Tricks of this type were used earlier in \cite{Cavaglia:2018lxi} by two of the authors and A.~Cavaglia in the AdS/CFT context.}. Consider two different transfer matrix eigenstates labelled as $A$ and $B$, so that
\beq
    \hat O^A Q_1^A=0 \ , \ \ \ \ \  \hat O^B Q_1^B=0 \ ,
\eeq
then as a consequence of \eq{fgO} we have
\beq
\label{odiff}
    \bl Q_1^B(\hat O^A-\hat O^B)Q_1^A\br_\alpha=0\;.
\eeq
The only difference between operators $\hat O^A$ and $\hat O^B$ comes from the transfer matrices which have the form
\beq
    \tau_1^A=2\cos\phi \  u^L+\sum_{\alpha=0}^{L-1}I_\alpha^A u^\alpha
\eeq
where $I_\alpha^A$ are eigenvalues of the integrals of motion. Thus \eq{odiff} gives a linear system of equations on the differences $I_\alpha^A-I_\alpha^B$,
\beq\la{linsys}
    \sum_{\beta=0}^{L-1}\bl Q_1^AQ_1^Bu^\beta\br_\alpha (I_\beta^A-I_\beta^B) = 0 \ , \ \ \ \alpha=1,\dots,L\;.
\eeq
As the set of coefficients of  $\tau_1$ distinguishes the spin chain state uniquely, at least one of the differences $I_\alpha^A-I_\alpha^B$ has to be nonzero. This means that for $A\neq B$ the determinant of the linear system~\eq{linsys} should vanish
\beq
\label{orthsl2}
    {\rm det}\left|\bl Q_1^A Q_1^B u^{\alpha-1}\br_\beta \right|_{\alpha,\beta=1,\dots L}\propto \;\delta_{AB}\;.
\eeq
One can consider this identity as an orthogonality relation between 
the SoV wave functions, since the above determinant has the correct form!
In the next section we clarify more precisely the link with the explicit construction of the SoV basis from section \ref{sec:sl2}.

\subsubsection{Comparison to the SoV basis construction}
\la{sec:compsl2}

Let us  demonstrate that  the orthogonality property \eq{orthsl2}  is directly related to the SoV basis we constructed in section \ref{sec:sl2}. Since the determinant \eq{orthsl2} vanishes when $A$ and $B$ label different transfer matrix eigenstates, we expect to identify it with the scalar product
\beq
    \bra{\Psi_A}\Psi_B\rangle=\cN\times {\rm det}\left|\bl Q_1^A Q_1^B u^{\alpha-1}\br_\beta \right|_{\alpha,\beta=1,\dots L}
\eeq
up to an overall state-independent constant factor $\cN$ (which we can always introduce by rescaling the integration measure \eq{muj}). Let us relate the determinant to the SoV basis representation of this scalar product given in \eq{PsiABsl2}.
We see that each of the brackets in \eq{orthsl2} is an integral over the real line, where the integrands have asymptotics dictated by the measure and by the Q-functions which have the form $\lambda_1^{iu}\times[{\rm polynomial}]$. In section \ref{sec:sl2} we assumed that $|\lambda_1|> 1$ in order to ensure that the states we constructed are actually inside our Hilbert space (see the discussion after \eq{sl2omega}), and here this condition also plays a key role as it allows us
to close the integration contour in the upper half-plane. This means that the integral reduces to a sum over poles of the measure at $u=\theta_\alpha+i\bs+in$, \  $n=0,1,2,\dots$. As a result, we find
\beqa
\label{detexp2}&&
\cN\times{\rm det}\left|\bl Q_1^A Q_1^B u^{\alpha-1}\br_\beta \right|_{\alpha,\beta=1,\dots L}=\\ \nn &&
      \sum_{n_1,\dots,n_L} \cN\times M'_{n_1,n_2,\dots,n_L}\(
    \prod_{\alpha=1}^LQ_1^A(\svx_\alpha)\)\(\prod_{\beta=1}^LQ_1^B(\svx_\beta)\)
\eeqa
where the sum is over integers $n_\alpha\geq 0$ with $\svx_\alpha=\theta_\alpha+i\bs+in_\alpha$, and the $M'_{n_1,\dots,n_L}$ coefficients are some combination of residues of the integration measure. We see that the arguments of Q-functions in \eq{detexp2} are precisely the eigenvalues of the separated variables given by \eq{xvalsl2} from section \ref{sec:sl2}.
We also see that the expressions in the brackets match the SoV wavefunctions \eq{sovwfsl2}. It is clear now that \eq{detexp2} indeed has exactly the same form as the scalar product between two transfer matrix eigenstates $\bra{\Psi_A}\Psi_B\rangle$ we gave in \eq{PsiABsl2} above, with $\cN\times M'_{n_1,\dots n_L}$ appearing in the place of the measure $M_{n_1,\dots,n_L}$ in \eq{PsiABsl2}.  Thus we identify the coefficients in \eq{detexp2}, following from the evaluation of integrals, with the (inverse) overlaps of the SoV basis elements given by $M_{n_1,\dots,n_L}$,
\beq
\label{Mmu}
    M_{n_1,\dots,n_L}=\cN\times M'_{n_1,\dots n_L}\;.
\eeq
In fact, when we consider all different eigenstates $A,B$ of the transfer matrix, \eq{detexp2} has to vanish and thus we get a (infinite) system of linear equations that should be satisfied both by $\cN\times M'_{n_1,\dots,n_L}$ and by the inverse overlaps $M_{n_1,\dots,n_L}$. One can expect that its solution is unique up to an overall normalisation, which leads to  \eq{Mmu}. 
Since the overlap of two SoV vacua is $\bra{0}0\rangle=1$ in our conventions, we must have $M_{0,\dots,0}=1$ and this fixes the coefficient $\cN$ in terms of a combination of residues,
\beq
    \cN=\frac{1}{M'_{0,\dots,0}} \;.
\eeq
Now we can compute $M'_{n_1,\dots,n_L}$ quite directly again in terms of residues, leading to 
the correct result
\eq{mucompact} presented in section \ref{sec:sl2}.

\subsection{The $\sl(3)$ case}
\label{sec:intsl3}

In this section we generalise the derivation of the integral form for the scalar product and orthogonality relations for the Q-functions to the $\sl(3)$ case.

For $\sl(3)$ we have two Baxter equations, \eq{Bax1sl3} and \eq{Bax12s} that were discussed in section~\ref{sec:sl3int}. Like in the previous section we rewrite them in terms of two difference operators,
\beq
\label{Osl3}
   \hat O Q_1\equiv 
   \frac{Q_\theta^{[+2\bs+1]}Q_\theta^{[+2\bs-1]}}{Q_\theta^{[-2\bs+1]}}Q_1^{[+3]}
   - 
    \tau_{2}^{+}\frac{Q_\theta^{[+2\bs-1]}}{Q_\theta^{[-2\bs+1]}}Q_1^{+}
    +
     \tau_1^{-}Q_1^{-}
      -
Q_\theta^{[-2\bs-1]}Q_1^{[-3]}=0
\eeq
and
\beq
\label{Bax12s2}
	\hat { O}^\dagger Q_{12}=Q_\theta^{[-2\bs]}Q_{12}^{[-3]}-\tau_2Q_{12}^- + \tau_1Q_{12}^+ - Q_\theta^{[+2\bs]}Q_{12}^{[+3]}=0\;.
\eeq
We recall that $\tau_1$ and $\tau_2$ are related to the eigenvalues of the transfer matrices in the fundamental and twice antisymmetric representations defined in \eq{tau1Q}, \eq{tau2Q}.

The difference with the $\sl(2)$ case is that like in~\cite{Cavaglia:2019pow} we now have two operators
 $\hat O$ and $\hat { O}^\dagger$, which become related to each other under a scalar product of the form~\eq{scalargf}.
More precisely we will require that the measure $\mu_\alpha$
is such that
\beq
\label{OOb}
 \bl Q_1 \hat{ O}^\dagger f\br_\alpha = 0\;
\eeq
where $f$ has the large-$u$ asymptotics similar to $Q_{12}$ or $Q_{13}$, i.e. $\sim (\lambda_1\lambda_2)^{iu}u^t$ 
or $\sim (\lambda_1\lambda_3)^{iu}u^t$ for some powers $t$. 
In this section we will show that the measure factors $\mu_\alpha$ are in fact the same as in the $\sl(2)$ case i.e. \eq{defvarep} and \eq{muj}, in analogy with the case $\bs=1/2$~\cite{Cavaglia:2019pow}.
To verify that we have to move the contours of integration so that 
there are no shifts in the argument of $f$ in the l.h.s. of \eq{OOb},
    \beqa
\label{weakd3}
    \bl Q_1\hat {O}^\dagger  f\br_\alpha&=&
\int_{-\infty}^{+\infty} \mu_\alpha(u) Q_1(u)\( Q_\theta^{[-2\bs]}f^{[-3]} - \tau_2f^- + \tau_1f^+ -Q_\theta^{[+2\bs]}f^{[+3]} \)\frac{du}{2\pi i} \\ \nn
    &=&\int_{-\infty}^{+\infty} 
    \[
{\mu_\alpha^{[+3]}}
{Q_\theta^{[-2\bs+3]}}Q_{1}^{[+3]} - 
\mu_\alpha^+
\tau_2^+Q_{1}^+ + 
\mu_\alpha^-
{\tau_1^-}Q_{1}^- -
\mu_\alpha^{[-3]}
{Q_\theta^{[+2\bs-3]}}Q_{1}^{[-3]} \]f\;\frac{du}{2\pi i}
    \\ \nn
    &+& {\text{residues from poles}}\;,
\eeqa
where we indicate that there may be extra terms coming from poles of $\mu_\alpha$ which we will consider later. We would like the expression in square brackets in the second line to be proportional to $\hat O Q_1$ as then the result will simply be zero. Notice that we have  to match several terms in the operator $\hat O$ (defined in \eq{Osl3}) with only one function $\mu_\alpha$, so it is not trivial that a way to do this exists at all. However, from \eq{muj} we get
\beq
\label{arel}
    {\mu_\alpha^{[+3]}}=
    \frac{Q_\theta^{[+2\bs+1]}}{Q_\theta^{[-2\bs+3]}}
    \frac{Q_\theta^{[+2\bs-1]}}{Q_\theta^{[-2\bs+1]}}\mu^-_\alpha\;\;,\;\;
    {\mu_\alpha^{+}}=\frac{Q_\theta^{[+2\bs-1]}}{Q_\theta^{[-2\bs+1]}}\mu^-_\alpha\;\;,\;\;
    {\mu^{[-3]}_\alpha}=\frac{Q_\theta^{[-2\bs-1]}}{Q_\theta^{[+2\bs-3]}}{\mu_\alpha^{-}}\;
\eeq
which indeed gives $\mu_\alpha^- \hat OQ_1=0$ for the expression in the square brackets in \eq{arel}.

We now have to verify that there are no additional contributions from the poles, which one could potentially pick up when moving the integration contours around. Let us remind that $\mu_\alpha$ has simple poles
at $u=\theta_\beta-i\bs-in,\;n\ge 0$ for all $\beta=1,\dots,L$
and in addition has poles 
at $u=\theta_\alpha+i\bs+in,\;n\ge 0$ (with $\alpha$ fixed).
Since $\mu_\alpha$ is the only source of poles here, we see that for large enough $\bs$ there will be no poles at all to pick and thus the l.h.s. of \eq{arel} is indeed zero. At the same time, as all terms under the integral are analytic functions of $\bs$, so should be the integral. Thus we conclude that the poles should cancel when they are present.
In particular, for the case $\bs=1/2$ the cancellation of the poles was explicitly verified in~\cite{Cavaglia:2019pow}. We extended this consideration to the case $\bs>0$ in Appendix \ref{app:nopoles}.

Lastly, let us comment on convergence of the integrals in \eq{arel} at large $u$. As we already mentioned, below we will use this equation for the case when $f$ can have one of the two types\footnote{We also remind that in our conventions $\lambda_3=1/(\lambda_1\lambda_2)$.} of large $u$ behavior: either $\sim (\lambda_1\lambda_2)^{iu}u^t$ 
or $\sim (\lambda_1\lambda_3)^{iu}u^t$. Similarly to the $\sl(2)$ case (see the discussion above \eq{muj}) we will assume for definiteness that
\beq
\label{convsl3}
    0<\arg\lambda_2-\arg\lambda_1<\pi \ , \ \  0<\arg\lambda_3-\arg\lambda_1<\pi \;.
\eeq
These conditions ensure that the integral in \eq{arel} will be convergent for both choices of asymptotics of $f$. Also, like for $\sl(2)$, if e.g. the first inequality in \eq{convsl3} is violated, we should redefine $\mu_\alpha$ for the case when $f$ has asymptotics $f\sim (\lambda_1\lambda_2)^{iu}u^t$, by flipping the sign in the exponent in the denominator of \eq{muj}. Similarly, when $f\sim (\lambda_1\lambda_3)^{iu}u^t$ we redefine the measure in the same way when the second inequality in \eq{convsl3} is violated.

\paragraph{Integral orthogonality relation for the Q-functions.}

Now we are ready to derive the orthogonality relations for Q-functions following in analogy with what was done for $\sl(2)$ in section~\ref{sec:intsl2}.
 Let us consider again two different spin chain states labelled by $A$ and $B$ and take the combination
\beq
\label{Odif}
    \bl Q^A_1\({ \hat O^{\dagger B}}-{ \hat O^{\dagger A}}\)Q^B_{1,a+1}\br_\alpha =0 \ , \ \ \ a=1,2
\eeq
where we use the Q-functions $Q_{12}$ and $Q_{13}$ for the state $B$
and $Q_1$ for the state $A$. This expression is equal to zero due to \eq{Bax12s2} and \eq{OOb}. The operators  $\hat{\bar O}^A$ and $\hat{\bar O}^B$ differ only due to the different values of the transfer matrices $\tau_a^A$ and $\tau_a^B$ in their definition \eq{Osl3}, \eq{Bax12s2}  which encode the integrals of motion,
\beq
\label{IMs}
\tau_a(u) = u^L \chi_{a}(\lambda) + \sum_{\alpha=1}^{L} u^{\alpha-1} \, I_{a, \alpha-1}   ,\,\,\,\, a=1,2 \;,
\eeq
where $\chi_a(\lambda)$ is the state-independent coefficient that is simply the character of $\sl(3)$ in the $a$-th antisymmetric representation with eigenvalues given by the twists $\lambda_1,\lambda_2,\lambda_3$. With this notation, \eq{Odif} gives 
\beq\la{AB31}
\sum_{(b,\beta)=(1,1)}^{(2,L)}\bl Q_1^A \;u^{\beta-1}\; {\cal D}^{3-2b} Q_{1,a+1}^B \br_\alpha \;
\times\;(-1)^b\(I_{b,\beta-1}^A-I_{b,\beta-1}^B\)
= 0\; ,
\eeq
with $a=1,2$ and we
introduced the multi-index $(b,\beta)$, which takes $2L$ different values. Since for two different states at least one of the differences $I_{b,\beta-1}^A-I_{b,\beta-1}^B$ has to be zero, we find that the determinant of the linear system for these quantities in \eq{AB31} should vanish,
\beq\label{detsl3}
\det_{(a,\alpha),(b,\beta)}\bl Q_1^A \;u^{\beta-1}\; {\cal D}^{3-2b} Q_{1(a+1)}^B \br_\alpha=0\;.
\eeq

This is the key orthogonality relation, which generalizes the $\sl(2)$ relation \eq{orthsl2} to the $\sl(3)$ case.

This time, however, it is less obvious that \eq{detsl3}
has the form of the SoV product with some measure factor.
In section~\ref{sec:expmeasure} we show that indeed \eq{detsl3}
takes the exact form one gets for the scalar product of two wavefunctions in the SoV basis we built above in section \ref{sec:SoV}. 
Like in the $\sl(2)$ case, one can also argue that the number of the orthogonality relations in \eq{detsl3} is large enough to guarantee that we can actually deduce from it any element of the SoV overlap matrix $\langle \svy | \svx\rangle$. Indeed, since the entries of the matrix $\langle \svy | \svx\rangle^{-1}$ are rational functions\footnote{This follows from the fact that both $\langle\svx|$ and $|\svy\rangle$ are rational functions of $\bs$ which was demonstrated for example in \eqref{sl3x}.} of the spin $\bs$ we can explicitly solve for each block of fixed SoV charge $\langle \svy | \svx\rangle^{-1}$ by considering the finite-dimensional case $\bs\in\{0,-\frac{1}{2},-1,\dots\}$ with $-\bs$ large enough. Then we can simply analytically continue the result for that block to general values of $\bs$. We will do this calculation in section~\ref{sec:expmeasure}.

\subsection{General $\sl(N)$ case}\label{generalsln}
The integral form of the scalar product we obtained above for $\sl(2)$ and $\sl(3)$ spin chains can be generalized quite directly to any $\sl(N)$. In this section we will just present the result for the orthogonality relation, while the details of the derivation are given in Appendix~\ref{app:slNbax}.  The result is almost identical to that obtained for the $\bs=1/2$ case in~\cite{Cavaglia:2019pow} and reads
\beq\la{mdef0}
 \det_{(a,\alpha),(b,\beta)}
 \bl Q_1^A \;u^{\beta-1}\; {\cal D}^{-2b+N}\circ Q^{B,a+1} \br_\alpha \propto\; \delta^{AB}
\;.
\eeq
Here the indices take values $a,b=1,\dots,N$ and $\alpha,\beta=1,\dots,L$. We remind that here $\cD$ is the shift operator defined in \eq{ddef}. 
The only place where the $\bs$-dependence enters into this expression is through the definition of the double-brackets~
\eq{scalargf}, which contains the $\bs$-dependent factor $\mu_\alpha$ given in~\eq{muj}. In \eq{mdef0} we also introduced the notation for Q-functions with upper indices,
\beq
\label{Qupdef}
    Q^a=\epsilon^{b_1\dots b_{N-1}a}Q_{b_1\dots b_{N-1}} \ \  \ \ \text{(no summation over repeated indices)}  \ , 
\eeq
where $\epsilon$ is the fully antisymmetric tensor and $\epsilon^{12\dots N}=1$, while the indices are chosen as $\{b_1,\dots,b_{N-1}\}=\{1,\dots,N\} \backslash \{a\}$.  For example in the $\sl(3)$ case the functions $Q^a$ appearing in \eq{mdef0} are 
$Q^2=-Q_{13}$ and $Q^3=Q_{12}$, so that 
it reduces to the $\sl(3)$ result we gave above in \eq{AB31}.

In the next section we show how the relation~\eq{mdef0}
leads to an explicit expression for the SoV measure $M_{\svy,\svx}$.

\section{Explicit formula for the SoV measure}
\la{sec:expmeasure}
In this section we establish the relation between the operatorial SoV approach, discussed in section~\ref{sec:SoV}
on the example of $\sl(3)$, and the integral orthogonality relation we derived in the previous section~\ref{sec:measureint}.
As a result we will derive an explicit\footnote{The measure has implicitly been obtained in \cite{Cavaglia:2019pow,Gromov:2019wmz} and then later in \cite{Maillet:2020ykb}.} formula for the SoV measure for general $\sl(N)$.

\subsection{Comparison with the SoV construction for $L=2$ case}

In order to see how the relation with the SoV approach works, we first study the case of short length $L=2$ for the $\sl(3)$ spin chain explicitly in detail. In the next section 
we discuss arbitrary length spin chains and then consider the general $\sl(N)$ case.

We start from the integral orthogonality relation given by the determinant \eq{detsl3} which for $L=2$ reads
\beq\la{Det2}
 d_2\equiv    \left| \begin{array}{cccc}
       \bl Q_1 Q_{12}^-\br_1  & \bl Q_1 Q_{12}^- u\br_1 & \bl Q_1 Q_{12}^+\br_1  & \bl Q_1 Q_{12}^+ u\br_1 \\
      \bl Q_1 Q_{12}^-\br_2  & \bl Q_1 Q_{12}^- u\br_2 & \bl Q_1 Q_{12}^+\br_2  & \bl Q_1 Q_{12}^+ u\br_2 \\
      \bl Q_1 Q_{13}^-\br_1  & \bl Q_1 Q_{13}^- u\br_1 & \bl Q_1 Q_{13}^+\br_1  & \bl Q_1 Q_{13}^+ u\br_1 \\
       \bl Q_1 Q_{13}^-\br_2  & \bl Q_1 Q_{13}^- u\br_2 & \bl Q_1 Q_{13}^+\br_2  & \bl Q_1 Q_{13}^+ u\br_2 
    \end{array} \right|\;.
\eeq
Here as well as below we omitted the indices $A$ and $B$, indicating the state, for clarity. In order to make the connection with the operatorial SoV approach we rewrite the determinant as 
\beq\la{d2ints}
d_2 =
 \int t_2(\{u_{\alpha,a}\})\prod_{\alpha=1}^L\prod_{a=1}^2\frac{{\rm d}{u}_{\alpha,a}}{2\pi i} Q_1({u}_{\alpha,a})\mu_\alpha({u}_{\alpha,a})
\eeq
where 
\beq
t_2(\{u_{\alpha,a}\})= \left| \begin{array}{cccc}
       Q_{12}^-(u_{11})  &  u_{11} Q_{12}^-( u_{11}) & Q_{12}^+( u_{11})  &  u_{11} Q_{12}^+( u_{11}) \\
       Q_{12}^-( u_{21})  &  u_{21} Q_{12}^-( u_{21}) & Q_{12}^+( u_{21})  &  u_{21} Q_{12}^+( u_{21}) \\
       Q_{13}^-( u_{12})  &  u_{12} Q_{13}^-( u_{12}) & Q_{13}^+( u_{12})  &  u_{12} Q_{13}^+( u_{12}) \\
       Q_{13}^-( u_{22})  &  u_{22} Q_{13}^-( u_{22}) & Q_{13}^+( u_{22})  &  u_{22} Q_{13}^+( u_{22}) \\
    \end{array} \right|\;.
\eeq
Note that all terms in the integral, except $t_2(\{u_{\alpha,a}\})$, are symmetric under the permutations of $u_{\alpha,a}$ for each $\alpha$ separately.
Computing the determinant explicitly we observe that up to permutation we have the following
equation
\beqa\la{t2FF}
 t_2(\{u_{\alpha,a}\})&\simeq&
(u_{11}-u_{21})  (u_{12}-u_{22} )
F_1^{0,-1} F_2^{0,-1}\\
\nn &-&\frac{(u_{11}-u_{12} )
    (u_{21}-u_{22} )}{4} 
    \(F_1^{0,0}F_2^{-1,-1} 
    + F_1^{-1,-1}F_2^{0,0}\)\;
\eeqa
where $\simeq$ indicates that the equality holds up to the permutations.
We also introduced the notation
\beq\la{Fass}
F_\alpha^{s_1,s_2}=
Q_{12}(u^\alpha_1 +i s_1+\tfrac{i}{2})
Q_{13}(u^\alpha_2 +i s_2+\tfrac{i}{2})
-
Q_{13}(u^\alpha_1 +i s_1+\tfrac{i}{2})
Q_{12}(u^\alpha_2 +i s_2+\tfrac{i}{2})\;.
\eeq
We assume that the twists satisfy
\beq\la{cindition}
|\lambda_1|>|\lambda_3|\;\;,\;\;
|\lambda_1|>|\lambda_2|\;,
\eeq
as in section \ref{sec:SoV} (where this condition ensured that the states we built actually lie in the Hilbert space). This means that
we can close the integration contour in the upper half plane for all the integrals in~\eq{d2ints}, and evaluate them by picking the poles of $\mu_\alpha$ (defined in \eq{muj}
and \eq{defvarep}) at
\beq\la{polesu}
    u=\theta_\alpha+i\bs+in\;\;,\;\;n\ge 0\;.
\eeq
At these points the factor $\mu_\alpha$
has  simple poles, with  residues
given by a product of Pochhammer
functions defined in \eq{resm}. For example, consider the first term
in \eq{t2FF}, which gives the following contribution 
to the result in \eq{d2ints}:
\beq\la{smR}
d_2^{I} =
 \int (u_{11}-u_{21})  (u_{12}-u_{22} )
F_1^{0,-1} F_2^{0.-1}\prod_{\alpha=1}^L\prod_{a=1}^2\frac{{\rm d}{u}_{\alpha,a}}{2\pi i} Q_1({u}_{\alpha,a})\mu_\alpha({u}_{\alpha,a})
\eeq
which we can now write as a sum over residues 
\beq
d_2^I=
\sum_{n_{11}=0}^\infty
\sum_{n_{12}=0}^\infty
\sum_{n_{21}=0}^\infty
\sum_{n_{22}=0}^\infty
R^I_{n_{11}n_{12}n_{21}n_{22}}
\eeq
where
\beqa\la{Rdef}
R^I_{n_{11}n_{12}n_{21}n_{22}}&=&(\svx_{11}-\svx_{21})(\svx_{12}-\svx_{22})
\prod_{\alpha,a}\[ r_{\alpha,n_{\alpha a}} Q_1(\svx_{\alpha a})\]\\
&\times&\[Q_{12}(\svx_{11} + \tfrac{i}{2})
Q_{13}(\svx_{12}- \tfrac{i}{2})
-
Q_{13}(\svx_{11} + \tfrac{i}{2})
Q_{12}(\svx_{12}-\tfrac{i}{2})\]\\
&\times&\[Q_{12}(\svx_{21} + \tfrac{i}{2})
Q_{13}(\svx_{22}- \tfrac{i}{2})
-
Q_{13}(\svx_{21} + \tfrac{i}{2})
Q_{12}(\svx_{22}-\tfrac{i}{2})\]
\eeqa
and
$\svx_{\alpha a}=\theta_\alpha+i\bs+i n_{\alpha,a}$.
This already has a form familiar from the SoV approach~\eq{oversov} if we identify
$\svy_{\alpha 1}=\svx_{\alpha 1},\;
\svy_{\alpha 2}=\svx_{\alpha 2}-i$, which gives
\beqa\la{Myx}
\left. M_{\svy,\svx}\right|_{\svy_{\alpha 1}=\svx_{\alpha 1},\svy_{\alpha 2}=\svx_{\alpha 2}-i}=\frac{1}{{\cal N}}(\svx_{11}-\svx_{21})(\svx_{12}-\svx_{22})
\prod r_{\alpha,n_{\alpha,a}} \ .
\eeqa
The normalisation factor can be fixed by requiring
that for $n_{\alpha,a}=0$ the r.h.s. gives identity.
This results in
\beq
{\cal N}=(\svx_{10}-\svx_{20})^2
\prod_{\alpha=1}^L r_{\alpha,0}^2\;.
\eeq
This is already highly non-trivial as we should be able to reproduce all diagonal elements of the matrix of \eq{cherge01} and \eq{cherge2}. For example taking
$n_{2,1}=2$ and all other $n_{\alpha,a}=0$ we get
from \eq{Myx}
\beq
M^{0020}_{0020}=
\frac{\theta_{1}-\theta_{2}-2i}{\theta_{1}-\theta_{2}}
\frac{ r_{2,2} }{ r_{2,0}}
=
-\frac{\bs (2 \bs+1) \left(2 \bs+i \theta _{12}\right) \left(i \theta _{12}+2
   \bs+1\right)}{\theta _{12} \left(\theta _{12}-i\right)}
\eeq
which perfectly reproduces the $(3,3)$ element of the matrix \eq{cherge2}!
We introduced the notation $M^{m_{11}m_{12}m_{21}m_{22}}_{n_{11}\;n_{12}\;n_{21}\;n_{22}}=M_{\svy,\svx}$ where $\svy$ and $\svx$
are associated to $n$'s and $m$'s in the usual way
\eq{ntox} and \eq{mtoy}.

Now notice that for SoV eigenvalues we have to impose inequalities \eq{neq}
whereas \eq{smR} has the sum running over all positive $n's$. To account for that let us split the sum in \eq{smR} into $4$ parts depending on $n_{\alpha 1}\geq n_{\alpha 2}$ or $n_{\alpha 1}< n_{\alpha 2}$ for $\alpha=1,2$.
Then only one of four parts $n_{\alpha 1}\geq n_{\alpha 2}$
actually corresponds to \eq{Myx}.
Now consider the case $n_{11}< n_{12}$ and $n_{21}\geq n_{22}$. As in this case we violate the inequality \eq{neq}
we better replace the names of the summation labels $n_{12}\leftrightarrow n_{11}$.
After that replacement and slight rearrangements in \eq{Rdef}  we get
\beqa
R^I_{n_{12}n_{11}n_{21}n_{22}}&=&-(\svx_{12}-\svx_{21})(\svx_{11}-\svx_{22})
\prod\[ r_{\alpha,n_{\alpha,a}} Q_1(\svx_{\alpha a})\]\\
&\times&\[
Q_{12}(\svx_{11}-\tfrac{i}{2})
Q_{13}(\svx_{12} + \tfrac{i}{2})
-
Q_{13}(\svx_{11}- \tfrac{i}{2})
Q_{12}(\svx_{12} + \tfrac{i}{2})
\]
\\
&\times&\[Q_{12}(\svx_{21} + \tfrac{i}{2})
Q_{13}(\svx_{22}- \tfrac{i}{2})
-
Q_{13}(\svx_{21} + \tfrac{i}{2})
Q_{12}(\svx_{22}-\tfrac{i}{2})\]\;,
\eeqa
which under identification
$\svy_{11}=\svx_{11}-i$, $\svy_{12}=\svx_{12}$
and
$\svy_{11}=\svx_{11}-i$, $\svy_{12}=\svx_{12}$,
which it terms of $n$'s and $m$'s gives
$m_{11}=n_{11}-1$, $m_{12}=n_{12}+1$ and $m_{2a}=n_{2a}$
leads to
\beqa\la{M2t}
M^{
n_{11}-1,n_{12}+1,n_{21}n_{22}
}_{
n_{11},\;\;\;\;\,n_{12},\;\;\;\;\,n_{21}n_{22}}=-\frac{1}{\cal N}(\svx_{12}-\svx_{21})(\svx_{11}-\svx_{22})
\prod r_{\alpha,n_{\alpha,a}}\;.
\eeqa
For example taking $n_{\alpha,a}={2,0,0,0}$ we obtain from \eq{M2t}
\beq
M^{1100}_{2000}=\frac{\theta_{12}+2i}{\theta_{12}}\frac{r_{12}}{r_{10}}=+
\frac{\bs (2 \bs+1) \left(2 \bs-i \theta _{12}\right) \left(-i \theta _{12}+2
   \bs+1\right)}{\theta _{12} \left(\theta _{12}+i\right)}\;
\eeq
in agreement with
the element $2,5$ of the matrix~\eq{measuredata}.
There are two other orderings in $d_2^I$  
producing
$M^{
n_{11}n_{12}n_{21}-1,n_{22}+1
}_{
n_{11}n_{12}n_{21},\;\;\;\;\,n_{22}}$
and $M^{
n_{11}-1,n_{12}+1,n_{21}-1,n_{22}+1
}_{
n_{11},\;\;\;\;\,n_{12},\;\;\;\;\,n_{21},\;\;\;\;\,n_{22}}$
in an analogous way.

Finally let us consider a slightly different type of terms in \eq{t2FF} with $F_1^{0,0}F_2^{-1,-1}$.
It is clear that this time each ordering of $n$'s will contribute in the same way so we can remove the factor $1/4$
and assume that $n_{11}> n_{12}$
and $n_{21}> n_{22}$ (when $n_{11}=n_{12}$
or $n_{21}=n_{22}$ we simply get zero).
Repeating the same argument as before we deduce 
\beqa
M_{
n_{11}n_{12},\;\;\;\;\,n_{21},\;\;\;\;\,n_{22}}^{
n_{11}n_{12}+1,n_{21}-1,n_{22}
}=-\frac{1}{\cal N}(\svx_{12}-\svx_{21})(\svx_{11}-\svx_{22})
\prod r_{\alpha,n_{\alpha,a}}
\eeqa
and finally the last term in \eq{t2FF} gives
\beqa
M_{
n_{11},\;\;\;\;\,n_{12}n_{21},n_{22}}^{
n_{11}-1,n_{12}n_{21},n_{22}+1
}=-\frac{1}{\cal N}(\svx_{12}-\svx_{21})(\svx_{11}-\svx_{22})
\prod r_{\alpha,n_{\alpha,a}}\;.
\eeqa
We see that the structure of the result is very suggestive.
In the next section we will generalise the above derivation to general length $L$.

\subsection{General $L$ expression for $\sl(3)$ measure}\la{intmeasure}
In order to generalize the derivation in the previous section to any $L$ our starting point is again the determinant in the l.h.s. of \eq{detsl3},
\beq\label{detsl3:2}
d_L\equiv \det_{(a,\alpha),(b,\beta)}\bl Q_1 \;u^{\beta-1}\; {\cal D}^{3-2b}\; Q_{1,a+1} \br_\alpha\;.
\eeq
Then we drag out $2L$ integrations out of the determinant to obtain
\beq\la{d2intsL}
d_L =
 \int t_L(\{u_{\alpha,a}\})\prod_{\alpha=1}^L\prod_{a=1}^2\frac{{\rm d}{u}_{\alpha,a}}{2\pi i} Q_1({u}_{\alpha,a})\mu_\alpha({u}_{\alpha,a})
\eeq
where
\beq\la{tdet}
t_L(\{u_{\alpha,a}\})\equiv \det_{(a,\alpha),(b,\beta)}
\left[
\;u_{\alpha,a}^{\beta-1}\;  Q_{1,a+1}\(u_{\alpha,a}+i \frac{3-2b}{2}\) \right]\;.
\eeq
We use the following relation for the determinants
\beq\la{relation}
\det_{(a,\alpha),(b,\beta)}
H_{\alpha,a,\beta} G_{\alpha,a,b}=
\sum_{\sigma}(-1)^{|\bigcup\limits_{a}\sigma^{-1}(a)|}
\prod_{a}
\(
\det_{(\alpha,b)\in \sigma^{-1}(a),\beta}
H_{\alpha,b,\beta}
\prod_{\alpha}
G_{\alpha,a,\sigma_{\alpha,a}}
\)
\eeq
where the indices $a,b\in [1,\dots,K]$
and $\alpha,\beta\in [1,\dots,L]$,
$F$ and $G$ are two arbitrary tensors with $3$ indices.
In the r.h.s. we are summing over all
permutations $\sigma$ of the $L$ copies of ranges of numbers $1,\dots,K$ with $\sigma_{\alpha,b}$
denoting the number at the location
$b+(\alpha-1)K$. We also indicated that in the r.h.s.  the determinant is computed for the $L\times L$  matrix whose columns are labelled by  $\beta$, and whose rows are labelled by pairs $(\alpha,b)$ such that $\sigma_{\alpha,b}=a$ (there are $L$ such pairs).  The relation~\eq{relation} is easy to verify and we will use in the derivation below.

In application to our determinant~\eq{tdet} we get
\beq\la{relation1}
t_L(\{u_{\alpha,a}\})=
\sum_{\sigma}(-1)^{|\bigcup\limits_{a}\sigma^{-1}(a)|}
\Delta_{1}\Delta_{2}
\prod_{\alpha}
Q_{1,2}\(u_{\alpha,1}+i s_{\alpha,1}+\tfrac{i}{2}\)
Q_{1,3}\(u_{\alpha,2}+i s_{\alpha,1}+\tfrac{i}{2}\)\;\;,\;\;
\eeq
where $s_{\alpha,a}=1-\sigma_{\alpha,a}$
and $\Delta_b$ is the Vandermonde determinant, build out of $u_{\alpha,a}$ for which $\sigma_{\alpha,a}=b$.

Finally, in order to bring it close to the SoV
form we have to use that all terms
in \eq{d2intsL} (except for the $t_L$) 
are invariant under
$u_{\alpha,1}\leftrightarrow u_{\alpha,2}$.
For $t_L$, interchanging 
$u_{\alpha,1}\leftrightarrow u_{\alpha,2}$
is equivalent to interchanging 
$Q_{12}(u_{\alpha,1}+\dots)$ with $Q_{13}(u_{\alpha,1}+\dots)$
and $Q_{13}(u_{\alpha,2}+\dots)$ with $Q_{12}(u_{\alpha,2}+\dots)$
 and changing the overall sign of $t_L$,
 as it is clear from the initial determinant form of $t_L$.
I.e. that is equivalent to antisymmetrizing 
$Q_{12}$ and $Q_{13}$ in the last term under the product of \eq{relation1}, which will then allow us to rewrite the result in terms of $F_{\alpha}^{s_{\alpha,1},s_{\alpha,2}}$
\beqa\la{relation2}
&&\underset{{u_{\alpha,1}\leftrightarrow u_{\alpha,2}}}{\rm sym}t_L(\{u_{\alpha,a}\})=
\frac{1}{2^L}\sum_{\sigma}(-1)^{|\bigcup\limits_{a}\sigma^{-1}(a)|}
\prod_{a}
\Delta_a
\prod_{\alpha}
F_\alpha^{s_{\alpha,1},s_{\alpha,2}}\;.
\eeqa

Now the expression \eq{relation2} is ready to go under the integration \eq{d2intsL}. Closing the contour in the upper half plane 
and evaluating the integration by residues we  pick up poles
coming from the integration factors $\mu_\alpha$
at $u_{\alpha,a}=\theta_\alpha+i\bs+in_{\alpha,a}$
where $n_{\alpha,a}\geq 0$ and otherwise are unconstrained.
By construction the integrand is now invariant under $u_{\alpha,1}\leftrightarrow u_{\alpha,2}$ for every value of $\alpha=1,\dots,L$ and so the residues are also symmetric under $n_{\alpha,1}\leftrightarrow n_{\alpha,2}$.
Using this symmetry we can remove the factor $1/2^L$
and impose $0\leq n_{\alpha,2}\leq n_{\alpha,1}$.
The only potential problem could be that in this way we take the contributions with $n_{\alpha,2}=n_{\alpha,1}$ into account multiple times -- we will see in a moment that we do not. 

Thus we can read off the following expression for the measure factor $M_{\svy,\svx}=M^{\{m_{\alpha,a}\}}_{\{n_{\alpha,a}\}}$
\beq\la{Msl3res}
\left.M_{\svy,\svx}\right|_{m_{\alpha,a}=n_{\alpha,a}-\sigma_{\alpha,a}+\sigma^0_{\alpha,a}}
=(-1)^{|\sigma|}
\frac{
\Delta_1
\Delta_2
}{
\Delta_0^2
}
\prod \frac{r_{\alpha,n_{\alpha,1}}r_{\alpha,n_{\alpha,2}}}{r^2_{\alpha,0}}
\eeq
where $r_{\alpha,n}$ is defined in \eq{resm},
$\sigma^0_{\alpha,a}=a$pence  is a trivial permutation and $\Delta_0 and $ is the Vandermonde determinant build out of $\theta_\alpha$, as before is has to be added to 
ensure our normalisation with $M_{0,0}=1$ holds.
We also denote $|\sigma|$ is the number of elementary permutations needed to bring the set $\bigcup\limits_{a} u_{\sigma^{-1}(a)}$ to the canonical order $\bigcup\limits_{a} u_{\sigma_0^{-1}(a)}=u_{11},u_{12},\dots,u_{L1},u_{L2}$.

It remains to check that we have not over-counted the $n_{\alpha,1}=n_{\alpha,2}$ cases. Note that if at the same time $s_{\alpha,1}=s_{\alpha,2}$, then $F^{s_{\alpha,1},s_{\alpha,2}}=0$
and we do not have to worry about this case. Thus the dangerous situations are $\sigma_{\alpha,1}=1,\;\sigma_{\alpha,2}=2$ and $\sigma_{\alpha,1}=2,\;\sigma_{\alpha,2}=1$. But in the latter
case we get $m_{\alpha,1}=n_{\alpha,1}-2+1< m_{\alpha,2}=n_{\alpha,2}-1+2$ which is prohibited due to~\eq{sl3x} for example, so there is only one contribution coming from $n_{\alpha,1}=n_{\alpha,2}=m_{\alpha,1}=m_{\alpha,2}$ and thus \eq{Msl3res} is valid as is.

\subsection{General expression for the measure}
The general $\sl(N)$ case is almost completely clear after the previous two derivations. We start from the integral orthogonality relation~\eq{mdef0} and pull out $L(N-1)$ integrations,
factors of $Q_1(u_{\alpha,a})$ and the measure factors
$\mu_\alpha(u_{\alpha,a})$
\beq\la{d2intsN}
d_L =
 \int t_L(\{u_{\alpha,a}\})\prod_{\alpha=1}^L\prod_{a=1}^{N-1}\frac{{\rm d}{u}_{\alpha,a}}{2\pi i} Q_1({u}_{\alpha,a})\mu_\alpha({u}_{\alpha,a})\;.
\eeq
Then we apply the identity \eq{relation}
to the remaining integrand $t_L(\{u_{\alpha,a}\})$ and symmetrise in
permutations of $u_{\alpha,a}$ for each given $\alpha$,
which is equivalent to interchanging indices of $Q^a$.
Similarly to the $\sl(3)$ example we arrive to the following result
\beqa\la{relation2N}
&&\underset{\{{u_{\alpha,1},\dots ,u_{\alpha,N-1}\}}}{\rm sym}t_L(\{u_{\alpha,a}\})=
\sum_{\sigma}{(-1)^{|\sigma|}}
\[\prod_{a}
\Delta_a\]
\prod_{\alpha}
\frac{
F_\alpha^{s_{\alpha,1},s_{\alpha,2},\dots,s_{\alpha,N-1}}}
{(N-1)!}\;
\eeqa
where $\sigma$ is a permutation of $(N-1)L$ numbers
$1,2,\dots,N-1,1,2,\dots$.
We also introduced the generalisation of \eq{Fass} as a $(N-1)\times(N-1)$ determinant
\beq
F_\alpha^{s_{\alpha,1},s_{\alpha,2},\dots,s_{\alpha,N-1}}=
\det_{a,b} Q^{(1+a)}\(u_{\alpha,b}+i s_{\alpha,b}+i\frac{N-2}{2}\)\;
\eeq
where $s_{\alpha,a}=1-\sigma_{\alpha,a}\leq 0$.

Like in the $\sl(3)$ case we then have to close the contour in the upper half plane (which strictly speaking requires $|\lambda_1|>|\lambda_2|>\dots >|\lambda_N|$)
and rewrite the integral \eq{d2intsN}
as a sum over poles at $u_{\alpha,a}=\theta_\alpha+i\bs+i n_{\alpha,a}$ for $n_{\alpha,a}>0$.
At first let us assume that all $n_{\alpha,a}$ are 
all different for a given $\alpha$, then 
we can restrict ourselves to $n_{\alpha,1}\geq n_{\alpha,2}\geq\dots \geq n_{\alpha,N-1}\geq 0$ using the symmetry of the integrand by removing $(N-1)!$'s from the denominator of \eq{relation2N}.
If there are some $n_\alpha's$ which are equal among each other,
then the number of equivalent permutations of $n_{\alpha,a}$
is less then $N!$ and we have to compensate for the overcounting by dividing for each $\alpha$ by 
\beqa
M_\alpha\equiv \frac{N!}{\#{\rm perm}\{ n_{\alpha,a} \}_{a=1}^{N-1} }\;.
\eeqa
Thus up to an overall factor we get
\beqa\la{dlres}
d_L\propto &\sum\limits_{n_{\alpha,1}\geq\dots\geq n_{\alpha,N-1}\geq 0}&
\prod_{\alpha}\frac{1}{M_{\alpha}}
\prod_{\alpha,a} Q_1(\svx_{\alpha,a})r_{\alpha,n_{\alpha,a}}\\
\nn&\times&
\sum_{\sigma}{(-1)^{|\sigma|}}
\prod_{a}
\Delta_a
\prod_{\alpha}
\[\det_{a,b} Q^{(1+a)}\(\svy_{\alpha,0}+i n_{\alpha,b}+i s_{\alpha,b}+i\frac{N-2}{2}\)\]\;
\eeqa
where $\svx_{\alpha,a}=\theta_\alpha+i\bs+i n_{\alpha,a}$
and $\svy_{\alpha,0}=\theta_\alpha+i\bs$.
In analogy with $\sl(3)$ we define
\beq
\svy_{\alpha,a}=\svy_{\alpha,0}+i m_{\alpha,a}-i(a-1)\;\;,\;\;
m_{\alpha,1}\geq\dots m_{\alpha,N-1}\geq 0\;.
\eeq
We show in section~\ref{sec:secslN} that this in indeed the correct identification for the spectrum of the roots of the $\bC(u)$ operator. Note that for every combination of $n_{\alpha,a}$
and $s_{\alpha,a}$ there is a unique way to find $\svy_{\alpha,a}$
such that the expression in the brackets in \eq{dlres}
matches
\beq
\det_{a,b} Q^{(1+a)}\(\svy_{\alpha,b}+i\frac{N-2}{2}\)\;
\eeq
up to a sign. For that we have to find a permutation $\rho^\alpha$ of $[1,\dots,N]$ such that for each fixed $\alpha$
\beq
m_{\alpha,a}+s^0_{\alpha,a}=n_{\alpha,\rho^\alpha_a}+s_{\alpha,\rho^\alpha_a}\;\;,\;\;a=1,\dots,N
\eeq
where $s^0_{\alpha,a}=1-a$. For that we simply have to order the numbers
$n_{\alpha,a}+s_{\alpha,a}$. Some observations are in order:
Firstly, all $\{n_{\alpha,a}+s_{\alpha,a}\}_{a=1}^{N-1}$ must be distinct, as otherwise the determinant will vanish. This
means that their ordering produces a unique permutation $\rho_a$.
Secondly, the number in the l.h.s.
satisfy strict inequality $m_{\alpha,1}+s^0_{\alpha,1}>m_{\alpha,2}+s^0_{\alpha,2}>\dots>
m_{\alpha,N-1}+s^0_{\alpha,N-1}\geq 2-N
$ at the same time
$n_{\alpha,a}+s_{\alpha,a}\geq 2-N$ meaning that we will always be able to find unique set $\{m_{\alpha,a}\}$, satisfying the required inequalities for any $n_{\alpha,a}$
and $s_{\alpha,a}$. Explicitly for each $\alpha$ we find $m_{\alpha,a}=\({\rm sort}\{n_{\alpha,b}+s_{\alpha,b}\}\)_{\alpha,a}-s_{\alpha,a}^0$. Where we introduced the notation ${\rm sort}$, for a function which implements the sorting permutation $\rho_{a}^\alpha$.
We need to keep track of the signature of the permutation $\rho^\alpha$,
as this would affect the sign of the result.
Thus we conclude that we get the following contribution to the measure
\beq\la{Msl3resN}
\left.M^\sigma_{\svy,\svx}\right|_{m_{\alpha,a}=
\({\rm sort}\{n_{\alpha,b}-\sigma_{\alpha,b}\}\)_{\alpha,a}+\sigma_{\alpha,a}^0
}
=(-1)^{|\sigma|}
\(\prod_{a=1}^{N-1}
\frac{
\Delta_a
}{
\Delta_0
}\)
\prod_{\alpha=1}^L
\frac{(-1)^{|\theta_\alpha|}}{M_{\alpha}}
\prod_{a=1}^{N-1} \frac{r_{\alpha,n_{\alpha,a}}}{r_{\alpha,0}}\;.
\eeq
In practice we will have to determine for given 
properly ordered sets
$\{n_{\alpha,a}\}$ and
$\{m_{\alpha,a}\}$
what is the value of the corresponding $M_{\svy,\svx}$.
For that we have to sum over all possible permutations $\sigma$'s
for which the relation between $m$'s and $n$'s holds. 
Finally, we can simplify our result a bit by noticing that 
when we have a degeneracy in $n_{\alpha,a}$ we also have several $\sigma$'s
which give the same result -- and their number is exactly $M_\alpha$.
So instead of summing over all $\sigma's$ it simpler
to sum over all inequivalent permutations of $n_{\alpha,a}$
(within each $\alpha$). Denoting such permutations by $k$ we then get
\beq\la{Msl3resNfinal}
M_{\svy,\svx}=
\sum_{k={\rm perm}_\alpha n}
\left.
{\rm sign}(\sigma)
\(\prod_{a=1}^{N-1}
\frac{
\Delta(\svx_{\sigma^{-1}(a)})
}{
\Delta(\{\theta_a\})
}\)
\prod_{a=1}^{N-1} \frac{r_{\alpha,n_{\alpha,a}}}{r_{\alpha,0}}\right|_{\sigma_{\alpha,a}=k_{\alpha,a}-m_{\alpha,a}+a}.
\eeq
To have all notations summarised in one place let us remind that
$\sigma_{\alpha,a}$ should be one of $\frac{(N-1)L!}{L!^{N-1}}$ permutations of the numbers $1,2,3,\dots,(N-1)$ repeated $L$ times, otherwise we define ${\rm sign}(\sigma)=0$; we also define
$\svx_{\sigma^{-1}(a)}=\{\svx_{\alpha,b}: \sigma_{\alpha,b}=a\}$.
The signature of the permutation 
${\rm sign}(\sigma)$ is $\pm 1 $ depending on the number of elementary permutations needed to bring the ordered set $u_{\sigma^{-1}(1)}\cup u_{\sigma^{-1}(2)}\dots \cup
u_{\sigma^{-1}(N-1)}$ to the canonical order $u_{1,1},u_{1,2},\dots,u_{L,N-1}$. Whereas ${\rm sign}(\sigma)$
could be ambiguous due to different possible orderings inside $\sigma^{-1}(a)$, the combination with the Vandermonds $\Delta(x_{\sigma^{-1}(a)})$
it is a well defined. Finally $r_{\alpha,n}$ is the only $\bs $-dependent factor which is defined in \eq{resm}.

In  appendix~\ref{app:code} we give a simple {\it Mathematica} code
which computes the measure element for given $n$'s and $m$'s, implementing \eq{Msl3resNfinal}.
\begin{figure}
    \centering
    \includegraphics{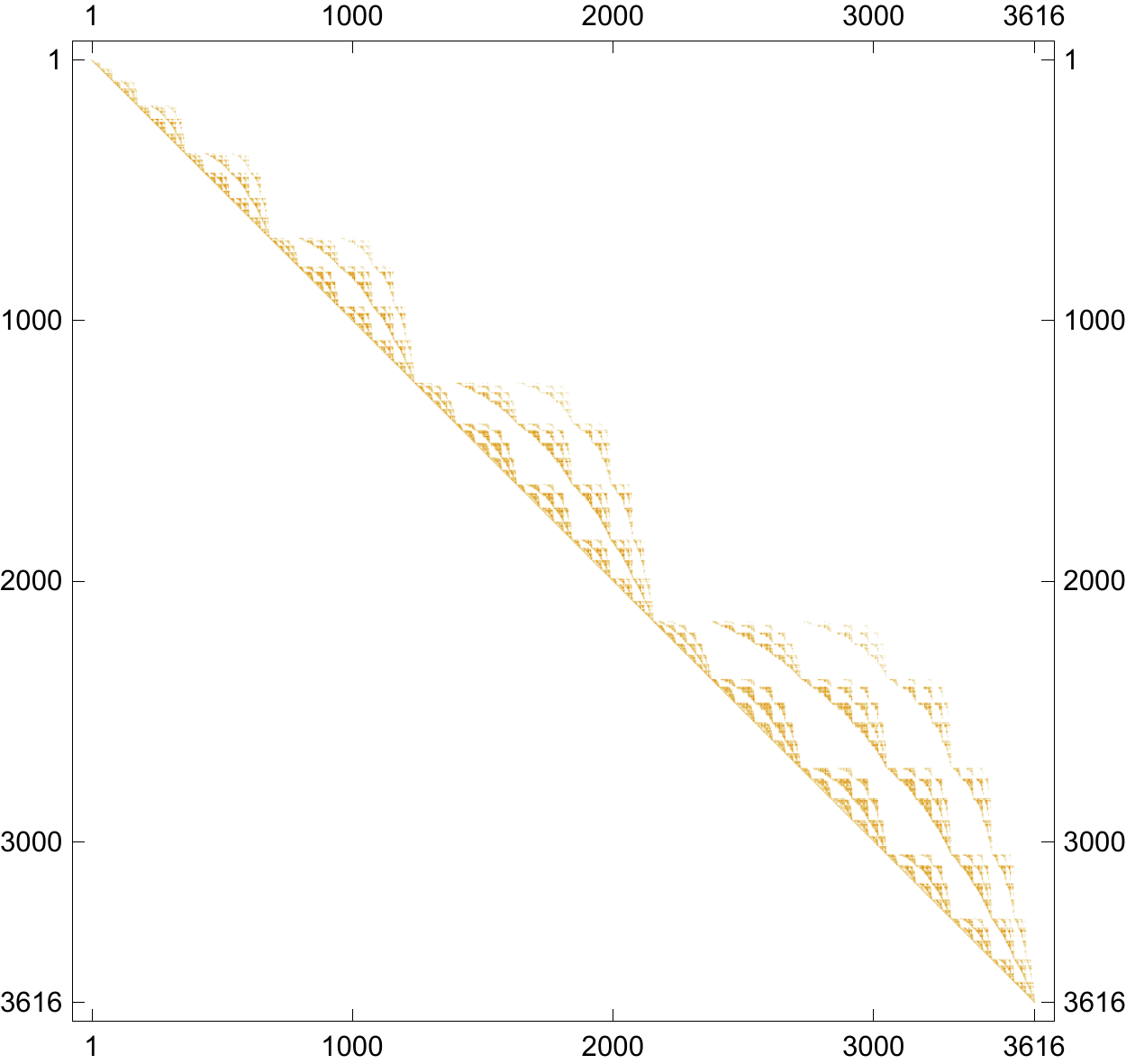}
    \caption{Structure of the measure matrix $M_{\svy,\svx}$ for $\sl(4)$ length $L=3$ spin chain with all $3616$ SoV states up to SoV charge ${\bf N}=10$. All non-zero elements denoted by a yellow pixel. Within each fixed SoV charge block the states $\svx$ and $\svy$ are ordered lexicographically according to the words $(n_{1,1},n_{2,1},n_{1,2},n_{2,2},n_{1,3},n_{2,3})$. The matrix is upper triangular and has a fractal-like self-repetitive structure.}
    \label{fig:my_label}
\end{figure}

\section{Extension to $\sl(N)$ spin chains}\la{sec:secslN}
The integral representation give a sharp suggestion on what the 
spectrum of the SoV operators in $\sl(N)$ case should be and how they should give rise to the measure factor, for which we produced a prediction in the previous section. Here we extend the results described in section \ref{sec:measureint} 
for $\sl(3)$ to $\sl(N)$. We will also fill some gaps in the previous discussions. In particular we demonstrate how to diagonalise the $\bB$ and $\bC$ operators by introducing new commutation relations between them and certain transfer matrices generalising the relation first obtained in \cite{Ryan:2018fyo}.

We begin by reviewing the key tools and relations and then proceed with the generalisation.

\subsection{Quantum minors}

A useful tool when dealing with the higher-rank $\sl(N)$ case are the so-called quantum minors which are certain anti-symmetric combinations of monodromy matrices where ``quantum" refers to the presence of extra shifts which disappear in the classical limit. The quantum minors $\bT\left[^{i_1 \dots i_a}_{j_1\dots j_a}\right](u)$, $n=1,2,\dots,N$, are defined as 
\begin{align}\label{qminors}
    \bT\left[^{i_1 \dots i_a}_{j_1\dots j_a}\right](u) & =\sum_{\sigma}(-1)^{{\rm deg}\,\sigma\,}\bT_{i_{1}j_{\sigma(1)}}(u+i(a-1))\bT_{i_{2}j_{\sigma(2)}}(u+i(a-2))\dots \bT_{i_{a}j_{\sigma(a)}}(u) \\
    &= \sum_{\sigma}(-1)^{{\rm deg}\,\sigma\,}\bT_{i_{\sigma(1)}j_1}(u)\bT_{i_{\sigma(2)}j_2}(u+i)\dots \bT_{i_{\sigma(a)}j_a}(u+i(a-1))
\end{align}
where the sum is over all elements $\sigma$ of the permutation group $S_a$ of $a$ elements. Note that these objects \eq{qminors} can also be identified as elements of the monodromy matrices in anti-symmetric representations. In other words in \eq{qminors} the fusion procedure~\cite{Zabrodin:1996vm} is performed directly at the level of the monodromy matrix instead of the Lax operators like in \eqref{La1}. The transfer matrices in anti-symmetric representations $\bbT_{a,1}(u)$ are then obtained as sums of quantum minors of a given size
\begin{equation}\la{Ta1minors}
    \bbT_{a,1}(u)=\sum_{1\leq i_1<\dots<i_a\leq N}\bT\left[^{i_1 \dots i_a}_{j_1\dots j_a}\right]\left(u-\frac{i}{2}\left(a-1\right)\right)\,.
\end{equation}

\begin{figure}[htb]
  \centering
  \includegraphics[width=70mm,scale=10]{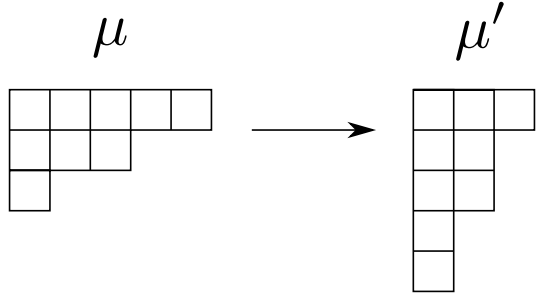}
  \caption{Transposition of Young diagrams}
  \label{fig:Youngtranspose}
\end{figure}

Transfer matrices in all other representations can be obtained through the recursive use of the Hirota identity if the representation is rectangular or by means of the CBR formula~\cite{Cherednik,Bazhanov:1989yk} for a representation corresponding to a generic Young diagram $\mu=\left(\mu_1,\dots,\mu_N\right)$
\begin{equation}\label{cbrformula}
   \bbT_{\mu}(u)=\det_{1\leq j,k\leq \mu_1}\bbT_{\mu'_j-j+k,1}\left(u-\frac{i}{2}\left(\mu'_1-\mu_1-\mu'_j+j+k-1\right)\right)
\end{equation}
where as usual we have the constraints $\mu_1\geq \mu_2\geq\dots\geq \mu_N$ and we remind the reader that the transpose Young diagram $\mu^\prime$ is defined as in Figure \ref{fig:Youngtranspose}. 

\subsection{Separated variables, $\bB(u)$ and $\bC(u)$}
We now turn to the construction of the SoV bases $\bra{\svx}$ and $\ket{\svy}$ obtained by diagonalising the operators $\bB(u)$ and $\bC(u)$ respectively. The $\bB$ operator is defined as~\cite{Sklyanin:1992sm,Smirnov2001,Gromov:2016itr}
\begin{equation}\label{Bdef}
    \bB(u)=\sum_{j_k}\bT\left[^{j_1}_{N}\right]\bT^{[-2]}\left[^{j_2}_{j_1,N}\right]\bT^{[-4]}\left[^{j_3}_{j_2,N}\right]\dots \bT^{[-2N+4]}\left[^{12\dots N-1}_{j_{N-2},N}\right]
\end{equation}
where $j_k=\{j_k^1,\dots,j_k^k\}$, $k=1,2,\dots,N-2$ is a multi-index and we sum over all configurations with $1\leq j_k^1<j_k^2<\dots <j_k^k\leq N-1$. Similarly, the $\sl(N)$ $\bC$ operator is defined as
\begin{equation}
    \bC(u)=\sum_{j_k}\bT\left[^{12\dots N-1}_{j_{N-2},N}\right]\dots \bT\left[^{j_3}_{j_2,N}\right]\bT\left[^{j_2}_{j_1,N}\right]\bT\left[^{j_1}_{N}\right]\;.
\end{equation}
We see that the only difference between $\bB$ and $\bC$ is the order in which the minors appear and the associated shifts and so the two operators coincide in the classical limit and constitute two different quantisations of the classical separated variables \cite{Sklyanin:1992eu}. We will see later how this definition of the $\bC$ operator, initially found in~\cite{Gromov:2019wmz} for $\su(3)$ case, comes about.

\paragraph{Untwisted operators and lowest-weight state.}
In order to progress we need to introduce the untwisted monodromy matrix elements $\lT_{ij}(u)$ i.e. 
\begin{equation}
    \bT_{ij}=\sum_{k=1}^N \lT_{ik}\Lambda_{kj}\,.
\end{equation}
If the representation is finite-dimensional i.e. $-2\bs\in {\mathbb N}$ there exists a lowest-weight state $\ket{\bar 0}$ in addition to the highest-weight state $\ket{0}$. The untwisted operators $\lT_{ij}$ have a simple action on both of these states, in particular 
\begin{equation}\label{LWprop}
    \bra{\bar{0}}\lT_{ij}(v)=0,\quad i>j,\quad \bra{\bar{0}}\lT_{kk}(v)=Q_\theta^{[2\bs]}\bra{\bar{0}},\ k=1,\dots,N-1,
\end{equation}
$$
\bra{\bar{0}}\lT_{NN}(v)=Q_\theta^{[-2\bs]}\bra{\bar{0}}
$$
and 
\begin{equation}\label{HWprop}
    \lT_{ij}(v)\ket{0}=0,\quad i>j,\quad \lT_{11}(v)\ket{0}=Q_\theta^{[-2\bs]}\ket{0}, \quad \lT_{kk}(v)\ket{0}=Q_\theta^{[2\bs]}\ket{0},\ k\geq 2\,.
\end{equation}

In \cite{Ryan:2018fyo} the $\bB$ operator was shown to have a very simple form in terms of the untwisted monodromy matrix elements $\lT_{ij}$ when we use the companion twist matrix~\eq{twistM}. In particular, the quantum minors appearing in \eqref{Bdef} were shown to have the form 
\begin{equation}
   \bT\left[^{j_n}_{j_{n-1},N}\right]=(-1)^{n+N-2}\det\Lambda\ \lT\left[^{j_n}_{1\ j_{n-1}+1}\right]
\end{equation}
which is easy to verify by direct calculation. Note that since we have chosen to put $\det\Lambda=1$ it follows immediately that $\bB$ (and also $\bC$) are independent of the twist eigenvalues and hence so are their eigenvectors, in a properly chosen normalisation. 
Explicitly, in terms of the untwisted operators $\lT_{ij}(u)$, $\bB$ and $\bC$ read 
\begin{equation}\label{untwistedB}
     \bB(u)=\sum_{j_k}\lT\left[^{j_1}_{1}\right]\lT^{[-2]}\left[^{j_2}_{1\ j_1+1}\right]\lT^{[-4]}\left[^{j_3}_{1\ j_2+1}\right]\dots \lT^{[-2N+4]}\left[^{12\dots N-1}_{1\ j_{N-2}+1}\right]
\end{equation}
and 
\begin{equation}
        \bC(u)=\sum_{j_k}\lT\left[^{12\dots N-1}_{1\ j_{N-2}+1}\right]\dots \lT\left[^{j_3}_{1\ j_2+1}\right]\lT\left[^{j_2}_{1\ j_1+1}\right]\lT\left[_1^{j_1}\right]\;.
\end{equation}
We can also verify that
\begin{equation}
    \bra{\bar{0}}\bB(u)= {(-1)^{\frac{1}{2}(N-1)(3N-4)}}\left(Q_\theta^{[2\bs]}(u)\right)^{N-1} \times \prod_{k=1}^{N-2}\left(Q_\theta^{[2\bs-2k]} \right)^{N-1-k}\bra{\bar{0}}
\end{equation}
and 
\begin{equation}
    \bC(u)\ket{0}={(-1)^{\frac{1}{2}(N-1)(3N-4)}}\prod_{k=1}^{N-1}Q_\theta^{[-2\bs +2(k-1)]}\ \times\ \prod_{k=1}^{N-2}\left(Q_\theta^{[2\bs+2(k-1)]}\right)^{N-1-k}\ket{0}
\end{equation}
as it follows immediately form \eqref{LWprop} and \eqref{HWprop}. 
Like in the $\mathfrak{sl}(3)$ case we subsequently define ${\bf b}(u)$ and ${\bf c}(u)$ by removing trivial ``non-dynamical" factors
\begin{equation}
    \bB(u) = {\bf b}(u) \times (-1)^{\frac{1}{2}(N-1)(3N-4)}  \prod_{k=1}^{N-2}\left(Q_\theta^{[2\bs-2k]} \right)^{N-1-k}\,,
\end{equation}
and
\begin{equation}
    \bC(u) =  {\bf c}(u) \times {(-1)^{\frac{1}{2}(N-1)(3N-4)}} \prod_{k=1}^{N-2}\left(Q_\theta^{[2\bs+2(k-1)]}\right)^{N-1-k}\;.
\end{equation}
respectively. Like in $\mathfrak{sl}(3)$ case the operators ${\bf c}$ and ${\bf b}$ are polynomials, as will be also clear from below. 

\paragraph{Commutation relation.} 
The key relation derived in \cite{Ryan:2018fyo} allowing us to show that $\bB$ indeed generates separated variables is its commutation relation with transfer matrices $\bbT_\mu(u)$ \footnote{Strictly speaking this relation as we have written it only holds when applied to states which are annihilated by $\lT_{j1}(v)$ for some $v$. This will not affect any of our arguments as we will only ever use it on such states. The precise details can be found in~\cite{Ryan:2018fyo}.}
\begin{equation}\label{BTcom}
    \bbT_\mu\left(v-\frac{i}{2}\left(\mu_1-\mu^\prime_1\right)\right)\bB(u)=f_\mu(u,v)\bB(u)\bbT_\mu\left(v-\frac{i}{2}\left(\mu_1-\mu^\prime_1\right)\right)+\lR_1(u,v)
\end{equation}
where the function $f_\mu(u,v)$ and the operator $\lR_1(u,v)$ are given by 
\begin{equation}
    f_\mu(u,v)=\prod_{k=1}^{h_\mu}\frac{u-v-i(k-1-\mu_a)}{u-v-i(k-1)},\quad \lR_1(u,v)= \sum_{j=1}^N \lT_{j1}(v)\times \dots
\end{equation}
where $h_\mu$ denotes the height (number of non-zero rows) of the Young diagram $\mu$  and ``$\dots$" in $\lR_1(u,v)$ refers to non-zero terms irrelevant for the rest of our discussion. The key idea is that if $\bra{\Lambda}$ is an eigenstate of $\bB$ and $\bra{\Lambda}\lT_{j1}(v)=0$, $j=1,\dots,N$ for some $v$ then the remainder term $\lR_1(u,v)$ in \eqref{BTcom} vanishes and so $\bra{\Lambda}\bbT_\mu(v)$ will be a new eigenstate of $\bB$. 

As a result of the properties \eqref{LWprop} the natural state to start with to build up the eigenvectors of $\bB$ is the lowest-weight state $\bra{\bar{0}}$. Unfortunately, for generic values of the spin $\bs$ we do not have a lowest-weight state meaning the relation \eqref{BTcom} is not directly applicable. On the other hand, as we will see, \eqref{BTcom} can be modified to allow us to diagonalise $\bB$ starting from the highest-weight state instead.  In order to gain some intuition for the required modifications we will first demonstrate how this relation can be used in the compact case $\bs\in\{0,-\frac{1}{2},-1,\dots,\}$ to diagonalise $\bB$ and then extend to general $\bs$.  

\paragraph{Vanishing of remainder term in the compact case.}
In this subsection we assume that $\bs\in\{0,-\frac{1}{2},-1,\dots,\}$ and hence a lowest-weight state $\bra{\bar{0}}$ with the properties \eqref{LWprop} exists. As we have just stated, we need a state $\bra{\Lambda}$ and a point $v$ such that $\bra{\Lambda}\lT_{j1}(v)=0$ and so we can take $\bra{\Lambda}=\bra{\bar{0}}$ and $v=\theta_\alpha-i\bs$. Hence, the commutation relation \eqref{BTcom} reduces to 
\begin{equation}
    \bra{\bar{0}}\bbT_\mu(\theta_\alpha -i\left(2\bs+\mu_1-\mu_1^\prime\right)) \bB(u) = f_\mu(u,\theta_\alpha-i\bs)\bra{\bar{0}}\bB(u) \bbT_\mu(\theta_\alpha -i\left(2\bs+\mu_1-\mu_1^\prime\right))
\end{equation}
We can then replace $\bB$ with ${\bf b}$ and use the fact that $\bra{\bar{0}}$ is an eigenvector of ${\bf b}$ with eigenvalue $\left(Q_\theta(u)^{[2\bs]}\right)^{N-1}$  to conclude that $\bra{\bar{0}}\bbT_\mu(\theta_\alpha -i\left(2\bs+\mu_1-\mu_1^\prime\right))$ is a new eigenvector of ${\bf b}$ with eigenvalue 
\begin{equation}\label{newBeigenvalue}
    f_\mu(u,\theta_\alpha-i \bs) \left(Q^{[2\bs]}_\theta(u)\right)^{N-1}\,.
\end{equation}
What is not clear from this construction is if $\bra{\bar{0}}\bbT_\mu(\theta_\alpha -i\left(2\bs+\mu_1-\mu_1^\prime\right))$ is actually non-zero. It was explained in \cite{Ryan:2020rfk}  how this requirement places strong constraints on the choice of Young diagram $\mu$. Specifically, $\mu$ must be contained in the Young diagram describing the physical space $[-2\bs,0,\dots,0]$ in which case it has no vanishing eigenvalues and so we conclude that $\bra{\bar{0}}\bbT_\mu(\theta_\alpha -i\left(2\bs+\mu_1-\mu_1^\prime\right))$ is non-zero. On the other hand, if $\mu$ is not contained in this diagram then $\bbT_\mu(\theta_\alpha -i\left(2\bs+\mu_1-\mu_1^\prime\right))$ vanishes identically. Hence we conclude that our new eigenstate should be of the form 
\begin{equation}
    \bra{\bar{0}}\bbT_{1,s}(\theta_\alpha -i\left(2\bs+s-1\right)) \quad \text{for any $s\in\{0,1,\dots,-2\bs \}$}\,.
\end{equation}
It was also demonstrated in \cite{Ryan:2018fyo} that we can construct a whole family of states in such a manner. In particular, the remainder term will continue to vanish as long as a transfer matrix associated to a given $\theta_\alpha$ is not used more that $N-1$ times \cite{Ryan:2018fyo}. The conclusion is that each vector of the form 
\begin{equation}\label{newBeigenstate}
    \bra{\bar{0}}\displaystyle\prod_{\alpha=1}^L \prod_{j=1}^{N-1}\bbT_{1,s_j^\alpha}(\theta_\alpha -i\left(2\bs+s_j^\alpha-1\right))
\end{equation}
is an eigenstate of ${\bf b}$ as long as we choose each $s_j^\alpha$ to satisfy $s_j^\alpha\leq -2\bs$. Since the transfer matrices commute we are also free to arrange the order of the product so that $s_1^\alpha \leq s_2^\alpha\leq \dots \leq s_{N-1}^\alpha$ for each $\alpha$, and so by \eqref{newBeigenvalue} we conclude that \eqref{newBeigenstate} is an eigenvector of ${\bf b}$ with eigenvalue 
\begin{equation}
    \prod_{\alpha=1}^L \prod_{j=1}^{N-1} (u-\theta_\alpha+i(2\bs - s_j^\alpha))\;.
\end{equation}
Finally, we make one more remark regarding the independence of the SoV states on the twist eigenvalues. Clearly, the state $\bra{\bar{0}}$ is independent of twist. Furthermore, it was demonstrated in \cite{Ryan:2018fyo} that every transfer matrix $\bbT_\mu(v)$ with our choice of twist is of the form 
\begin{equation}
    \bbT_\mu\left(v-\frac{i}{2}\left(\mu_1-\mu^\prime_1\right)\right)= \text{ twist independent part} + \sum_{j=1}^N \lT_{j1}(v) \times \text{ twist part} \;,
\end{equation}
see Appendix \ref{antipodeapp} for more details. 
The twist part drops out when we act on states which are killed by $\lT_{j1}(v)$, which is precisely the requirement that $\bbT_\mu$ generates a new eigenstate of $\bB$, and hence $\bB$ eigenstates constructed in this way are independent of the twist eigenvalues. Of course, we already know that this is the case since $\bB$ is independent of twist, but also eigenvectors constructed as in \eqref{newBeigenstate} does not introduce twist into the normalisation. 

In this section we needed the lowest weight state to construct the eigenstates of the ${\bf B}(u)$ operator. The existence of this state is only guaranteed for particular values of $\bs$. In the next section we will extend this construction to all values of $\bs$, circumventing the requirement of a lowest-weight state.

\subsection{$*$-map} 
Now that we have reviewed how to diagonalise $\bB$ in the compact case we turn our attention to the non-compact case and further explain how similar techniques can be used to diagonalise $\bC$. 

We introduce the following map, which we call $*$-map, which acts\footnote{This map is a composition of the Yangian anti-automorphism $\bT(u)\mapsto \bT(-u)$ together with a relabelling of the spectral parameter $-u\mapsto u$. Hence, terms such as $f(u)\bT(u)$ transform to $f(-u)\bT(u)$. } on monodromy matrix $\bT_{ij}(u)$ elements as 
\begin{equation}
    \bT_{ij}(u +a) \mapsto \bT_{ij}(u-a)
\end{equation}
and furthermore reverses the order of products
\begin{equation}
    \bT_{ij}(u +a)\bT_{kl}(u +b) \mapsto \bT_{kl}(u -b)\bT_{ij}(u-a)
\end{equation}
for any $a,b\in\mathbb{C}$. 

We will now discuss the key properties of this map, in particular how it acts on $\bB$ and the transfer matrices. First we will find how it acts on quantum minors since these are the building blocks for both of these objects. From the definition \eqref{qminors} we find 
\begin{equation}\label{minortransform}
    \bT\left[^{i_1\dots i_a}_{j_1\dots j_a}\right](u)\mapsto \bT\left[^{i_1\dots i_a}_{j_1\dots j_a}\right](u-i(a-1))\;
\end{equation}
which implies 
\begin{equation}\label{minortransform2}
    \bT\left[^{i_1\dots i_a}_{j_1\dots j_a}\right](u-i(a-1))\mapsto \bT\left[^{i_1\dots i_a}_{j_1\dots j_a}\right](u)\;,
\end{equation}
and hence relates $\bB(u)\mapsto \bC(u)$! This means that all results for the $\bB(u)$ operator can be translated into some new relations for $\bC(u)$ and will allow us to diagonalise $\bC(u)$ in the next seciton.

Next we examine the transfer matrices $\bbT_\mu(u)$. Let us denote their image under the $*$-map as $\bbT^*_\mu(u)$. We first start with transfer matrices in anti-symmetric representations $\bbT_{a,1}(u)$. Since these transfer matrices are defined by 
\begin{equation}
    \bbT_{a,1}(u) = \sum_{1\leq i_1<\dots<i_a\leq N}\bT\left[^{i_1 \dots i_a}_{j_1\dots j_a}\right]\left(u-\frac{i}{2}\left(a-1\right)\right)
\end{equation}
it follows immediately from \eqref{minortransform}, \eq{minortransform2} that $\bbT^*_{a,1}(u) = \bbT_{a,1}(u)$, and so the full set of conserved charges is invariant under the $*$-map. Since transfer matrices in other representations $\bbT_\mu(u)$ can be defined in terms of $\bbT_{a,1}$ using the CBR formula 
\begin{equation}
   \bbT_{\mu}(u)=\det_{1\leq j,k\leq \mu_1}\bbT_{\mu'_j-j+k,1}\left(u-\frac{i}{2}\left(\mu'_1-\mu_1-\mu'_j+j+k-1\right)\right)
\end{equation}
we can immediately read off that $\bbT_\mu^*(u)$ is given by 
\begin{equation}\la{Tstardef}
    \bbT_{\mu}^*(u)=\det_{1\leq j,k\leq \mu_1}\bbT_{\mu'_j-j+k,1}\left(u+\frac{i}{2}\left(\mu'_1-\mu_1-\mu'_j+j+k-1\right)\right)
\end{equation}
which explains the origin of the transfer matrices \eqref{Tstar}. The meaning of these transfer matrices in terms of representation theory is given in Appendix~\ref{basisproof}.

Finally, let us recall that the remainder term $\lR_1(u,v)$ appearing in the commutation relation \eqref{BTcom} is given by 
\begin{equation}
    \lR_1(u,v)= \sum_{j=1}^N \lT_{j1}(v)\times \dots 
\end{equation}
and so under the $*$-map it results in $\lR_1^*(u,v)= \sum_{j=1}^N \dots \times \lT_{j1}(v)$. Combining these results we immediately find that the commutation relation \eqref{BTcom} transforms under the $*$-map to produce a new commutation relation which reads
\begin{equation}\label{CTcom}
    \bC(u)\bbT_\mu^*\left(v+\frac{i}{2}\left(\mu_1-\mu^\prime_1\right)\right)=g_\mu(u,v)\bbT_\mu^*\left(v+\frac{i}{2}\left(\mu_1-\mu^\prime_1\right)\right)\bC(u)+\lR_1^*(u,v)
\end{equation}
where $g_\mu(u,v)$ is defined as 
\begin{equation}
    g_\mu(u,v)=\prod_{k=1}^{h_\mu}\frac{u-v+i(k-1-\mu_a)}{u-v+i(k-1)}\;.
\end{equation}
In the next subsection we will explain how to use this relation to construct right SoV states $\ket{\svy}$, the eigenstates of $\bC(u)$. 

\subsection{Constructing the SoV bases}

\paragraph{Diagonalising $\bC$.}

We will now explain how to diagonalise $\bC$ using the commutation relation \eqref{CTcom}. By \eqref{HWprop} we have that
\begin{equation}
    \lT_{j1}(\theta_\alpha+i\bs)\ket{0}=0
\end{equation}
which itself implies that $\lR_1^*(u,\theta_\alpha+i\bs)\ket{0}=0$. Applying \eqref{CTcom} to this state then produces the relation 
\begin{equation}
    \bC(u)\bbT_\mu^*\left(\theta_\alpha+\frac{i}{2}\left(2\bs+\mu_1-\mu^\prime_1\right)\right)\ket{0}=g_\mu(u,\theta_\alpha+i\bs)\bbT_\mu^*\left(\theta_\alpha+\frac{i}{2}\left(2\bs+\mu_1-\mu^\prime_1\right)\right)\bC(u)\ket{0}
\end{equation}
where $\bbT_\mu^*$ is defined in \eq{Tstardef}.
Replacing $\bC$ with ${\bf c}$ and using the fact that $\ket{0}$ is an eigenvector of ${\bf c}$ with eigenvalue $\prod_{k=1}^{N-1}Q_\theta^{[-2\bs +2(k-1)]}$ we find that $\bbT_\mu^*\left(\theta_\alpha+\frac{i}{2}\left(2\bs+\mu_1-\mu^\prime_1\right)\right)\ket{0}$ is a new eigenvector of ${\bf c}$ with eigenvalue 
\begin{equation}\label{newCeigenvalue}
    \prod_{k=1}^{N-1}Q_\theta^{[-2\bs +2(k-1)]}g_\mu(u,\theta_\alpha+i\bs)\,.
\end{equation}
In terms of the different states we can construct the situation can be shown to be identical to that of diagonalising $\bB$ for the case where the physical space carries symmetric powers of the anti-fundamental representation which was discussed in \cite{Ryan:2018fyo} but now using the transfer matrices $\bbT^*$. Since the situation is identical we simply state the final result which is that the set of vectors 
\begin{equation}\label{Ceigenstates}
    \prod_{\alpha=1}^L\bbT^*_{\mu^\alpha}\left(\theta_\alpha+\frac{i}{2}\left(2\bs+\mu_1^\alpha-\mu^{\alpha,\prime}_1\right)\right)\ket{0}
\end{equation}
are eigenvectors of $\bC$ for any choice of the Young diagrams $\mu^\alpha=(\mu^\alpha_1,\dots,\mu^\alpha_{N-1},0)$ and are always non-zero (see Appendix \ref{basisproof}). The eigenvalues of ${\bf c}$ on these excited states is then deduced immediately from \eqref{newCeigenvalue} and have the form
\begin{equation}
\prod_{\alpha=1}^L \prod_{j=1}^{N-1}(u-\svy_j^\alpha)
\end{equation}
where we have defined $\svy_{\alpha,j}=\theta_\alpha+i(\bs+\mu^\alpha_j+1-j)$, where $\mu^\alpha_j=m^\alpha_j$ in the notation of Section \ref{sec:SoV}

The operator $\bbT^*_{\mu^\alpha}\left(\theta_\alpha+i\bs+\frac{i}{2}\left(\mu_1^\alpha-\mu^{\alpha,\prime}_1\right)\right)$ is diagonalised by any eigenvector $\bra{\Psi}$ of the transfer matrix with eigenvalue given by
\begin{equation}
    \bra{\Psi}\bbT^*_{\mu^\alpha}\left(\theta_\alpha+i\bs+\frac{i}{2}\left(\mu_1^\alpha-\mu^{\alpha,\prime}_1\right)\right) \ \propto\  \frac{\displaystyle\det_{1\leq j,k\leq N-1} Q^j \left(\svy_k^\alpha+\frac{i}{2}\left(N-2\right)\right)}{\displaystyle \det_{1\leq j,k\leq N-1} Q^j\left(\svy_0^\alpha + \frac{i}{2}\left(N-2k\right)\right)}\bra{\Psi}
\end{equation}
where we have omitted certain $Q_\theta$-dependent factors which we reabsorb into the definition of the SoV basis, similar to what was done in the $\sl(3)$ case \eqref{sl3y}, and like for $\sl(3)$ we define
\begin{equation}
 \svy^\alpha_0=\theta_\alpha+i\bs\,.
\end{equation}

We can then choose to normalise $\bra{\Psi}$ so that 
\begin{equation}
    \langle\Psi|0\rangle = \prod_{\alpha=1}^L \displaystyle\det_{1\leq j,k\leq N-1} Q^j \left(\svy_0^\alpha + \frac{i}{2}\left(N-2k\right)\right)
\end{equation}
and hence obtain the following SoV wave function for $\bra{\Psi}$
\begin{equation}
     \langle\Psi|\svy\rangle = \prod_{\alpha=1}^L \displaystyle\det_{1\leq j,k\leq N-1} Q^j \left(\svy_k^\alpha+\frac{i}{2}\left(N-2\right)\right)\,
\end{equation}
which perfectly matches the wave functions appearing in the integral approach to scalar products, see section \ref{generalsln}.

\paragraph{Diagonalising $\bB$.}
We already mentioned we would have some trouble diagonalising $\bB$ since our commutation relation \eqref{BTcom} requires the presence of a lowest-weight state $\bra{\bar{0}}$. Let us return to the compact case one more time. We already explained that the set of vectors
\begin{equation}\label{lwBeigen}
    \bra{\bar{0}}\prod_{\alpha=1}^L \prod_{j=1}^{N-1} \bbT_{1,s_j^\alpha}\left(\theta_\alpha-\frac{i}{2}(2\bs+s_j^\alpha-1)\right),\quad 0\leq s_j^\alpha\leq -2\bs,\quad s_j^\alpha\leq s_{j+1}^\alpha\,
\end{equation}
diagonalise $\bB$. We can gain some further insight by rewriting these transfer matrices using $Q$-operators \cite{Bazhanov:1996dr,Krichever:1996qd,Kazakov:2010iu,Bazhanov:2010jq,Frassek:2011aa}. The $Q$-operators $\mathbb{Q}_i(u)$ and $\mathbb{Q}^i$ are defined as the operators which have the Q-functions $Q_i(u)$ and $Q^i(u)$ corresponding to the transfer matrix eigenstate $\ket{\Psi}$ as their eigenvalues 
\begin{equation}
    \mathbb{Q}_i(u)\ket{\Psi}=Q_i(u)\ket{\Psi},\quad \mathbb{Q}^i(u)\ket{\Psi}=Q^i(u)\ket{\Psi}\,.
\end{equation}
The main relation we will make use of is the following 
\begin{equation}
    \bbT_{1,s_j^\alpha}\left(\theta_\alpha-\frac{i}{2}(2\bs+s_j^\alpha-1)\right)\ \propto\  \frac{\mathbb{Q}_1(\theta_\alpha-i(\bs+s^\alpha_j))}{\mathbb{Q}_1(\theta_\alpha-i\bs)}
\end{equation}
where $\propto$ indicates that the equality holds up to non-zero multiples of $Q_\theta$ \cite{Ryan:2020rfk}. This relation allows us to rewrite the above set of vectors \eqref{lwBeigen} as
\begin{equation}
    \bra{\bar{0}}\prod_{\alpha=1}^L \prod_{j=1}^{N-1} \frac{\mathbb{Q}_1(\theta_\alpha-i(\bs+s^\alpha_j))}{\mathbb{Q}_1(\theta_\alpha-i\bs)}\,.
\end{equation}
The highest-weight state can then be obtained by choosing all $s_j^\alpha=-2\bs$
\begin{equation}
    \bra{0}\ \propto\ \bra{\bar{0}}\prod_{\alpha=1}^L \prod_{j=1}^{N-1} \frac{\mathbb{Q}_1(\theta_\alpha+i\bs)}{\mathbb{Q}_1(\theta_\alpha-i\bs)}\;.
\end{equation}
This simple rewriting has actually helped us quite a lot -- we see that we can start from the highest-weight state $\bra{0}$ and move back down towards the lowest-weight state by acting with operators of the form 
\begin{equation}\label{Qopratio}
    \frac{\mathbb{Q}_1(\theta_\alpha+i(\bs-s^\alpha_j))}{\mathbb{Q}_1(\theta_\alpha+i\bs)},\quad s^\alpha_j=0,1,\dots,-2\bs\,
\end{equation}
and we will be able to obtain all $\bB$ eigenvectors in this way. Explicitly, the $\bB$ eigenvectors can be written as 
\begin{equation}\label{Beigenvecs}
    \bra{0}\prod_{\alpha=1}^L \prod_{j=1}^{N-1} \frac{\mathbb{Q}_1(\theta_\alpha+i(\bs-s_\alpha^j))}{\mathbb{Q}_1(\theta_\alpha+i\bs)}
\end{equation}
where now $-2\bs\geq s_1^\alpha \geq \dots \geq s^\alpha_{N-1}\geq 0$. We can now analytically continue in the spin $\bs$ from $\{0,-\frac{1}{2},\dots\}$ to general values. Then in \eqref{Beigenvecs} the constraint $-2\bs\geq s_1^\alpha \geq \dots \geq s^\alpha_{N-1}\geq 0$ should reduce to $ s_1^\alpha \geq \dots \geq s^\alpha_{N-1}\geq 0$, with no upper limit on the value of $s_1^\alpha$, and we expect this set of vectors to exhaust all eigenvectors of $\bB$. In order to verify this it would be convenient to be able to generate the vectors \eqref{Beigenvecs} using transfer matrices instead of $Q$-operators since the former are usually easier to work with. Hence, we need some transfer matrix which at some value of the spectral parameter becomes the ratio \eqref{Qopratio}. Luckily, it is not hard 
to work out that the transfer matrix in question is given by 
\begin{equation}\label{goodBtransfer}
    \bbT_{N-1,s_j^\alpha}\left(\theta_\alpha+i \bs-\frac{i}{2}(N-s^\alpha_j-1)\right)
\end{equation}
and we derive a new commutation relation between $\bB$ and $\bbT_{N-1,s}$, alternate to \eqref{BTcom}, which reads 
\begin{equation}\label{newBTcom1}
    \bbT_{N-1,s}\left(v-\frac{i}{2}(N-s-1)\right)\bB(u)=h(u,v)\bB(u)\bbT_{N-1,s}\left(v-\frac{i}{2}(N-s-1)\right)+\sum_{j=1}^N \lT_{1,j}(v)\times\dots\;.
\end{equation}
This relation is a special case of a more general relation involving $\bB$ and the transfer matrices constructed from the \textit{inverse} monodromy matrix $\bT^{-1}(u)$, see Appendix \ref{antipodeapp} but for our purposes this relation is enough.  Now we can use \eqref{newBTcom1} directly to diagonalise $\bB$ starting from $\bra{0}$, avoiding analytic continuation. Like what was previously described when starting from the lowest weight state, it is possible to apply a transfer matrix corresponding to a given $\theta_\alpha$ at most $N-1$ times and the remainder term will still vanish, meaning all eigenvectors of ${\bf b}$ can be constructed as 
\begin{equation}
    \bra{0}\prod_{\alpha=1}^L\prod_{j=1}^{N-1}\bbT_{N-1,s_j^\alpha}\left(\theta_\alpha+i \bs-\frac{i}{2}(N-s^\alpha_j-1)\right),
\end{equation}
and the corresponding eigenvalue is given by 
\begin{equation}
    \prod_{\alpha=1}^L \prod_{j=1}^{N-1}(u-\svx_{\alpha,j}),\quad \svx_{\alpha,j}=\theta_\alpha+i(\bs+s^\alpha_j)
\end{equation}
with $s^\alpha_j=n_{\alpha,j}$ in the notation of Section \ref{sec:SoV}. The fact that this normalisation ensures that the states $\bra{\svx}$ are independent of the twist eigenvalues is clarified in Appendix \ref{antipodeapp}.

\subsection{Reduction of SoV bases to the compact case}
\label{sec:opcomp}
We close this section by demonstrating how the SoV bases we have constructed behave when we restrict ourselves to the compact case with  $\bs\in\{0,-\frac{1}{2},-1,\dots\}$. In this scenario our infinite-dimensional irreducible space becomes reducible with a finite-dimensional irreducible part. As we will see, the SoV basis vectors corresponding to the irreducible part remain non-zero and everything else vanishes. 

Since the SoV bases are polynomial functions in the spin $\bs$ when we use the differential operator realisation like in the $\sl(3)$ case there is no problem with simply setting $\bs$ to some negative half-integer value. Using the results of Appendix \ref{basisproof} only a finite number of transfer matrices will remain non-zero when we do this. For the right SoV basis $\ket{\svy}$ defined by 
\begin{equation}
    \ket{\svy}=\prod_{\alpha=1}^L\bbT^*_{\mu^\alpha}\left(\theta_\alpha+\frac{i}{2}\left(2\bs+\mu_1^\alpha-\mu^{\alpha,\prime}_1\right)\right)\ket{0}
\end{equation}
the state will vanish for any configuration with $\mu^\alpha_1>-2\bs$ but will be non-zero as long as $\mu^\alpha_1\leq -2\bs$, and hence the number of non-zero states precisely matches the dimension of the finite-dimensional irreducible part of the representation, see \cite{Ryan:2018fyo,Ryan:2020rfk}.

Similarly, for $\bB$, the transfer matrix \eqref{goodBtransfer} can be shown to be non-zero for any $s$ in the non-compact case, and if we reduce to the compact case it remains non-zero only if $0\leq s\leq -2\bs$ and we reach the same conclusion. 

\section{Determinant representations for overlaps and expectation values}\label{detrep}

In this section we will extend the previous results by deriving the SoV based determinant representations for overlaps and expectation values of various operators.

\subsection{Defining det-product and its relation to SoV}

Here we discuss the main tools for computing some physical observables with the help of the SoV approach we developed in the previous sections.
For simplicity we will mostly demonstrate the method on the $\sl(3)$
example but in all cases the generalisation to $\sl(N)$ is quite clear.

In particular, in this section we consider
the overlaps between the state of the chains with different twists.
Such overlaps were recently considered in the context of AdS/CFT correspondence~\cite{Cavaglia:2020hdb}
 and can be interpreted as $3$ point correlation functions involving so-called color twist operators.

The key observation is that the SoV states, in the non-diagonal frame~\eq{twistM}, are not sensitive to the twist of the monodromy matrix. In other words the same SoV basis will separate the wave function for any values of the twist eigenvalues $\lambda_a$. This implies that the integral representation we derived in the previous section for the states of the same spin chain can be, very non-trivially, used 
to compute overlaps between the eigenstates of different transfer matrices.

For what follows it will be convenient to introduce the notation
for what we call the {\it det-product},
\beq\la{detproduct}
\detl G_{\alpha,1+a}\middle|F_\alpha \detr= \frac{1}{N_0}\det_{(\alpha,a),(\beta,b)}\bl F_\alpha(u) \;u^{\beta-1}\; {\cal D}^{3-2b}\; G_{\alpha,a+1}(u) \br_\alpha
\eeq
where the notation with double brackets, which initially
referred to an integration~\eq{scalargf}, we understand now more generally as a sum over the poles of the factor $\mu_\alpha$. So in this section we re-define
\beq
\bl f(u) \br_\alpha \equiv  \sum_{n=0}^\infty r_{\alpha,n} f(\svx_{\alpha,0}+i n)\;.
\eeq
The normalisation factor is
\beqa
N_0&=&
   \Delta ^2_{\theta_\alpha}  \prod _{\alpha
   }^L r_{\alpha ,0}^2\;.
\eeqa
Thus for the case when $G$ and $F$ in \eq{detproduct} are  Q-functions describing two spin chain states, the det-product gives the overlap of these states we presented above in \eq{detsl3}. 
Notice that with the expression \eq{detproduct} one can follow all the same steps as in section~\ref{sec:expmeasure} to arrive to the following result:
\beqa\la{detsov}
\detl G_{\alpha,1+a}|F_\alpha\detr&=&
\sum_{\svx,\svy}
M_{\svy,\svx}
\prod_{\alpha,a}
F_\alpha(\svx_{\alpha,a})
\\&\times& \nn
\prod_{\alpha}
\[
G_{\alpha,2}(\svy_{\alpha,1}+\tfrac{i}{2})
G_{\alpha,3}(\svy_{\alpha,2}+\tfrac{i}{2})-
G_{\alpha,3}(\svy_{\alpha,1}+\tfrac{i}{2})
G_{\alpha,2}(\svy_{\alpha,2}+\tfrac{i}{2})
\]\;.
\eeqa
We will see that a number of correlators can be expressed in terms of the det-product. 
Even though we found an explicit expression for the SoV measure $M_{\svy,\svx}$, our ultimate goal is to bring the correlator to a determinant form, rather than to a sum over the SoV states. We will show how in some important cases the explicit form of the measure is not needed as the result takes the form of the det-product.

In order for two states $\Theta$ and $\Phi$ to have a scalar product which can be written in the det-product form, we have to require what we call {\it separability} property from these states, which can be expressed as
\beqa\la{PsiPhi}
\langle \svx|\Theta\rangle &=& \prod_{\alpha=1}^L
F_\alpha(\svx_{\alpha,1})F_\alpha(\svx_{\alpha,2})\;,\\
\langle \Phi |\svy \rangle &=& 
\prod_{\alpha=1}^L\[
{G_{\alpha,2}(\svy_{\alpha,1}+\tfrac{i}{2})
G_{\alpha,3}(\svy_{\alpha,2}+\tfrac{i}{2})-
G_{\alpha,3}(\svy_{\alpha,1}+\tfrac{i}{2})
G_{\alpha,2}(\svy_{\alpha,2}+\tfrac{i}{2})
}\]\;.
\eeqa
If that is the case, then as a consequence of the completeness of both SoV bases $\{\svx\}$ and $\{\svy\}$ and due to the relation 
\eq{detsov} we then immediately get
\beq
\langle \Phi |\Theta\rangle=
\detl G_{\alpha,1+a} |F_\alpha\detr\;.
\eeq
In what follows we explore a few examples when \eq{PsiPhi} does hold.
One immediate example is when both states are  transfer matrix eigenstates. In this case of course we simply have $F_{\alpha}=Q_1$ and $G_{\alpha,c}=Q_{1,c}$, so that
\beq\la{ABover}
\langle \Psi^A | \Psi^B\rangle = \detl Q^A_{1,1+a} |
Q^B_1\detr  \;\propto\; \delta^{AB}
\;.
\eeq
In the above expression the left and right wave functions are normalised according to our conventions from section~\ref{sec:SoV}. 

\subsection{Overlaps between wave functions with different twists}
Another quite obvious example where the separability property~\eq{PsiPhi}
is satisfied for both states but gives much less trivial overlap
than~\eq{ABover} is the case when both states are eigenstates of transfer matrices with {\it different} sets of twists eigenvalues $\lambda_a$ and $\tilde\lambda_a$. As we emphasised before, the SoV states do not depend on $\lambda's$ and thus should separate wavefunctions corresponding to the spin chains with arbitrary twist eigenvalues $\lambda_a$ (provided the twist matrix is of the form~\eq{twistM}).

Thus we conclude that the overlap between the states of the spin chains with different twist eigenvalues can we written in the form
\beq\la{overl}
\langle \Psi^{\tilde\lambda_a}|
\Psi^{\lambda_a}\rangle = 
\detl
\tilde Q_{12},
\tilde Q_{13}
\Big| Q_1\detr\;.
\eeq
In the above expression we still assume that the states are normalised in agreement with our conventions. However, we can also form a normalisation independent combination, for example
\beq
\frac{
\langle \Psi^{\tilde\lambda_a}|
\Psi^{\lambda_a}\rangle
\langle \Psi^{\lambda_a}|
\Psi^{\tilde\lambda_a}\rangle
}
{
\langle \Psi^{\tilde\lambda_a}|
\Psi^{\tilde\lambda_a}\rangle
\langle \Psi^{\lambda_a}|
\Psi^{\lambda_a}\rangle
}
=
\frac{
\detl
\tilde Q_{12},
\tilde Q_{13}
\Big| Q_1\detr
\detl
 Q_{12},
 Q_{13}
\Big|\tilde Q_1\detr
}{
\detl
\tilde Q_{12},
\tilde Q_{13}
\Big|\tilde Q_1\detr
\detl
 Q_{12},
 Q_{13}
\Big| Q_1\detr
}\;.
\eeq
\paragraph{Examples of the non-trivial overlaps.}
The simplest example of a non-trivial overlap is the overlap between two ground states corresponding to two different twists.
Since our twist is non-diagonal, the corresponding ground states can be obtained by acting with a suitable global rotation on the constant polynomial. 
This is discussed in detail in section~\ref{sec:SoV} where we explicitly found the ground states 
\eq{lvac} and \eq{rvac}. For the simplest length $1$ spin chain
we have
\beqa
|\Omega^{\lambda_a}_{L=1}\rangle &=& 
\lambda_1^{2i\svx_{1,0}}
\left(\frac{ x}{\lambda _1}+\frac{ y}{\lambda _1^2}+1\right)^{-2\bs}
\;\;,\\
\langle \Omega^{\tilde\lambda_a}_{L=1}| &=& 
\tilde\lambda_1^{i\svx_{1,0}+\frac{1}{2}}
(
\tilde\lambda_{3}
-
\tilde\lambda_{2}
)
\left(x \left(\tilde\lambda _2+\tilde\lambda_3\right)-\frac{y}{\tilde\lambda _1}+1\right)^{-2\bs}\;.
\eeqa
In order to demonstrate that equation~\eq{overl} holds, we first compute
the l.h.s. by expanding both functions up to some fixed order, computing the scalar product between two resulting polynomials and then sending the expansion order to infinity, like we did previously in~\eq{vacvac}
for two equal twists. We find that the generalisation of~\eq{vacvac} reads\footnote{for the limit of the truncated series to exists one  requires $|\lambda_1|>|\tilde\lambda_2|$ and $|\lambda_1|>|\tilde\lambda_3|$, which also coincides with the condition for the convergence of the sum over poles in the r.h.s. of \eq{overl} (generalising a similar condition for the equal twists case discussed in equation \eq{cindition}).}
\beq\la{vactvac}
\langle \Omega^{\tilde\lambda_a}_{1}|\Omega^{\lambda_a}_{1}\rangle=
\lambda_1^{2i\svx_{1,0}}
\tilde\lambda_1^{i\svx_{1,0}+\frac{1}{2}}
(
\tilde\lambda_{3}
-
\tilde\lambda_{2}
)
{(1-\tilde\lambda _2/\lambda _1)^{-2\bs} (1 -\tilde\lambda_3/\lambda _1)^{-2\bs}}\;.
\eeq

Now we can try to reproduce this result using the det-product of the corresponding Q-functions, i.e.
\beq
Q^{\lambda_a}_1=\lambda_1^{i u}\;\;,\;\;
Q^{\tilde\lambda_a}_{1,1+a}=\tilde\lambda_1^{i u}\tilde\lambda_{1+a}^{i u}\;.
\eeq
Evaluating the sum over residues we get
\beq
\bl \lambda_1^{i u} \; 
\tilde\lambda_1^{i u+b-\frac{3}{2}}\tilde\lambda_{1+a}^{i u+b-\frac{3}{2}}
\br_\alpha = 
-
\frac{\Gamma (2 \bs)}{2\pi}\lambda _1^{i\theta_1-\bs}  \(\tilde\lambda _1
\tilde\lambda_{a+1}
\)^{b+i\theta_1-\bs-\frac{3}{2}}  \left(1-\frac{1}{\lambda _1 \tilde\lambda _1 \tilde\lambda
   _{a+1}}\right)^{-2 \bs} \ .
\eeq
Then we plug this expression into the determinant and divide by the normalisation constant, obtaining
\beq\la{detproduct2}
\detl Q^{\tilde\lambda_a}_{1,1+a}\middle|Q_1^{\lambda_a} \detr= \frac{1}{N_0}\det_{a,b}\bl \lambda_1^{i u} \; 
\tilde\lambda_1^{i u+b-\frac{3}{2}}\tilde\lambda_{1+a}^{i u+b-\frac{3}{2}}
\br_\alpha\;,
\eeq
which perfectly reproduces \eq{vactvac}!

\paragraph{Probing the transition matrix.}
The overlap between two eigenstates of the transfer matrix in different frames is $\sl(N)$ invariant.
This means that one can diagonalise either one of the two twist matrices~\eq{twistM}.
The matrix which relates the two frames that diagonalise one of these two twist matrices has the following general form, valid for $\sl(N)$:
\beq\la{Sab}
S_{ab}=\prod_{i\neq a}\frac{\lambda_i-\tilde\lambda_b}{\lambda_i-\lambda_a}\;\;,\;\;
S_{ab}^{-1}=\prod_{i\neq a}\frac{\tilde\lambda_i-\lambda_b}{\tilde\lambda_i-\tilde\lambda_a}\;.
\eeq
Let us show that the above transformation 
is hard-wired into the SoV construction and into the det-product in particular.
Consider the normalisation independent combination
of the scalar products of two twisted vacua,
\beq\la{vactvac2}
\frac{
\langle \tilde\Omega|\Omega\rangle
\langle \Omega|\tilde\Omega\rangle
}
{
\langle\Omega|\Omega\rangle
\langle  \tilde\Omega|\tilde\Omega\rangle
}
=
\frac{
{(\lambda _1-\tilde\lambda _2)^{-2\bs} (\lambda _1 -\tilde\lambda_3)^{-2\bs}}
{(\tilde\lambda _1-\lambda _2)^{-2\bs} (\tilde\lambda _1 -\lambda_3)^{-2\bs}}
}
{
{(\lambda _1-\lambda _2)^{-2\bs} (\lambda _1 -\lambda_3)^{-2\bs}}
{(\tilde\lambda _1-\tilde\lambda _2)^{-2\bs} (\tilde\lambda _1 -\tilde\lambda_3)^{-2\bs}}
}
\;.
\eeq
Let's now focus on the fundamental representation, i.e. $\bs=-1/2$. Let's assume that $|\Omega\rangle$
is in the diagonalised frame. We know that for the diagonal twist the ground is simply the highest weight state $|\Omega\rangle = \vec e_1$, whereas the other state reads $|\tilde\Omega\rangle = S^{-1}|\Omega\rangle = S_{11}^{-1} \vec e_1+ S_{21}^{-1}\vec e_2+ S_{31}^{-1}\vec e_3$. Similarly for the left
states $\langle\Omega| = \vec e_1$
and $\langle\tilde\Omega| = \langle\Omega|S = 
S_{11} \vec e_1+ S_{21}\vec e_2+ S_{31}\vec e_3$, from where we would expect that for $\bs=-1/2$ we should get
\beq\la{vactvac3}
\frac{
\langle \tilde\Omega|\Omega\rangle
\langle \Omega|\tilde\Omega\rangle
}
{
\langle\Omega|\Omega\rangle
\langle  \tilde\Omega|\tilde\Omega\rangle
}
=S_{11} S^{-1}_{11}
\;,
\eeq
which is indeed the case as we see from \eq{vactvac2}. Note that one can further interchange the order of the eigenvalues, changing the vacua accordingly, to deduce  any combination of the form $S_{ab}S_{ba}^{-1},\;a,b=1,2,3$.
One can invert the logic and verify that
the knowledge of all $S_{ab}S_{ba}^{-1},\;a,b=1,2,\dots,N$
allows one to reconstruct $S_{ab}$ modulo the transformation $S\to D_1.S.D_2$, where $D_1,D_2$
are two independent diagonal matrices. The diagonal matrices will commute with the twist matrices and they reflect the freedom in the definition of $S$ in the first place.

\subsection{On-shell off-shell overlap}
In this section we explore the effect of the action by  ${\bf B}(u)$ or ${\bf C}(u)$ operators
on the states. Assuming the state $\ket{\Theta}$
is separated by the SoV basis
like in \eq{PsiPhi}, we have
\beq\la{PsiPhiB}
\langle \svx|{\bf b}(w)|\Theta\rangle = 
\langle \svx|\Psi\rangle
\prod_{\alpha=1}^L (w-\svx^\alpha_{1})(w-\svx^\alpha_{2})
= \prod_{\alpha=1}^L (u-x^\alpha_{1})(u-\svx^\alpha_{2})F_\alpha(\svx^\alpha_1)F_\alpha(\svx^\alpha_2)
\eeq
where ${\bf b}(w)$ is the nontrivial part of the ${\bf B}(w)$ operator defined in \eq{Bb}. We see that the action by ${\bf b}(w)$
simply translates into the replacement $F_\alpha(u)\to (w-u)F_\alpha(u)$. It is clear that there is a potential to generalising this further. We can define a ``local" ${\bf b}_\alpha$ operator so that
\beq
{\bf b}(u)=\prod_{\alpha=1}^L {\bf b}_\alpha(u)
\eeq
where ${\bf b}_\alpha(u)$ is a polynomial of degree $N-1$ in $u$, 
diagonalised by $\langle \svx|$,
with the spectrum $\prod_{a=1}^{N-1}(u-\svx_{\alpha,a})$.
Repeating the same calculation as in \eq{PsiPhiB}
we see that ${\bf b}_\beta(w)$
acts on $F_\alpha$ as\footnote{One should be careful with the $\circ$ notation, as there is no linearity in the first argument, e.g. sum of two operators does not necessarily produce a factorisable state and thus does not have any well defined action on individual $F_\alpha$. However, $\circ$ is an associative operation and does support an action by several operators.}
\beq
{\bf b}_\beta(w)\;\;\circ\;\; F_\alpha(u) = 
 (w-u)^{\delta_{\beta\alpha}}F_\alpha(u)\;.
\eeq
To summarise, this means that multiple action of 
any ${\bf b}_\beta(w)$ operators does not spoil
the separability property of the wave function.
This means that we can compute a set of rather non-trivial form factors in a determinant form,
\beq\la{FBBT}
\frac{\langle \Phi|{\bf b}_{\beta_1}(v_1)\dots
{\bf b}_{\beta_K}(v_K)|\Theta\rangle}{
\langle \Phi|\Theta\rangle
}
=
\frac{\detl G_{\alpha,a} \middle|
\prod_{i=1}^K(v_i-u)^{\delta_{\beta_i\alpha}}
 F_\alpha
\detr}
{\detl G_{\alpha,a} |F_\alpha \detr}\;.
\eeq
A particularly important case is the following state
\beq\la{offdef}
|\Psi\rangle_{\rm off\;shell}\equiv{\bf b}(v_1)\dots {\bf b}(v_k)|\Omega\rangle\;,
\eeq
which in analogy with $\sl(2)$ one could call the
{\it off-shell} Bethe state.
To distinguish it from some other {\it off-shell}
Bethe states existing in the literature, one could call it
{\it algebraic off-shell} Bethe states as opposed to the hybrid coordinate-algebraic way of building eigenstates of transfer matrix in nested Bethe ansatz approach. It follows immediately from \eq{FBBT}
that the overlap between \eq{offdef} and any separable state, and in particular with an eigenstate $\bra{\Phi}$ of the transfer matrix, is of a determinant form
\beq
{\langle \Phi|\Psi\rangle_{\rm off\; shell}}
=
{\detl Q_{1,a+1} \middle|\lambda_1^{iu}\prod_{k=1}^K(u-v_k)\detr}\;.
\eeq
Note that for that to be true it is not required that $\{v_k \}$ are the Bethe roots, solving Bethe ansatz equations.
As we described before in section~\ref{sec:wfs}, when the parameters
$\{v_k \}$ do satisfy the Bethe ansatz equations
the state $|\Psi\rangle_{\rm off\;shell}$ does actually become an eigenstate of the transfer matrix.

In analogy with ${\bf b}_\alpha(u)$ we can also define
${\bf c}_{\alpha}(u)$, containing only those roots of ${\bf c}(u)$ that are associated with $\theta_\alpha$.
For the insertion of this operator we can use the relation
\beqa\la{PsiPhiC}
\langle \Phi|{\bf c}_\beta(w) |\svy \rangle &=&
\langle \Phi|\svy \rangle  
(w-\svy_{\beta,1})
(w-\svy_{\beta,2})\\
\nn&=&
(w-\svy_{\beta,1})(w-\svy_{\beta,2})
\prod_{\alpha=1}^L \(
G_{\alpha,2}(\svy_{\alpha 1}+\tfrac{i}{2})G_{\alpha,3}(\svy_{\alpha 2}+\tfrac{i}{2})-
G_{\alpha,3}(\svy_{\alpha 1}+\tfrac{i}{2})G_{\alpha,2}(\svy_{\alpha 2}+\tfrac{i}{2})
\)
\eeqa
implying that $G_{\beta,a}(u)\to (w-u+\tfrac{i}{2})G_{\beta,a}(u)$, leaving other $G_{\alpha,a}(u)$ with $\alpha\neq\beta$ unchanged.
Therefore we can generalise the result~\eq{FBBT}
as follows:
\beqa
&&\frac{\langle \Phi|{\bf c}_{\gamma_1}(v_1)\dots
{\bf c}_{\gamma_K}(v_K)
{\bf b}_{\beta_1}(w_1)\dots
{\bf b}_{\beta_J}(w_J)
|\Theta\rangle}{
\langle \Phi|\Psi\rangle
}\\
\nn &=&
\frac{\detl G_{\alpha,a+1}(u) \prod_{k=1}^K(v_k-u+\tfrac{i}{2})^{\delta_{\gamma_k\alpha}}
|F_\alpha(u)
\prod_{k=1}^J(w_k-u)^{\delta_{\beta_k\alpha}}
\detr}
{\detl G, H |F\detr}\;.
\eeqa

\subsection{Form factors of derivatives of the transfer matrices}

In this section we show how our integral SoV approach leads to determinant representations for a large class of diagonal form factors, extending the results of \cite{Cavaglia:2019pow} from $\bs=1/2$ to generic $\bs$. While the extension is almost straightforward, we present here the key steps to make the discussion self-contained. We first consider the $\sl(3)$ case, but generalization to $\sl(N)$ is immediate as we will explain shortly. We also show how to compute matrix elements of some $\textit{local}$ operators from this data.

The form factors we consider are the diagonal matrix elements of the derivatives of integrals of motion (coefficients of the transfer matrices) defined in \eq{IMs},
\beq
\label{ffac}
    \frac{\bra{\Psi}\frac{\d \hat I_{b,\beta-1}}{\d p}|\Psi\rangle}{\bra{\Psi}\Psi\rangle}=\frac{\d I_{b,\beta-1}}{\d p}
\eeq
where $p$ is a parameter of the model (either an inhomogeneity $\theta_\alpha$ or a twist $\lambda_a$). While the spectrum of the model is under good control and one could in principle compute the derivative in the r.h.s. of \eq{ffac} directly (as a ratio of finite differences), here we rather wish to express it in terms of Q-functions evaluated at one fixed value of $p$, and it is nontrivial that such an expression exists at all. We will see that the result has a rather natural form of a ratio of two determinants, with 
the denominator corresponding to the norm \eq{detsl3} and the numerator given by the same expression with an extra insertion that we interpret as describing the operator $\d_p\hat I_{a,\alpha-1}$ we consider.  In the AdS/CFT context correlators of this kind are also important as they correspond to 3-pt functions with marginal operators \cite{Costa:2010rz}. 

 If we consider a small variation of our parameter $p\to p+\delta p$, the Q-functions $Q^{a+1}$ as well as the operator $\hat O^\dagger$ in the Baxter equation \eq{Bax12s2} will change, but the equation will remain satisfied, so that $(\hat O^\dagger+\delta\hat O^\dagger)(Q^{a+1}+\delta Q^{a+1})=0$. Using that the original Q-function satisfies $\hat O^\dagger Q^{a+1}=0$, and dropping the terms quadratic in variations, we have
\beq
    0=\bl Q_1(\hat O^\dagger+\delta\hat O^\dagger)(Q^{a+1}+\delta Q^{a+1})\br_\alpha=\bl Q_1\hat O^\dagger\delta Q^{a+1}\br_\alpha+\bl Q_1\delta\hat O^\dagger Q^{a+1}\br_\alpha \ .
\eeq
Using now the self-adjoint property in the form of \eq{OOb}, we see that the first term vanishes so that we get
\beq
\label{Ovar}
  \bl Q_1\d_p\hat O^\dagger Q^{a+1}\br_\alpha=0 \;.
\eeq
Explicitly, the variation of $\hat O^\dagger$ reads
\beq
\d_p \hat { O }^\dagger = \sum_{(b,\beta)}(-1)^{b+1}\d_p I_{b,\beta-1} u^{\beta-1}{\cal D}^{-2b+3}
-\hat Y_p
\eeq
with
\beq
\label{Yvar}
\hat Y_p=
-\[\d_p Q_\theta^{[-2\bs]} {\cal D}^{-3}+(-1)^3\d_p Q_\theta^{[+2\bs]} {\cal D}^{+3}\]
-\sum_{b}(-1)^{b+1}\d_p I_{b,L} u^{L}{\cal D}^{-2b+3} \;.
\eeq
Here we denoted by $I_{b,L}$ the leading coefficient in the transfer matrices of \eq{IMs} so that $I_{b,L}=\chi_b(\lambda_1,\lambda_2,\lambda_3)$. 
Plugging this into \eq{Ovar} we get a linear system for the variations $\d_p I_{b,\beta-1}$, of the form
\beq\la{ABN1}
\sum_{(b,\beta)}m_{(a,\alpha),(b,\beta)}(-1)^{b+1}\d_p I_{b,\beta-1} =
y_{(a,\alpha)}\;\;,\;\;y_{(a,\alpha)}\equiv \bl Q_1\;\hat Y_p\circ Q^{a+1} \br_\alpha\;
\eeq
where
\beq
    m_{(a,\alpha),(b,\beta)}\equiv \bl Q_1 \;u^{\beta-1}\; {\cal D}^{-2b+3}\circ Q^{a+1} \br_\alpha
\eeq
is the same matrix appearing in the $\sl(3)$ scalar product \eq{detsl3} with the two states taken to be the same. We can write the solution of \eq{ABN1} using Cramer's formulas as
\beq
\label{Ivar}
\d_p I_{b',\beta'-1}=(-1)^{b'+1}\frac{\det_{(a,\alpha),(b,\beta)}\tilde m_{(a,\alpha),(b,\beta)}}{\det_{(a,\alpha),(b,\beta)}m_{(a,\alpha),(b,\beta)}} \;,
\eeq
where $\tilde m_{(a,\alpha),(b,\beta)}$ is the matrix
$m_{(a,\alpha),(b,\beta)}$ with the column $(b',\beta')$ replaced with $y_{(a,\alpha)}$ defined in \eq{ABN1}. This gives a determinant representation for the variation of integrals of motion and the form factor \eq{ffac}. 

From the discussion in appendix \ref{app:slNbax} it is clear that the  result \eq{Ivar} extends immediately to the $\sl(N)$ case provided that instead of \eq{Yvar} we use
\beq
\hat Y_p=
{-\[\d_p Q_\theta^{[-2\bs]} {\cal D}^{-N}+(-1)^N\d_p Q_\theta^{[+2\bs]} {\cal D}^{+N}\]
-\sum_{b=1}^{N-1}(-1)^{b+1}\d_p \chi_b u^{L}{\cal D}^{-2b+N}} \;,
\eeq
and the indices $a,b,\dots$ take now values from $1$ to $N-1$ while the matrix $m_{(a,\alpha),(b,\beta)}$ is given by \eq{mdef0}. Let us also note that we can derive a similar determinant representation for the values of $I_{a,\alpha}$ themselves. For that we simply repeat the steps above starting now from the identity
\beq
\label{Ozer}
  \bl Q_1\hat O^\dagger Q^{a+1}\br_\alpha=0 \;
\eeq
rather than \eq{Ovar}.   This gives a linear system for the set of $I_{a,\alpha-1}$ with $\alpha=1,\dots,L$  whose solution reads
\beq
\label{Idet}
I_{b',\beta'-1}=(-1)^{b'+1}\frac{\det_{(a,\alpha),(b,\beta)}\tilde m_{(a,\alpha),(b,\beta)}}{\det_{(a,\alpha),(b,\beta)}m_{(a,\alpha),(b,\beta)}} \;,
\eeq
where now $\tilde m_{(a,\alpha),(b,\beta)}$ is the matrix
$m_{(a,\alpha),(b,\beta)}$ with the column $(b',\beta')$ replaced by $z_{(a,\alpha)}$ defined (similarly to \eq{ABN1}) as
\beq
    z_{(a,\alpha)}\equiv \bl Q_1\;\hat Z\circ Q^{a+1} \br_\alpha\;
\eeq
with
\beq
\hat Z=
{-\[ Q_\theta^{[-2\bs]} {\cal D}^{-N}+(-1)^N Q_\theta^{[+2\bs]} {\cal D}^{+N}\]
-\sum_{b=1}^{N-1}(-1)^{b+1} \chi_b u^{L}{\cal D}^{-2b+N}} \;.
\eeq
Notice that the determinant in the denominator of the result \eq{Idet} for the quantities $I_{a,\alpha-1}$ is the same as the one appearing for their variations given by \eq{Ivar} (and is the overlap of the state with itself).

\subsubsection{Example: local spin expectation value}\la{sec:detloc}

One of the key quantities of interest in spin chains are correlators of ``local" operators, i.e. those that act on a particular spin chain site in contrast to ``global" operators such as the transfer matrix. While certain maps from local to global operators are well known (see the reviews \cite{Slavnov:2018kfx,Slavnov:2019hdn}), here we will demonstrate that our approach offers yet another way to access local quantities.  Namely, there is a remarkable relation between a subset of local operators and derivatives of the integrals of motion ${\d\hat I_{a,\beta}}/{\d\theta_\alpha}$, whose expectation values we computed in the previous section.

The main idea is that when taking the derivative in $\theta_\alpha$ we can single out the $\alpha$-th spin chain site. To make it precise, let us write explicitly the large $u$ expansion of the transfer matrix with fundamental representation in the auxiliary space defined in \eq{Tdefab}, \eq{Tdef} using the form of the Lax matrix from \eq{LaxGen},
\beqa
\label{Tuexp}
    {\mathbb T}(u)&=& u^L\Tr\;(\Lambda)+u^{L-1}\sum_{\alpha=1}^L\(i\Tr({\mathbb E}^{(\alpha)t}\Lambda)-\theta_\alpha\Tr(\Lambda)\)+O(u^{L-2}) \ .
\eeqa
The trace here is taken over the auxiliary space, and ${\mathbb E}^{(\alpha)}$ is an $N\times N$ matrix whose element at position $(a,b)$ is the operator ${\mathbb E}_{a,b}$ (the $\sl(N)$ generator) acting on the $\alpha$-th site of the spin chain. Note that $\mE$ in this expression  is transposed w.r.t. the indices $a,b$ as we indicated with the superscript $t$ (this is due to the form of the Lax matrix \eq{LaxGen}).  We see that in \eq{Tuexp} we have a sum of local operators over all sites of the spin chain. Now we notice that when we differentiate the transfer matrix in $\theta_\gamma$, the Lax operator at position $\gamma$ in its definition will be simply replaced by minus the identity matrix, so as a result we will get the transfer matrix for the spin chain with the $\gamma$-th site removed. This means that the derivative will be given by the same result \eq{Tuexp} but with sum taken over all sites except one,
\beq
     \frac{\d{\mathbb T}(u)}{\d\theta_\alpha}=
     -u^{L-1}\Tr(\Lambda)+u^{L-2}\sum_{\beta\neq \alpha}\(i\Tr({\mathbb E}^{(\beta)t}\Lambda)-\theta_\beta\Tr(\Lambda)\)+O(u^{L-3}) \;.
\eeq
By combining this with \eq{Tuexp} we can therefore extract the contribution from the site $\gamma$ only,
\beq
    {\mathbb T}(u)+u \frac{\d{\mathbb T}(u)}{\d\theta_\alpha}=u^{L-1}\(i\Tr({\mathbb E}^{(\alpha)t}\Lambda)-\theta_\alpha\Tr(\Lambda)\)+O(u^{L-2})\;.
\eeq
Taking the coefficient of $u^{L-1}$ in this relation, we finally get
\beq
\label{EErel2}
    \Tr({\mathbb E}^{(\alpha)t}\Lambda)=-i\frac{\d \hat I_{1,L-2}}{\d\theta_\alpha}-i\hat I_{1,L-1}-i\theta_\alpha\Tr(\Lambda) \ .
\eeq
We remind that $\hat I_{a,\alpha}$ are the operator coefficients in the expansion of the transfer matrices in \eq{IMs}. We see that \eq{EErel2} is a relation between a local operator acting on the $\alpha$-th site (in the l.h.s.) and a global operator acting on all sites (in the r.h.s.).  Sandwiching this relation between left and right transfer matrix eigenstates $\ket{\Psi}$ and $\bra{\Psi}$, we find that the expectation value is given by
\beq
\label{EEE}
    \frac{
    \bra{\Psi}\Tr({\mathbb E}^{(\alpha)t}\Lambda)
    \ket{\Psi}
    }{{\bra{\Psi}\Psi\rangle}}=
    -i\frac{
    \bra{\Psi}\frac{\d \hat I_{1,L-2}}{\d\theta_\alpha}\ket{\Psi}
    }
    {\bra{\Psi}\Psi\rangle}
    -i I_{1,L-1}-i\theta_\alpha\Tr(\Lambda)\;.
\eeq
Let us note that this expression does not depend on normalisation of the states $\ket{\Psi}$. The only nontrivial correlator in the r.h.s. is the first term, which is given by the determinant \eq{Ivar} we derived above in the SoV approach. Thus we find a compact result for the expectation value of the local operator $\Tr({\mathbb E}^{(\alpha)t}\Lambda)$. 

We can also repeat a similar argument starting from the transfer matrices in $a$-th antisymmetric $\sl(N)$ representation in the auxiliary space. Using the results of appendix~\ref{oscrep}, we find that the generalization of \eq{EErel2} reads
\beq\label{acorr}
\sum_{j=1}^a
{\rm Tr}\(({\mathbb E}^{(\alpha)t}-\bs)(-\Lambda)^j\)\;\chi_{a-j}
=
\(i\theta_\alpha+ \bs\)\hat I_{a,L}+i \hat I_{a,L-1}+i\d_{\theta_\alpha} \hat I_{a,L-2}\;
\eeq
where $a=1,2,\dots,N-1$ and we recall that $\chi_j$ is the character defined in \eq{comptwist}. This gives a (lower triangular) system of $N-1$ linear equations from which we can extract the expectation values of the local operators ${\rm Tr}\({\mathbb E}^{(\alpha)t}\Lambda^j\)$ for $j=1,2,\dots,N-1$.

To illustrate the structure of this linear system, let us give as an example the expectation values of the operators in the r.h.s. of \eq{acorr} for the  $\sl(3)$ case with $a=2$ and $L=2$, taking $\alpha=1$. The first term is simply the character, $I_{2,2}=\chi_2$, while the other two are given by the determinants presented in \eq{Ivar} and \eq{Idet},
\beq
    I_{2,1}
=
    \frac{1}{\cN}\begin{vmatrix}
    \bl uQ_1Q_{12}^-\br_1 &
   \bl Q_1Q_{12}^-\br_1
    &\bl Q_1R_{12}\br_1
    &\bl Q_1Q_{12}^+\br_1
    \\ 
    \bl uQ_1Q_{13}^-\br_1 &
   \bl Q_1Q_{13}^-\br_1
     &\bl Q_1R_{13}\br_1
    &\bl Q_1Q_{13}^+\br_1
    \\
    \bl uQ_1Q_{12}^-\br_2 &
   \bl Q_1Q_{12}^-\br_2
    &\bl Q_1R_{12}\br_2
    &\bl Q_1Q_{12}^+\br_2
    \\ 
    \bl uQ_1Q_{13}^-\br_2 &
   \bl Q_1Q_{13}^-\br_2
    &\bl Q_1R_{13}\br_2
    &\bl Q_1Q_{13}^+\br_2
    \end{vmatrix}
\eeq
and 
\beq
    \d_{\theta_1}I_{2,0}=
     \frac{1}{\cN}\begin{vmatrix}
    \bl uQ_1Q_{12}^-\br_1 &
   \bl Q_1Q_{12}^-\br_1
    &
    \bl uQ_1Q_{12}^+\br_1
    &\bl Q_1S_{12}\br_1
    \\ 
    \bl uQ_1Q_{13}^-\br_1 &
   \bl Q_1Q_{13}^-\br_1
    &
    \bl uQ_1Q_{13}^+\br_1
    &\bl Q_1S_{13}\br_1
    \\
    \bl uQ_1Q_{12}^-\br_2 &
   \bl Q_1Q_{12}^-\br_2
    &
    \bl uQ_1Q_{12}^+\br_2
    &\bl Q_1S_{12}\br_2
    \\ 
    \bl uQ_1Q_{13}^-\br_2 &
   \bl Q_1Q_{13}^-\br_2
    &
    \bl uQ_1Q_{13}^+\br_2
    &\bl Q_1S_{13}\br_2
    \end{vmatrix}
\eeq
where $\cN={\rm det}_{(a,\alpha),(b,\beta)}\bl Q_1 \;u^{\beta-1}\; {\cal D}^{-2b+3}\circ Q^{a+1} \br_\alpha$ and
\beq
    R_A=Q_\theta^{[-2\bs]} Q_{A}^{[-3]}-Q_\theta^{[+2\bs]} Q_{A}^{[+3]}+u^2\chi_1Q_{A}^{[+1]}-u^2\chi_2Q_{A}^{[-1]} \ ,
\eeq
\beq
    S_A=(\d_{\theta_1}Q_\theta^{[-2\bs]}) Q_{A}^{[-3]}-(\d_{\theta_1}Q_\theta^{[+2\bs]}) Q_{A}^{[+3]}
     \ .
\eeq

Let us describe more explicitly the local operator $\Tr({\mathbb E}^{(\alpha)t}\Lambda)$ which enters \eq{EEE}. Notice that this equation holds as long as $\ket{\Psi}$ and $\bra{\Psi}$ are eigenstates of the transfer matrix constructed as in \eq{Tdefab}, \eq{Tdef} with the twist given by $\Lambda$. In our current notation $\Lambda$ is the non-diagonal companion twist matrix \eq{twistM}, so the operator $\Tr({\mathbb E}^{(\alpha)t}\Lambda)$ is quite nontrivial. At the same time, the corresponding states $\ket{\Psi}$ are somewhat unusual as they are built starting from the nontrivial ground state (e.g. \eq{sl2omega} for $\sl(2)$). Alternatively, we can go to a more standard frame by performing a global rotation that diagonalizes the twist matrix $\Lambda\to\Lambda_{diag}$. Then the states 
$\ket{\Psi}$ will also get rotated $\ket{\Psi}\to\ket{\Psi'}$, so that e.g. the ground state will become simply the trivial state $\ket{0}$ (given by \eq{00vacsl2} for $\sl(2)$). The value of the r.h.s. of \eq{EEE} is the same in either frame\footnote{since the eigenvalues of the transfer matrices do not change under this rotation
}, so we have 
\beq
    \frac{
    \bra{\Psi}\Tr({\mathbb E}^{(\alpha)T}\Lambda)
    \ket{\Psi}
    }{{\bra{\Psi}\Psi\rangle}}
    =
    \frac{
    \bra{\Psi'}\Tr({\mathbb E}^{(\alpha)T}\Lambda_{diag})
    \ket{\Psi'}
    }{{\bra{\Psi'}\Psi'\rangle}}\;.
\eeq
In the new frame, our local operator will be a simple combination of the Cartan elements. As an example, for $\sl(2)$ it will be
\beq
    \Tr({\mathbb E}^{(\alpha)t}\Lambda_{diag})=\lambda_1{\mathbb E}^{(\alpha)}_{11}+\frac{1}{\lambda_1}{\mathbb E}^{(\alpha)}_{22}=
    \frac{\lambda_1-1/\lambda_1}{2}\({\mathbb E}^{(\alpha)}_{11}-{\mathbb E}^{(\alpha)}_{22}\)    
\eeq
so it reduces to the usual spin projection operator ${\mathbb E}_{11}-{\mathbb E}_{22}$. Here we used that for $\sl(N)$ in our representation the operator $\sum_{a=1}^N{\mathbb E}_{aa}$ acts as a scalar,
\beq
\label{Eaasc}
    \sum_{a=1}^N{\mathbb E}_{aa} =(N-2)\bs \ .
\eeq
For $\sl(N)$ we can compute in this way the expectation values of all $N$  operators $\mE_{aa}$ on a given site by considering \eq{acorr} for $a=1,\dots,N-1$ together with the condition \eq{Eaasc}.

We note that form factors of exactly the type we can compute here are important e.g. in Landau-Lifshitz models \cite{Gerotto:2017sat}, and it would be interesting to further explore their properties.
Let us also point out that the expectation values of operators like $\d T(u)/\d \theta$ are not straightforwardly accessible by traditional methods of the algebraic Bethe ansatz, but appear to be natural objects in the SoV approach. We believe that exploring the interrelations between the SoV and more standard methods should open the way to computing a still larger class of correlators in the future.

\section{Outlook}
In this paper for the first time we found an explicit expression for the SoV measure for spin chains in highest-weight representation of $\sl(N)$ with general spin $\bs$. This was done by carefully analysing and bringing together two different approaches -- the operatorial SoV approach~\cite{Sklyanin:1992sm,Gromov:2016itr,Maillet:2018bim,Ryan:2018fyo,Liashyk:2018qfc} 
and the functional SoV approach~\cite{Cavaglia:2019pow}. 

The knowledge of the SoV measure has unlocked for us the possibility of computing a number of nontrivial scalar products, overlaps and form factors
for which we derived new determinant representations.
These results include in particular overlaps of states with different twists and on-shell/off-shell type overlaps.

Having direct access to the elements of the measure opens the way to an in-depth study of a great variety of important quantities in higher rank $\sl(N)$ models.
Some future applications may include form factors, correlators and g-functions of the types studied via SoV in  \cite{Smirnov:1998kv,Caetano:2020dyp,Kitanine:2015jna,Kitanine:2016pvg} for rank-one cases. 
Our results should also facilitate studying the thermodynamic limits for higher rank spin chains in the SoV framework, extending the existing $\su(2)$ results  (see e.g. \cite{Niccoli:2020zla}).

Another class of objects which can be computed using the functional SoV approach, generalising the initial observation in~\cite{Cavaglia:2019pow}, are the form factors of derivatives of the transfer matrices w.r.t. external parameters like twists or inhomogeneities. As we discussed in section~\ref{sec:detloc}, in particular this type of form factors includes local spin operators -- Cartan generators acting on one site of the chain. These are still to be fully understood within the operatorial SoV approach and could be relevant for the exact calculation of correlation functions in ${\cal N}=4$ SYM. 

Most of these results seem to be highly nontrivial to get  within traditional algebraic nested Bethe ansatz methods~e.g. \cite{Slavnov:2020ngt,Belliard:2019bfz}. Trying to merge these methods together could promise a fruitful interplay. 

Our results have already allowed us to compute rather exotic overlaps involving states with different twists. We are hopeful that they could find applications in ${\cal N}=4$ SYM where similar objects emerged already~ \cite{Cavaglia:2018lxi,McGovern:2019sdd,Cavaglia:2020hdb}. Our results could also give further clues about the type of algebraic structures that may appear in the $\cN=4$ SYM context.

We leave for the future investigation the generalisation of our results to non highest-weight representations. One should keep in mind that there are certain complications on this way -- none of the $Q$-functions $Q_i$, the constituent blocks of the SoV wave function, are polynomial anymore, furthermore the spectrum of conserved charges will no longer be a discrete set of points. These additional features should have certain effect on the SoV construction as well, and there are still some mysteries to uncover. This direction is particularly important as it has applications for the Fishnet/Fishchain theories~\cite{Gurdogan:2015csr,Gromov:2019aku}.
SoV-type methods adapted to the principal series representations have already led to a variety of interesting results in this context \cite{Derkachov:2018rot,Derkachov:2019tzo,Basso:2019xay,Derkachov:2020zvv} (see also \cite{Derkachov:2018ewi}). 

Another interesting direction is developing SoV for higher rank spin chains with different symmetry groups such as ${\mathfrak{ so}}(N)$ where progress was made recently in \cite{Ferrando:2020vzk, Ekhammar:2020enr}. 
Other natural extensions include the super-symmetric case (see \cite{Maillet:2019ayx}, \cite{Gromov:2018cvh} for related results), boundary problems and $q$-deformations (see \cite{Maillet:2018rto,Pei:2020ljw} for recent work). We hope to come back to some of these problems in future work.

\section*{Acknowledgements}
We thank D. Volin for collaboration at an initial stage of this project and for discussions. We are also grateful to A.~Cavaglia, G.~Ferrando, V.~Kazakov, G.~Korchemsky, A.~Liashyk and D.~Serban for helpful discussions. The work of N.G. was supported by European Research Council (ERC) under the European Union’s Horizon 2020 research and innovation programme (grant agreement No. 865075)  EXACTC. The work of P.R. is supported in part by a Nordita Visiting PhD Fellowship and by SFI and Royal Society grant UF160578.

\appendix

\section{Transfer matrices and antipode}\label{antipodeapp}

In this section we will derive the relation \eqref{newBTcom1} -- actually we will derive a more general form of it. Our main tool for doing this will be the so-called Yangian \textit{antipode} map $S$ which sends the monodromy matrix $\lT(u)$ to its inverse 
\begin{equation}
    S\left(\lT(u)\right)=\lT^{-1}(u)\,.
\end{equation}
For ease of notation we will also denote $\mathcal{S}(u)=\lT^{-1}(u)$. Note that this map extends in an obvious way to the twisted case: if, as before, $\bT(u)$ denotes the twisted monodromy matrix $\bT(u)=\lT(u)\Lambda$ then the antipode sends $\bT(u)\mapsto  \bS(u):=\Lambda^{-1}\lS(u)$. We will now derive a new commutation relation which intertwines $\bB$ and $\bbS_\mu$, which are transfer matrices constructed from $\bS(u)$ in a similar way as to how $\bbT_\mu$ is constructed from $\bT(u)$, and reduces to \eqref{newBTcom1} for special choice of $\mu$. 

In order to derive this relation we will need some properties of the antipode map, which we now describe. 
We will need to perform fusion with the inverted monodromy $\bS(u)$. The original monodromy matrix $\bT$ satisfies the RTT relation 
\begin{equation}
    R_{ab}(u-v)\bT_a(u)\bT_b(v)=\bT_b(v)\bT_a(u)R_{ab}(u-v)
\end{equation}
which acts on two copies $a$ and $b$ of the auxiliary space $\mathbb{C}^N$ and the physical Hilbert space, and the $R$-matrix $R_{ab}(u)$ is given by \begin{equation}
    R_{ab}(u)=u\, 1_{ab}+i\,P_{ab}
\end{equation}
where $1_{ab}$ and $P_{ab}$ denote the identity operator and the permutation operator on $\mathbb{C}^N\otimes \mathbb{C}^N$, respectively. Note that when $v=u+i$ the $R$-matrix $R_{ab}(u-v)=R_{ab}(-i)$ becomes an antisymmetriser, and it is this reason why the quantum minors \eqref{qminors} are constructed by taking products with the shift in each subsequent $\bT(u)$ increased by $i$. Inverting the RTT relation, we obtain
\begin{equation}
    R_{ab}^{-1}(u,v)\bT_a^{-1}(u)\bT_b^{-1}(v)=\bT_b^{-1}(v)\bT_a^{-1}(u)R_{ab}^{-1}(u-v)
\end{equation}
or
\begin{equation}
    R_{ab}^{-1}(u,v)\bS_a(u)\bS_b(v)=\bS_b(v)\bS_a(u)R_{ab}^{-1}(u-v)
\end{equation}
Since $R_{ab}(u)=u\, 1_{ab}+i P_{ab}$ we have that $R^{-1}(u)= u\, 1_{ab}-i P_{ab}$ up to a scalar factor, so fusion for $\bS$ is performed in exactly the same way as for $\bT$ up to changing the sign of the shifts, which leads to the following definition for quantum minors constructed from $\bS$
\begin{equation}\label{Sqminors}
    \bS\left[^{i_1 \dots i_n}_{j_1\dots j_n}\right](u)=\sum_{\sigma}(-1)^{{\rm deg}\,\sigma\,}\bS_{i_{\sigma(1)}j_1}(u)\bS_{i_{\sigma(2)}j_2}(u-i)\dots \bS_{i_{\sigma(n)}j_n}(u-i(n-1))\,.
\end{equation}
Since the transfer matrices $\bbT_\mu(u)$ corresponding to generic Young diagrams $\mu$ can be expressed in terms of $\bbT_{a,1}$ using the CBR formula 
\begin{equation}
   \bbT_{\mu}(u)=\det_{1\leq j,k\leq \mu_1}\bbT_{\mu'_j-j+k,1}\left(u-\frac{i}{2}\left(\mu'_1-\mu_1-\mu'_j+j+k-1\right)\right)
\end{equation}
it then follows that all $\bbS_\mu(u)$ can be expressed in terms of $\bbS_{a,1}(u)$ as 
\begin{equation}\label{Scbr}
   \bbS_{\mu}(u)=\det_{1\leq j,k\leq \mu_1}\bbS_{\mu'_j-j+k,1}\left(u+\frac{i}{2}\left(\mu'_1-\mu_1-\mu'_j+j+k-1\right)\right)\,.
\end{equation}
It is important to stress that the transfer matrices $\bbS_\mu(u)$ do not give us a different set of conserved charges, or even any new ones. All transfer matrices constructed from $\bS(u)$ can be written as simple expressions in transfer matrcies obtained from $\bT(u)$. Denoting by $\mathcal{I}=\{i_1,i_2,\dots,i_N\}$ and $\mathcal{J}=\{j_1,j_2,\dots,j_N \}$ two permutations of $\{1,2,\dots,N\}$ then the following relation can be shown to be true \cite{molev2007yangians}
\begin{equation}
    \bbT_{N,1}\left(u+\frac{i}{2}(N-1)\right)\, \bS\left[^{j_{m+1}\dots j_N}_{i_{m+1}\dots i_N}\right](u+i(N-1))={\rm sgn}(\mathcal{I}){\rm sgn}(\mathcal{J}) \bT\left[^{i_1\dots i_m}_{j_1 \dots j_m}\right](u)\,.
\end{equation}
This relation then implies the following relation for transfer matrices
\begin{equation}\label{STrelation}
    \bbT_{N,1}\left(u+\frac{i}{2}(N-1)\right)\bbS_{a,1}\left(u+i(N-1)-\frac{i}{2}(a-1)\right) = \bbT_{N-a,1}\left(u+\frac{i}{2}(N-a-1)\right)\,.
\end{equation}
Since all $\bbS_{\xi}$ and all $\bbT_{\xi}$ respectively can be constructed as polynomials in these anti-symmetric transfer matrices it follows that 
\begin{equation}
    [\bbS_{\mu_1}(u_1),\bbT_{\mu_2}(u_2)]=0
\end{equation}
for any Young diagrams $\mu_1$ and $\mu_2$.

We will introduce one final map, which is usually denoted by $\omega$ \cite{molev2007yangians}, obtained by first applying the $*$-map and then the antipode $S$. It is trivial to check using the definitions that 
\begin{equation}
    \omega\left(\bT\left[^{i_1\dots i_a}_{j_1\dots j_a}\right](u)\right)=\bS\left[^{i_1\dots i_a}_{j_1\dots j_a}\right](u)\,.
\end{equation}
This holds if we are twisted or not, so in particular 
\begin{equation}\label{sminorrel}
    \omega\left(\lT\left[^{i_1\dots i_a}_{j_1\dots j_a}\right](u)\right)=\lS\left[^{i_1\dots i_a}_{j_1\dots j_a}\right](u)\,.
\end{equation}
We will now consider the commutation relation \eqref{BTcom} which we repeat here for convenience 
\begin{equation}\label{BTcomagain}
    \bbT_\mu\left(v-\frac{i}{2}\left(\mu_1-\mu^\prime_1\right)\right)\bB(u)=f_\mu(u,v)\bB(u)\bbT_\mu\left(v-\frac{i}{2}\left(\mu_1-\mu^\prime_1\right)\right)+\lR_1(u,v)\,.
\end{equation}
As was stated earlier in the paper the relation \eqref{BTcomagain} is strictly speaking not correct as we have written it. The correct version is given by 
\begin{equation}\label{BTcorrect}
    \bbT^{\lN}_\mu\left(v-\frac{i}{2}\left(\mu_1-\mu^\prime_1\right)\right)\bB(u)=f_\mu(u,v)\bB(u)\bbT^{\lN}_\mu\left(v-\frac{i}{2}\left(\mu_1-\mu^\prime_1\right)\right)+\lR_1(u,v)\,.
\end{equation}
The objects $\bbT^{\lN}_\mu$ are \textit{null twist} transfer matrices, obtained from $\bbT_\mu(u)$ by sending all twist eigenvalues\footnote{In this section for transparency we do not impose $\det\Lambda=1$ and leave all $\lambda_1,\dots,\lambda_N$ free.} $\lambda_i$ to $0$. It was demonstrated in \cite{Ryan:2018fyo} that all $\bbT_\mu(u)$ have the form 
\begin{equation}
    \bbT_\mu\left(v-\frac{i}{2}\left(\mu_1-\mu^\prime_1\right)\right) = \bbT^{\lN}_\mu\left(v-\frac{i}{2}\left(\mu_1-\mu^\prime_1\right)\right)+\displaystyle \sum_{j=1}^N \lT_{j1}(v)\times \text{twist}
\end{equation}
and so when applied to an eigenstate of $\bB$ which is annihilated by $\lT_{j1}(v)$ \eqref{BTcorrect} is equivalent to \eqref{BTcomagain}. 

The new relation we will derive then reads 
\begin{equation}\label{SBcom}
    \bbS^{\infty}_\mu\left(v+\frac{i}{2}\left(\mu_1-\mu^\prime_1+2\right)\right)\bB(u)=g_\mu(u,v)\bB(u)\bbS^{\infty}_\mu\left(v+\frac{i}{2}\left(\mu_1-\mu^\prime_1+2\right)\right)+\lR_3(u,v) 
\end{equation}
where 
\begin{equation}
    g_\mu(u,v)=\displaystyle\prod_{a=1}^{h_\mu}\frac{u-v+i(a-1-\mu_a)}{u-v+i(a-1)},\quad \lR_3(u,v)=\sum_{j=1}^N T_{1j}(v)\times\dots
\end{equation}
and $\bbS^{\infty}_\mu$ are transfer matrices obtained from $\bbS_\mu(u)$ by sending the twist eigenvalues $\lambda_i\rightarrow \infty$, in analogy with the null twist transfer matrices. 

Let us recall that the twist matrix $\Lambda$ has the explicit form $\Lambda_{ij}=(-1)^{j-1}\chi_j\delta_{i1}+\delta_{i,j+1}$, and lets define $\Lambda^{\lN}_{ij}=\delta_{i,j+1}$ so that 
\begin{equation}
    \Lambda_{ij}=\Lambda_{ij}^{\lN}+(-1)^{j-1}\\chi_j\delta_{i1}\,.
\end{equation}
Next, it is easy to work out the inverse matrix $\Lambda^{-1}$ is given by
\begin{equation}
    \Lambda^{-1}_{ij}=(-1)^j\frac{\chi_{j-1}}{\chi_N}\delta_{iN}+\delta_{i+1,j}
\end{equation}
and in the same way we define $\Lambda^{\infty}_{ij}=\delta_{i+1,j}$. The main property we will use below is that $\Lambda^{\lN}$ and $\Lambda^{\infty}$ are related by a simple change of basis. Let $K$ denote the matrix with $K_{ij}=\delta_{N+1-i,j}$. Then 
\begin{equation}
    K \Lambda^{\lN}K^{-1}=\Lambda^{\infty}\,.
\end{equation}

Now lets consider the transfer matrix in the fundamental representation, $\mathbb{T}^{\lN}_{1,1}={\rm tr}\left(\lT(u)\Lambda^{\lN}\right)$. Unfortunately we cannot apply the antipode map to $\bbT^{\lN}_{1,1}$ as the twist $\Lambda^{\lN}$ is not invertible. On the other hand, we can bring $\bbT^{\lN}_{1,1}$ to $\bbS^{\infty}_{1,1}$ by first performing a change of basis by $K$ and then applying the $\omega$ map to the untwisted monodromy. Let consider the change of basis $\lT_{ij}\rightarrow \lT_{N+1-i,N+1-j}$ which can be done with $K$. More precisely, we transform 
\begin{equation}
    {\rm tr}\left(\lT(u)\Lambda^{\lN}\right)\rightarrow {\rm tr}\left(K\lT(u)K^{-1}\Lambda^{\lN}\right)
\end{equation}
which, because of the cyclicity of the trace, is equivalent to sending $\Lambda^{\lN}\rightarrow K\Lambda^{\lN} K^{-1}=\Lambda^{\infty}$! If we then apply $\omega$ to the untwisted monodromy $\lT$ we will obtain precisely $\bbS^{\infty}_{1,1}$. It is a straightforward calculation to show that the procedure can be done for all transfer matrices $\bbT^{\lN}_\mu$, by first considering transfer matrices in anti-symmetric representations $\bbT_{a,1}$ and then considering $\bbT_\mu$ together with the CBR formula \eqref{cbrformula}, being careful to take in account various shifts.

To summarise, the net effect of the change of basis $K$ followed by $\omega$ is 
\begin{equation}
    \bbT^{\lN}_\mu(u)\rightarrow \bbS^{\infty}_\mu(u)\,.
\end{equation}

It now remains to see what happens to $\bB$ when we apply $K$ followed by $\omega$. For simplicity of the calculation we will consider the $\sl(3)$ case, with higher rank following immediately. 

Recall the explicit form of $\bB$ in terms of untwisted monodromy entries $\lT_{ij}$ given in \eqref{untwistedB}: 
\begin{equation}
    \bB(u)=\lT_{11}\lT^{[-2]}\left[^{12}_{12}\right]+\lT_{21}\lT^{[-2]}\left[^{12}_{13}\right]\,.
\end{equation}
Now we perform the change of basis given by $K$ which sends $\lT_{ij}\rightarrow \lT_{N+1-i,N+1-j}$ which results in 
\begin{equation}
    \bB\mapsto \lT_{33}\lT^{[-2]}\left[^{23}_{23}\right]+\lT_{23}\lT^{[-2]}\left[^{23}_{13}\right]\,.
\end{equation}
By applying $\omega$ we obtain
\begin{equation}
    \omega\left(\lT_{33}\right)\omega\left(\lT^{[-2]}\left[^{23}_{23}\right]\right)+\omega\left(\lT_{23}\right)\omega\left(\lT^{[-2]}\left[^{23}_{13}\right]\right)
\end{equation}
after which we apply the relation \eqref{sminorrel} to obtain, up to an overall factor of the scalar quantum determinant, $\bB^{[-2]}$, and the same computation goes through for higher rank in exactly the same way\footnote{Actually, one finds that the order of minors will be reversed. For $\sl(3)$ this doesn't matter since all of the minors commute. For higher-rank this is no longer the case, but it can be easily checked, using the method of \cite{Ryan:2018fyo} that the commutation relation \eqref{BTcom} is invariant under reversing the order of minors in $\bB$, meaning one could apply the mentioned sequence of transformations to that commutation relation and the end result will contain precisely $\bB$.}.

Now lets examine the remainder term $\lR_1(u,v)=\sum_{j=1}^N \lT_{j1}(v)\times$ in \eqref{BTcomagain}. Applying the same transformations results in $\sum_{j=1}^3 \omega\left(\lT_{4-j,3}(v)\right)\times$ which works out to be, up to quantum determinant factors, 
\begin{equation}
    \lT\left[^{12}_{12}\right]^{[-4]}\times\dots+\lT\left[^{12}_{13}\right]^{[-4]}\times\dots+\lT\left[^{12}_{23}\right]^{[-4]}\times\dots
\end{equation}
which can be recast in the form $\lT_{1j}^{[-2]}(v)\dots$ by expanding out the minors. Putting all of this together we hence obtain 
\begin{equation}\label{BScom}
    \bbS^{\infty}_\mu\left(v+\frac{i}{2}\left(\mu_1-\mu^\prime_1+2\right)\right)\bB(u)=g_\mu(u,v)\bB(u)\bbS^{\infty}_\mu\left(v+\frac{i}{2}\left(\mu_1-\mu^\prime_1+2\right)\right)+\lR_3(u,v) 
\end{equation}
with $\lR_3(u,v)=\sum_{j=1}^N \lT_{1j}(v)$.

Next, if we recall that the transfer matrices $\bbT_\mu(v)$ satisfied 
\begin{equation}
   \bbT_\mu \left(v-\frac{i}{2}\left(\mu_1-\mu^\prime_1\right)\right)= \bbT^\lN_\mu\left(v-\frac{i}{2}\left(\mu_1-\mu^\prime_1\right)\right)+\sum_{j=1}^N \lT_{j1}(v)\times \dots
\end{equation}
and so, on eigenstates $\bra{\Lambda} $of $\bB$ which satisfy $\bra{\Lambda}\lT_{j1}(v)=0$ we can replace $\bbT^{\lN}_\mu$ in \eqref{BTcorrect} with $\bbT_\mu$, resulting in \eqref{BTcomagain}. In a similar style, we can show that 
\begin{equation}
 \bbS_\mu\left(v+\frac{i}{2}\left(\mu_1-\mu^\prime_1+2\right)\right) =\bbS^{\infty}_\mu\left(v+\frac{i}{2}\left(\mu_1-\mu^\prime_1+2\right)\right)+\sum_{j=1}^N \lT_{1j}(v)\times \dots
 \end{equation}
 and so on eigenstates $\bra{\Lambda} $of $\bB$ which satisfy $\bra{\Lambda}\lT_{1j}(v)=0$ we can replace $\bbS^{\infty}_\mu$ in \eqref{BScom} with $\bbS_\mu$, resulting in
 \begin{equation}\label{newBSrel}
         \bbS_\mu\left(v+\frac{i}{2}\left(\mu_1-\mu^\prime_1+2\right)\right)\bB(u)=g_\mu(u,v)\bB(u)\bbS_\mu\left(v+\frac{i}{2}\left(\mu_1-\mu^\prime_1+2\right)\right)+\lR_3(u,v)\,.
 \end{equation}
Now in order to derive \eqref{newBTcom1} we will specialise to Young diagrams $\mu$ with a single row, so that $\mu_1=s$ and $\mu_1^\prime=1$ and in this case the transfer matrix $\bbS_\mu$ is denotes $\bbS_{1,s}$, and so \eqref{newBSrel} becomes 
\begin{equation}
    \bbS_{1,s}\left(v+\frac{i}{2}\left(s+1\right)\right)\bB(u)=g_\mu(u,v)\bB(u)\bbS_{1,s}\left(v+\frac{i}{2}\left(s+1\right)\right)+\lR_3(u,v)\,.
\end{equation}
By using the CBR formula \eqref{Scbr} together with \eqref{STrelation} we can easily derive the following relation 
\begin{equation}
    \bbS_{1,s}\left(v+\frac{i}{2}(s+1)\right)=\frac{\bbT_{N-1,s}\left(v-\frac{i}{2}(N-s-1)\right)}{\displaystyle \prod_{k=1}^s \bbT_{N,1}\left(v-\frac{i}{2}(N-2k-1)\right)}\,.
\end{equation}
Since the $\bbT_{N,1}$ factors are just scalar multiples of the identity operator we can then replace $\bbS_{1,s}\left(v+\frac{i}{2}(s+1)\right)$ in \eqref{newBSrel} with $\bbT_{N-1,s}\left(v-\frac{i}{2}(N-s-1)\right)$ to finally obtain \eqref{newBTcom1}.

\section{Eigenstates of $\bB$ and $\bC$ form a basis}\label{basisproof}
In this appendix we prove that the eigenstates $\bra{\svx}$ of $\bB$ and $\ket{\svy}$ of $\bC$ indeed form a basis of the Hilbert space which we remind the reader is the space of functions analytic at the origin. We will demonstrate this in two parts. First, we show that all of the vectors $\bra{\svx}$ and $\ket{\svy}$ constructed in the main text are non-zero. The fact that they are linearly independent follows from the fact that each $\bra{\svx}$ and $\ket{\svy}$ corresponds to a unique eigenvalue of $\bB$ and $\bC$. Then, we will show that every element of the Hilbert space admits a series representation in $\bra{\svx}$ and $\ket{\svy}$. 

\subsection{$\bra{\svx}$ and $\ket{\svy}$ are non-zero}
In the main text we constructed the SoV bases by action of certain transfer matrices evaluated at special points. We need to check that the resulting states are not identically zero. To do this we need to check that the required transfer matrices do not having vanishing eigenvalues at the required point. In order to do this we use the method developed in \cite{Ryan:2020rfk}, which is to check that the transfer matrices in question do not have vanishing eigenvalues for a specific value of the twist and hence to not in general. We use the fact that, for length $L=1$, when the twist matrix is the identity matrix all transfer matrices are just scalar multiples of the identity operator where we can easily verify that the required transfer matrices are non-vanishing. At higher values of $L$ the statement can be shown to reduce to the $L=1$ case by taking limit where inhomogeneities are largely separated \cite{Ryan:2020rfk}. 

In order to prove the above claims we will use so-called quantum semi-standard Young tableaux \cite{Kuniba:1994na,Kuniba:1995gi,Tsuboi:1998ne,Tsuboi:1998sc,Ryan:2020rfk}. A semi-standard Young tableaux $\lT$ of shape $\mu$ is a Young diagram $\mu$ filled with numbers from $\{1,2,\dots,N\}$ such that the numbers in each row weakly decrease and the numbers in each column strictly decrease. Let $g\in\mathsf{GL}(N)$ have eigenvalues $\lambda_1,\dots,\lambda_N$. Then the character $\chi_\mu(g)$ of $g$ in the representation $\mu$ can be written as a sum over all tableaux $\lT$ of shape $\mu$:
\begin{equation}
    \chi_\mu(g) = \displaystyle\sum_{\lT}\prod_{(a,s)\subset \mu} \lambda_{\#(a,s)}
\end{equation}
where $\#(a,s)$ denotes the number in position $(a,s)$ of the tableaux $\lT$.

\begin{figure}[htb]\label{reducedYD1}
  \centering
  \includegraphics[width=70mm,scale=10]{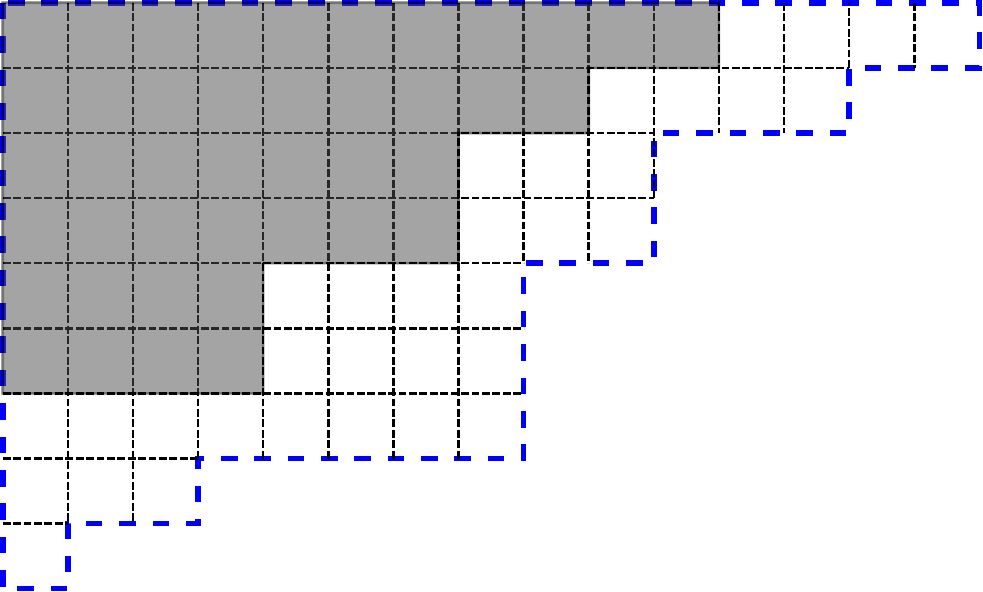}
  \caption{The reduced Young diagram $\bar{\nu}$ is denoted by the dotted exterior and the shaded region denotes the Young diagram $\mu$.}
\end{figure}

Similar expressions exist for transfer matrices. In order to describe them, let us consider generic highest-weight representations with $\gl(N)$ highest-weight $\nu_1,\dots,\nu_N$. Denote by $\lR_k(u)= (u-\theta+i \bs+i \nu_k)$, $k=1,\dots,N$. Then the transfer matrix $\bbT_\mu(u)$ can be also written as a sum over tableaux of shape $\mu$ according to the rule 
\begin{equation}\label{qtableauxsum}
    \bbT_\mu\left(u-\frac{i}{2}(\mu_1-\mu^\prime_1)\right) = \displaystyle\sum_{\lT}\prod_{(a,s)\subset \mu} \lR_{\#(a,s)}(u+i(a-s))\,.
\end{equation}
In \cite{Ryan:2020rfk} $\bbT_\mu\left(\theta_\alpha-i\bs-i\nu_N-\frac{i}{2}(\mu_1-\mu^\prime_1)\right)$ was proven to be non-zero under certain conditions on $\mu$, which were precisely the cases needed for constructing the SoV basis. Namely, let $\bar{\nu}$ denote the "reduced" diagram of $\nu$, i.e. $\bar{\nu}$ is the Young diagram $[\bar{\nu}_1,\dots,\bar{\nu}_N]$ where $\bar{\nu}_j:=\nu_j-\nu_N$. Then $\bbT_\mu\left(\theta_\alpha-i\bs-i\nu_N-\frac{i}{2}(\mu_1-\mu^\prime_1)\right)$ is non-zero if any only if $\mu\subset \bar{\nu}$, i.e. if and only if the diagram of $\mu$ can be inscribed inside the diagram of $\bar{\nu}$, see Figure \ref{reducedYD1}.

Let us examine how this rule changes under action of the $*$-map, i.e. when are the transfer matrices $\bbT^*_\mu(v)$ necessary for constructing $\ket{\svy}$ non-zero? To understand this we should first understand what happens with transfer matrices in antisymmetric representations $\bbT_{a,1}$. By writing the sum \eqref{qtableauxsum} we obtain
\begin{equation}
    \bbT_{a,1}\left(u+\frac{i}{2}(a-1)\right) = \displaystyle\sum_{i_{a}>\dots>i_1} \lR_{i_a}(u)\dots \lR_{i_1}(u+i(a-1))
\end{equation}
Applying the $*$-map we find that 
\begin{equation}
    \bbT_{a,1}\left(u+\frac{i}{2}(a-1)\right) \mapsto \bbT_{a,1}\left(u-\frac{i}{2}(a-1)\right)
\end{equation}
which can be written as 
\begin{equation}
    \displaystyle\sum_{i_1<\dots<i_a} \lR_{i_1}(u)\dots \lR_{i_a}(u-i(a-1))
\end{equation}
and hence we can deduce from the CBR formula \cite{Cherednik,Bazhanov:1989yk} that all of the transfer matrices $\bbT_\mu^*(u)$ can be written as 
\begin{equation}\label{tstartableaux}
    \bbT^*_\mu \left(u+\frac{i}{2}(\mu_1-\mu_1^\prime)\right)=\displaystyle\sum_{\lT^\prime}\prod_{(a,s)\subset \mu} \lR_{\#(a,s)}(u-i(a-s))
\end{equation}
where $\lT^\prime$ is a tableaux of shape $\mu$ but now the entries strictly increase (instead of decrease) in each column and weakly increase in each row, which is equivalent to permuting the weights $\{\nu_1,\dots,\nu_N\}\mapsto \{\nu_N,\dots,\nu_1\}$. Notice also the difference in the sign of the shift in $\lR_{\#(a,s)}$ in \eqref{tstartableaux} compared to \eqref{qtableauxsum}. Hence, $\bbT_\mu^*$ has the interpretation of being a Young diagram describing the lowest-weight of the representation instead of the highest. For the classical character $\chi_\mu(g)$ there is no difference, but $\bbT_\mu$ and $\bbT_\mu^*$ correspond to two different quantizations of this character. Alternatively, instead of swapping the weights and signs of shifts we can interpret both of these as having flipped the Young diagram upside down and backwards, while keepting the same rules for associating shifts to boxes after assigning zero shift to the bottom right corner. 

\begin{figure}[htb]\label{reducedYD2}
  \centering
  \includegraphics[width=70mm,scale=10]{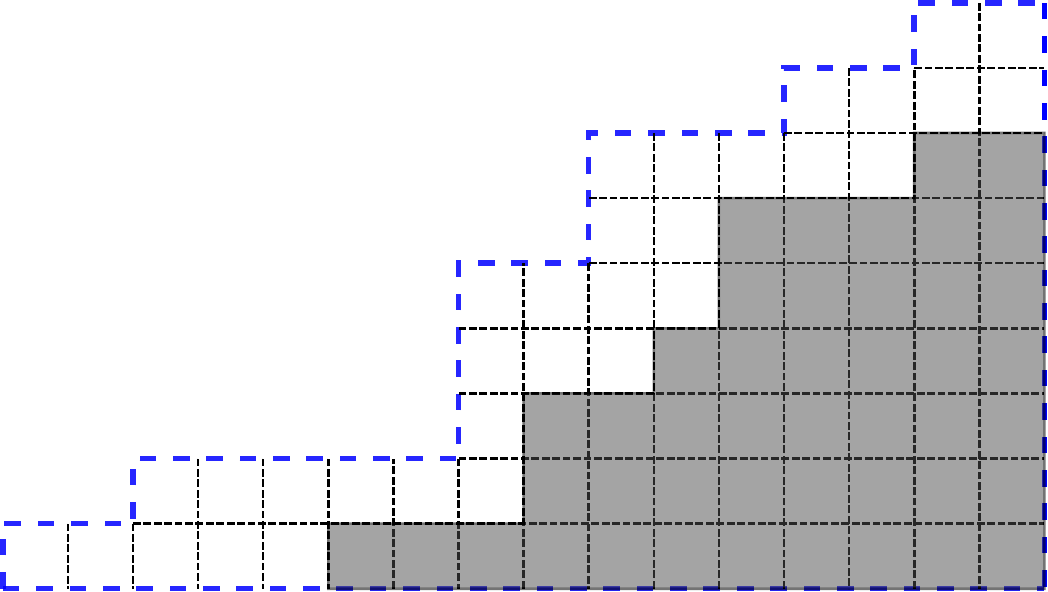}
  \caption{The reduced Young diagram $\bar{\nu}^*$ is denoted by the dotted exterior and the shaded region denotes the Young diagram $\mu$.}
\end{figure}

Now consider the $*$-reduced diagram $\bar{\nu}^* = \{\bar{\nu}^*_1,\dots, \bar{\nu}^*_N\}$ where we define $\bar{\nu}^*_j:= \nu_j-\nu_1$. Since all of the entries of $\bar{\nu}^*$ are either zero or negative we can view the diagram as having been flipped upside down and backwards. Then the requirement on $\mu$ for $\bbT_\mu^*\left(\theta_\alpha-i\bs-i\,\nu_1+\frac{i}{2}(\mu_1-\mu_1^\alpha)\right)$ to be nonzero is totally analogous to the original case, that is $\mu$ should be a subdiagram of $\bar{\nu}^*$ after $\mu$ has been flipped upside down and backwards as well, see Figure \ref{reducedYD2}. 

Restricting to the compact case of interest where $\nu_1=-2\bs,\nu_2=0=\dots=\nu_N$, this then implies that the transfer matrix 
\begin{equation}
    \bbT_\mu^*\left(\theta_\alpha+i\bs+\frac{i}{2}(\mu_1-\mu_1^\alpha)\right)
\end{equation}
will be non-zero as long as we restrict to $\mu$ being height at most $N-1$ and width at most $-2\bs$. When we extend to general $\bs$ this latter restriction is no longer present, and the only condition is that $\mu$ is of height at most $N-1$, and hence all $\ket{\svy}$ are non-zero.

Finally, we examine $\bra{\svx}$ which is constructed using the transfer matrix 
$$\bbT_{N-1,s}\left(\theta_\alpha+i\bs-\frac{i}{2}(N-s-1)\right)\,.$$ 
This transfer matrix is the same as $\bbT_{N-1,s}\left(\theta_\alpha+i\bs+\frac{i}{2}(N-s-1)\right)$, which admits the sum over tableaux \eqref{qtableauxsum}, together with an overall shift of $i(s-N+1)$. If we write 
$$\bbT_{N-1,s}\left(\theta_\alpha-i\bs+\frac{i}{2}(N-s-1)\right)$$
using the sum over tableaux then any tableaux containing $1$ must contain it in the bottom right corner, which comes with a shift of $-i(s-N+1)$. Hence, the transfer matrix $\bbT_{N-1,s}\left(\theta_\alpha+i\bs+\frac{i}{2}(N-s-1)\right)$ when expanded in a sum over tableaux will always contribute a factor $\lR_1(\theta_\alpha+i\bs)=0$ if that tableaux contains $1$, and so only tableaux which don't contain $1$ can contribute -- in fact there is a unique such tableaux. In the compact case, it can easily be worked out that this single contribution is non-vanishing as long as $s\leq -2\bs$, and in the non-compact case it is always non-vanishing. Hence, all $\bra{\svx}$ are non-zero. 

\subsection{Series representation}
Now that we have demonstrated that all $\bra{\svx}$ and $\ket{\svy}$ are non-zero, we need to show that every element $f$ of the Hilbert space can be written as 
\begin{equation}
    f=\sum_{\svx}c_\svx \bra{\svx}
\end{equation}
where $c_{\svx}$ are some finite coefficients and similarly with $\ket{\svy}$. 

The key point is then that both the orthonormal basis of monomials and $\bra{\svx}$ are eigenvectors of the SoV charge operator \eqref{SoVcharge}. The charge operator gives the Hilbert space $\mathcal{H}$ the structure of a graded space, that is $\lH=\bigoplus_{s\geq 0} \lH_{s}$ where $\lH_{s}$ is the subspace of SoV charge $s$. It is then a trivial counting exercise to verify that the number of $\bra{\svx}$ contained in $\lH_s$ precisely matches the number of basis monomials, and each is contained in precisely one $\lH_s$. Hence, $\bra{\svx}$ with charge $s$ form a basis of the finite-dimensional space $\lH_s$. 

By definition $f$ admits a series expansion in the orthonormal basis of monomials and this series is absolutely convergent so can be rearranged in any order we like. We can then arrange it in order of increasing charge and so can write $f$ as 
\begin{equation}
    f=\displaystyle\sum_{s\geq 0} f_s
\end{equation}
where $f_s$ denotes the projection of $f$ onto $\lH_s$. Then, each $f_s$ can be written as a finite linear combination of $\bra{\svx}$ with charge $s$ and precisely the same argument goes through for $\ket{\svy}$, completing the proof.

\section{Measure for $\sl(N)$ from the Baxter equations}
\label{app:slNbax}

In this section we extend the results of sections \ref{sec:intsl2} and \ref{sec:intsl3} for the integral form of the orthogonality relation to the generic $\sl(N)$ case. We present the derivation in a more algebraic way, largely following the one done in \cite{Cavaglia:2019pow} for the $\bs=1/2$ case.

In order to concisely obtain the transfer matrix eigenvalues in the $\sl(N)$ case, it is convenient to use the generating functional described \cite{Krichever:1996qd} (see \cite{Gromov:2010kf} for a review) which in our case it can be written as
\beq
\label{Wsln}
    {\cal W}=Q_\theta^{[+2\bs]}(1-\Lambda_1 {\cal D}^{-2})(1-\Lambda_2 {\cal D}^{-2})\dots (1-\Lambda_N {\cal D}^{-2}) \;,
\eeq
where $\Lambda_n$ are `quantum eigenvalues' that are given in terms of Q-functions,
\beq
\label{Lamsln}
\Lambda_1=\frac{Q^{[-2\bs]}_{\theta}}{
Q_{\theta}^{[+2\bs]}
}
\frac{
Q_{1}^{--}
}{Q_{1}}
\;\;,\;\;\Lambda_i=\frac{
Q_{J_{i-1}}^{[+i]}
}{Q^{[-2+i]}_{J_{i-1}}}
\frac{
Q_{J_{i}}^{[-3+i]}
}{Q^{[-1+i]}_{J_{i}}} \ \ , \ \ i=2,\dots,N \;,
\eeq
where $J_i\equiv 12\dots i$ is a multi-index, so that e.g. $Q_{J_2}=Q_{12}$ (recall also that $Q_{12\dots N}=1$ in our conventions). The Q-functions which appear here are all twisted polynomials, i.e. polynomials times exponents. 
This functional provides the (nontrivial part of the) eigenvalues of transfer matrices $\tau_k$ with $k$-th antisymmetric representation of $\sl(N)$ in the auxiliary space as coefficients of powers $\cD^n$ in its expansion,
\beq
\label{Wtau}
   \cW=
    \sum_{a=0}^N\tau_a(-1)^a\cD^{-2a} \ .
\eeq
Let us note also that the first and last terms are state-independent,
\beq
    \tau_0=Q_\theta^{[+2\bs]} \ , \ \ \ \tau_N=Q_\theta^{[-2\bs]} \ .
\eeq
All the other $\tau_k$ are also polynomials (of degree $L$) as long as the Bethe ansatz equations are satisfied. This can be shown by following the same argument as in \cite{Cavaglia:2019pow}. Their relation to $T_{a,1}$ (eigenvalues of $\mathbb{T}_{a,1}$ defined in section \ref{sec:heis}) is
\beq
\label{Tvstau}
T_{a,1}=\tau_a^{[+a-1]}\prod_{k=1}^{a-1}Q_\theta^{[+2\bs-2k+a-1]} \ .
\eeq

Using this functional we can write the Baxter equations in a very compact form. Following \cite{Gromov:2019wmz} let us introduce the notation for action of the shift operators to the left and to the right,
\beq
    f\overleftarrow{\cD}=f^- \ , \ \ \ \overrightarrow{\cD}f=f^+\;.
\eeq
Then we see from the last factor in \eq{Wsln} that $\overrightarrow{\cW}Q^{N[+N]}=0$. Similarly to \cite{Cavaglia:2019pow} one can show that this is true for all twisted polynomial Q's with one upper index (which we remind are defined in \eq{Qupdef}),
\beq
\label{WQup}
    \overrightarrow{\cW}Q^{(a+1)[+N]}=0 \ , \ \ a=1,\dots,N-1 \ .
\eeq
Moreover, from the form of the first factor in \eq{Wsln} we find that when acting to the left $\cW$ annihilates $Q_1$ if we multiply it by the same function $\varepsilon$ we used before in \eq{defvarep},
\beq
\label{WQdn}
     Q_1 \varepsilon\overleftarrow{\cW}=0 \; .
\eeq
As a result we can write the $N$-th order Baxter equations satisfied by $Q^a$ and $Q_1$ as \eq{WQup} and \eq{WQdn}.
Due to \eq{Wtau}, the first of these equations can be written in a more explicit form as
\beq
\hat { O}^\dagger Q^{a+1}
=0 \ ,  \ \ \ \ a=1,\dots,N-1
\eeq
where we introduced the difference operator $\hat { O}^\dagger$,
\beq
\hat { O}^\dagger Q^a\equiv\tau_0 Q^{a[+N]} - \tau_1Q^{a[+N-2]} + \dots +  (-1)^{N-1}\tau_{N-1}Q^{a[-N+2]} +    (-1)^N\tau_N Q^{a[-N]}=0 \; .
\eeq
This makes it clear that in particular for $N=3$ we get the Baxter equation \eq{Bax12s2} we described before.

\subsection{Orthogonality}

Like for the $\sl(3)$ case, in order to derive the orthogonality relations for Q-functions we will prove the key relation
\beq
\label{sasln}
   \bl Q_1\hat {O}^\dagger f\br_\alpha=0
\eeq
where we take the measure $\mu_\alpha$ in the definition of the bracket (see \eq{scalargf}) to be the same function \eq{muj} as for $\sl(3)$ and $\sl(2)$. Here $f$ is a regular function with the same large $u$ asymptotics as any of the $Q^{a+1}$ functions ($a=1,\dots,N-1$). To demonstrate \eq{sasln}, we start from its l.h.s. and use that $\hat{O}^\dagger=\overrightarrow{\cW}\cD^{N}$, then we transfer the shifts of the argument away from $f$ by moving the integration contour,
\beqa
\label{saprsln}
    \bl Q_1\hat {O}^\dagger f\br_\alpha&=&\frac{1}{2\pi i}\int du\;\mu_\alpha Q_1 \overrightarrow{\cW} f^{[+N]}
    \\ \nn
    &=&
    \frac{1}{2\pi i}\int du\;\(\mu_\alpha Q_1  \overleftarrow{\cW}\)f^{[+N]}+\text{[poles contributions]} \;.
\eeqa
We see that as a result we get the operator $\overleftarrow{\cW}$ acting to the left on the combination $\mu_\alpha Q_1$, and this gives zero due to \eq{WQdn}, thus proving \eq{sasln}. In \eq{saprsln} we have also indicated that when we move the integration contour there could be extra contributions from poles of $\mu_\alpha$ (located at $u=\theta_\beta-i\bs-in$ and $u=\theta_\alpha+i\bs+in$ with $\beta=1,\dots,L$ and $n=0,1,\dots$). However, we see that when $\bs$ is positive and large enough, there are no poles in the region where the contour is being moved. Since all the terms under the integral are analytic as functions of $\bs$, so should be the whole expression and thus the poles do not give any extra contribution\footnote{One can also verify directly that the poles contributions cancel by adapting the derivation from \cite{Cavaglia:2019pow}.} .

Let us also comment on convergence of the integrals.  Since we take $f$ to have the same asymptotics as any of the $Q^{a+1}$, the integrals in \eq{saprsln} will be finite if
\beq
    0<\arg \lambda_c-\arg\lambda_1<2\pi \ , \ \ \ c=2,\dots,N\; .
\eeq
Here for definiteness we assume that these inequalities hold. If that is not the case, one should modify the integration measure in the same way as for the $\sl(2)$ and $\sl(3)$ spin chains (see the discussion after \eq{convsl3}).

Having the property \eq{sasln} we can derive orthogonality relations for different states exactly like for $\sl(2)$ and $\sl(3)$ cases. 
For $\sl(n)$ the transfer matrix eigenvalues $\tau_a$ have the form
\beq
    \tau_a=\chi_a(\lambda_1,\dots,\lambda_N)u^L+\sum_{\alpha=1}^LI_{a,\alpha}u^{\alpha-1}
\eeq
where the leading term is the character of the twist matrix in the $a$-th antisymmetric representation, and the $I_{a,\alpha}$ are eigenvalues of the integrals of motion. Using \eq{sasln} we have for two different states $A$ and $B$
\beq
    \bl Q_1^A (\hat{ O}^{\dagger A}-\hat{ O}^{\dagger B})Q^{B,a+1}\br_\alpha=0
\eeq
where the difference of the operators comes from the transfer matrices,
\beq\la{AB3gen}
(\hat{ O}^{\dagger A}-\hat{ O}^{\dagger B})\circ Q^{B,a+1}=
\sum_{\beta=1}^{L}\sum_{b=1}^{N-1}
(-1)^b(I_{b,\beta-1}^A-I_{b,\beta-1}^B)u^{\beta-1}\;{\cal D}^{[-2b+N]}\circ Q^{B,a+1} \;  .
\eeq
Requiring that this linear system has a nontrivial solution leads to
\beq\la{mdef}
 \det_{(a,\alpha),(b,\beta)}m_{(a,\alpha),(b,\beta)}=0\;\;,\;\;
m_{(a,\alpha),(b,\beta)}\equiv \bl Q_1^A \;u^{\beta-1}\; {\cal D}^{-2b+N}\circ Q^{B,a+1} \br_\alpha \;.
\;
\eeq
This is the orthogonality condition that we presented in the main text in \eq{mdef0}.

\section{Scalar product for compact $\su(N)$ spin chains and analytic continuation in the spin}

\label{app:ancont}

In this section we discuss how our results for the scalar product can be analytically continued from $\bs>0$ to negative values of $\bs$, as well as the reduction to the compact spin chain case.

\subsection{Analytic continuation to $\bs<0$}

Let us recall that when deriving the integral form of the scalar product (e.g. \eq{mdef0} for $\sl(N)$) we assumed that $\bs>0$. All these scalar products are written in terms of the bracket \eq{scalargf} which is an integral along the real line with the measure \eq{muj} which has poles at $u=\theta_\beta-i\bs+in$ (for all $\beta=1,\dots,L$) and $u=\theta_\alpha+i\bs+in$ with integer $n\geq 0$. One potential  possibility to define the analytic continuation of this integral to $\bs<0$ would be to keep the contour always slightly below the pole at $u=\theta_\alpha+i\bs$, but still we see that when we go from $\bs>0$ to $\bs<0$ the poles at $u=\theta_\beta-i\bs$ cross the integration contour and thus one should also subtract their contribution. As we further decrease the value of $\bs$, more and more poles will cross the contour, making the result somewhat cumbersome. A simpler approach is to first rewrite the integral (for $\bs>0$) as a sum of poles in the upper half plane at $u=\theta_\alpha+i\bs+in$ with $n=0,1,\dots$ as we discussed in sections \ref{sec:compsl2} and \ref{sec:expmeasure} by closing the contour in the upper half plane. The sum over these poles itself is analytic in $\bs$ and thus can be directly used for $\bs<0$.

Thus, for $\bs<0$ we understand the bracket in the scalar product to be the sum over the poles at $u=\theta_\alpha+i\bs+in$. Being analytic in $\bs$, it retains all its key features and ensures that the scalar product \eq{mdef0} vanishes for different states $A,B$.

\subsection{Reduction to the compact $\su(N)$ case}

\label{app:comp}

Having clarified the construction for $\bs<0$, we can now explore the particularly interesting case when $\bs$ takes negative half-integer values, $\bs=-1/2,-1,-3/2,\dots$. In this case the representation of $\sl(N)$ on the spin chain sites becomes reducible and acquires a finite-dimensional irreducible subspace, corresponding to the $(-2\bs)$'th symmetric power of the fundamental irrep of $\sl(N)$. As discussed in the end of section \ref{sec:secslN}, our construction of the SoV basis then provides the basis precisely for this subspace, so that now our model reduces to a finite-dimensional compact $\su(N)$ spin chain.

We expect that accordingly the scalar products (defined in terms of the sum over poles as we just discussed) should reduce from an infinite to a finite sum, over the values that label the SoV basis. Nicely, for the $\sl(2)$ case we can see at once that this sum truncates for $\bs$ a negative half-integer, 
as all but the first several elements of the SoV measure vanish since \eq{mucompact} gives zero for $n_\beta\geq -2\bs$. The same is true for the $\sl(N)$ case, as one can see from the explicit result for the SoV measure \eq{Msl3resNfinal} since the factor $r_{\alpha,n}$ defined in \eq{resm} vanishes for $n\geq -2\bs$. As a result, we see that our SoV measure works perfectly for the finite-dimensional case as well.

For various applications it is interesting to still write the scalar products in an integral form for the finite-dimensional case as well. This was done for $\su(3)$ originally in \cite{Gromov:2019wmz} for the case $\bs=-1/2$ (i.e. fundamental representation). We can now almost immediately extend this result to any $\su(N)$ and any $\bs=-1/2,-1,\dots$.

Let us note that for these values of $\bs$ the measure \eq{muj} we used so far in the integral form of the scalar product simplifies and becomes
\beq
\label{mure}
    \mu_\alpha(u)=\frac{{\rm const}}{1+e^{2\pi(u-\theta_\alpha)}}\times \frac{1}{\prod\limits_{k=-\bs}^{\bs}Q_\theta(u+ik)}
\eeq
where the product goes over $k=-\bs,-\bs+1,\dots,\bs$. Let us redefine it by multiplying with an $i$-periodic function that removes the inifnite set of poles coming from the first factor, and also removes the poles of the $Q_\theta$ factors associated to all $\theta_\beta$ with $\beta\neq \alpha$, so that we define
\beq
\label{muc}
    \mu_\alpha= \frac{\prod_{\beta\neq \alpha}(1-e^{2\pi(u-\theta_\beta-i\bs)})}{\prod\limits_{k=-\bs}^{\bs}Q_\theta(u+ik)}\;.
\eeq
We see that now we only have poles at $u=\theta_\alpha+ik$ with $k=\bs,-\bs+1,\dots,\bs$. We should change the definition of the bracket accordingly to pick up these poles, so we define
\beq
\label{brc}
    \bl f\br_\alpha=\frac{1}{2\pi i}\oint du\;\mu_\alpha(u)f(u)
\eeq
where the integral goes around a large circle enclosing all the poles of the measures $\mu_\alpha$ (i.e. all the points $u=\theta_\beta+ik$ for $k=-\bs,-\bs+1,\dots,\bs$ and  $\beta=1,\dots,L$). Since the measure \eq{muc} differs from the one we used previous just by an $i$-periodic factor, the same argument as in section \ref{app:slNbax} leads to the key `self-adjoint' property\footnote{Let us note that now we do not have to worry about any extra poles contributions (i.e. the terms in square brackets in the second line of appendx C11 \eq{saprsln}) when proving this property, since the integration contour is now far away from all the poles.
} \eq{sasln}, 
and consequently to the orthogonality relation in the form \eq{mdef0}, the only difference being that the bracket is now given by \eq{brc}. Of course, this integral can be evaluated as the sum over poles at $u=\theta_\alpha+ik$ (with $k=\bs,-\bs+1,\dots,\bs$) as expected. This provides an integral representation for the scalar products for the compact $\su(N)$ spin chains in the finite-dimensional\footnote{For instance, in our notation for $\sl(2)$ the fundamental representation corresponds to $\bs=-1/2$, the 3-dimensional vector representation corresponds to $\bs=-1$, etc.} representation with spin $\bs$.

\section{Direct proof of poles cancellation for $\sl(3)$}
\la{app:nopoles}

Here we will verify explicitly  that there is no additional contribution from the poles in \eq{weakd3} when we shift the contour to transform the integral in the first line to the one in the second line. For convenience we repeat this equation here,
    \beqa
\label{weakd32}
    \bl Q_1\hat {O}^\dagger  f\br_\alpha&=&
\int_{-\infty}^{+\infty} \mu_\alpha(u) Q_1(u)\( Q_\theta^{[-2\bs]}f^{[-3]} - \tau_2f^- + \tau_1f^+ -Q_\theta^{[+2\bs]}f^{[+3]} \)\frac{du}{2\pi i} \\ \nn
    &=&\int_{-\infty}^{+\infty} 
    \[
{\mu_\alpha^{[+3]}}
{Q_\theta^{[-2\bs+3]}}Q_{1}^{[+3]} - 
\mu_\alpha^+
\tau_2^+Q_{1}^+ + 
\mu_\alpha^-
{\tau_1^-}Q_{1}^- -
\mu_\alpha^{[-3]}
{Q_\theta^{[+2\bs-3]}}Q_{1}^{[-3]} \]f\;\frac{du}{2\pi i}
    \\ \nn
    &+& {\text{residues from poles}}\;.
\eeqa
We assume for simplicity that $\bs>0$ and the inhomogeneities $\theta_\beta$ are real.

The poles of the integrand are located at
$    u=\pm(i\bs+in)+\theta_\beta$ with $\beta\neq \alpha$ and at $u=i\bs+in+\theta_\alpha$, with $n\geq 0$. Let us first consider the  case $0<\bs<1/2$. We start with the first type of poles, i.e. those associated with $\theta_\beta$ where $\beta\neq \alpha$. The first term in \eq{weakd32} (the term with $f^{[-3]}$) has no pole at potentially dangerous points $u=i\bs+\theta_\beta$ and  $u=i\bs+i+\theta_\beta$ which we are crossing, and thus gives no contribution. The second term in \eq{weakd32} (the term with $f^-$) also has no pole at $u=i\bs+\theta_\beta$. The third term (the term with $f^+$) gives a contribution from the pole at $u=\theta_\beta-i\bs$ that we denote as $r_1$ with
\beqa
    r_1&=&f(\theta_\beta-i\bs+i/2)Q_1(\theta_\beta-i\bs)\tau_1(\theta_\beta-i\bs)\frac{ie^{2\pi \theta_\alpha+4i\pi \bs}}{(e^{2\pi \theta_\alpha+4i\pi \bs}-e^{2\pi \theta_\beta})\Gamma(1-2\bs)}
    \\ \nn &\times&
    \prod_{\gamma\neq \beta}\frac{\Gamma(i(\theta_\gamma-\theta_\beta))}{\Gamma(-2\bs+i(\theta_\gamma-\theta_\beta)+1)}\;.
\eeqa
The fourth term (the term with $f^{[+3]}$) has no pole at $u=\theta_\beta-i\bs$ but gives a contribution from the pole at $u=\theta_\beta-i\bs-i$ that we denote as $ r_2$ with
\beqa
    r_2&=&f(\theta_\beta-i\bs+i/2)Q_1(\theta_\beta-i\bs-i)Q_\theta(\theta_\beta-i)\frac{ie^{2\pi \theta_\alpha+4i\pi s}}{(e^{2\pi \theta_\alpha+4i\pi \bs}-e^{2\pi \theta_\beta})\Gamma(-2\bs)}
    \\ \nn
    &\times&
    \prod_{\gamma\neq \beta}\frac{\Gamma(i(\theta_\gamma-\theta_\beta)-1)}{\Gamma(-2\bs+i(\theta_\gamma-\theta_\beta))}\;.
\eeqa
Notice that as a consequence of \eq{tau1Q}, \eq{tau2Q} we have
\beq
\label{t1app}
    \tau_1(-i\bs+\theta_\beta)=\frac{Q_1(-i\bs-i+\theta_\beta)}{Q_1(-i\bs+\theta_\beta)}Q_\theta(-2i\bs+\theta_\beta)
\eeq
and
\beq
\label{t2app}
    \tau_2(i\bs+\theta_\beta)=\frac{Q_1(i\bs+i+\theta_\beta)}{Q_1(i\bs+\theta_\beta)}Q_\theta(2i\bs+\theta_\beta)\;.
\eeq
Nicely, using these relations we find that $r_1+r_2=0$ ! Thus, the poles we considered so far give no contribution.

Next we consider the poles associated with $\theta_\alpha$. The first term in \eq{weakd32} has no pole at $u=i\bs+\theta_\alpha$ but has a pole at $u=i\bs+i+\theta_\alpha$ which contributes $ r_3$ with
\beqa
    r_3&=&
    -\frac{1}{2\pi}f(\theta_\alpha+i\bs-i/2)Q_1(\theta_\alpha+i\bs+i)Q_\theta(\theta_\alpha+i)
    \Gamma(2\bs+1)
    \\ \nn
    &\times&
    \prod_{\gamma\neq\alpha}\frac{\Gamma(2\bs-i\theta_\alpha+i\theta_\gamma+1)}{\Gamma(-i\theta_\alpha+i\theta_\gamma+2)}\;.
\eeqa
The second term in \eq{weakd32} gives a contribution $ r_4$ from the pole at $u=\theta_\alpha+i\bs$ with
\beqa
    r_4&=&
    \frac{1}{2\pi}
    f(\theta_\alpha+i\bs-i/2)Q_1(\theta_\alpha+i\bs)\tau_2(\theta_1+i\bs)
    \Gamma(2\bs)
    \\ \nn
    &\times&
    \prod_{\gamma\neq\alpha}\frac{\Gamma(2\bs-i\theta_\alpha+i\theta_\gamma)}{\Gamma(-i\theta_\alpha+i\theta_\gamma+1)}\;.
\eeqa
Using \eq{t2app} we find that $r_3+r_4=0$ so these contributions cancel against each other. Finally, we can check in a similar way that the contributions from the 3rd and 4th terms in \eq{weakd32} also cancel each other.

Thus we see that when $0<\bs<1/2$ the poles give no contribution. The case when $\bs=1/2$ was considered in \cite{Cavaglia:2019pow} and the poles similarly cancel. Finally, let us consider the case $\bs>1/2$. Then the 2nd and 3rd terms in \eq{weakd32} do not have poles in the relevant region at all. For the 1st and 4th terms the poles trivially vanish due to their accompanying $Q_\theta$ factors. Thus there are no contributions from any poles. Let us also mention that when $\bs>3/2$ we see at once that all potential poles are absent from the relevant strip $-3/2\leq{\rm Im}\; u \leq3/2$ where we are moving the contour.

As a result, we have shown that all the contributions from poles cancel nontrivially once we invoke relations between various transfer matrices and Q-functions.

\section{Oscillator representation for $\sl(N)$, and relations for generators}\label{oscrep}
In the main text we noticed that the transfer matrices in anti-symmetric representations have trivial non-dynamical factors \eq{Ttau3} and \eq{Wtau}. This can be traced back to some simple relations satisfied by the generators of $\sl(N)$ in our specific representation. 

Firstly, there are linear and quadratic Casimirs, which are easy to determine by acting on the HW state
\beq
{\mathbb E}_{aa}=(N-2) \bs\;\;,\;\;
{\mathbb E}_{ab}{\mathbb E}_{ba}=N \bs^2-2 N \bs+2 \bs\;,
\eeq
where the repeated indices $a$ and $b$ are summed over. 
Another relation, which we found to be very useful, is
\beqa\la{EErel}
\mathbb{E}_{a,c}\mathbb{E}_{b,d}-\mathbb{E}_{b,c}\mathbb{E}_{a,d}&=&\bs\;\(
   \mathbb{E}_{a,c} \delta _{b,d}-
    \mathbb{E}_{b,c}\delta _{a,d}\)
    +(\bs-1) \(
   \mathbb{E}_{b,d}\delta _{a,c}
    - \mathbb{E}_{a,d} \delta _{b,c}\)\\
\nn   &+&(\bs-1) \bs \(\delta _{a,d} \delta _{b,c}- \delta
   _{a,c} \delta _{b,d}\)\;.
\eeqa
This is easy to verify for $\sl(3)$, using the explicit form of the generators \eqref{sl3raise},\eqref{sl3lower},\eqref{sl3cartan}. For general $\sl(N)$ we used the oscillator representation of \cite{Okuboosc, Palev:1996kh}.

For completeness we write it below in our conventions
\beqa
{\mathbb E}_{i,j}&=&b^+_{i-1}b^-_{j-1}+\bs\; \delta_{i,j}\;\;,\;\;i,j\neq 1\\
{\mathbb E}_{i,1}&=&b^+_{i-1}\sqrt h\;\;,\;\;i\neq 1\\
{\mathbb E}_{1,i}&=&\sqrt h\; b^-_{i-1}\;\;,\;\;i\neq 1\\
{\mathbb E}_{1,1}&=&h+\bs 
\eeqa
where $h\equiv -2\bs-\sum_{i=1}^{N-1}b^+_{i}b^-_{i}$ and $\[b^-_i,b^+_j\]=\delta_{ij}$.

One of the consequences of the relation \eq{EErel}
is that the Lax operators $\mathcal{L}^{a,1}$ \eqref{La1} is linear in generators and produces trivial scalar factors. For example for $\mathcal{L}^{2,1}$ we find
\beqa
&&
{\cal L}^{d_1}_{\;\;\;\;[e_1}\(
u+\tfrac{i}{2}\)
{\cal L}^{d_2}_{\;\;\;\;e_2]}\(
u-\tfrac{i}{2}\)=(u+i\bs-\tfrac{i}{2})\((u-i\bs+\tfrac{i}{2}) \delta^{d_1}_{[e_1}
\delta^{d_2}_{e_2]}+
2i\, {\mathbb E}_{[e_1}^{\;\;\;\;[d_1}\delta_{e_2]}^{d_2]}\)
\eeqa
where we raised some indices to indicate antisymmetrisation more easily. The general expression, which can be deduced as a consequence of \eq{EErel}, takes the form
\beqa
&&
{\cal L}^{d_1}_{\;\;\;\;[e_1}\(
u+i\frac{a-1}{2}\)
\dots
{\cal L}^{d_a}_{\;\;\;\;e_a]}\(
u-i\frac{a-1}{2}\)=\\
\nn&&
\(\(u-\frac{1}{2} i (a-1) (2 \bs-1)\) \delta^{d_1}_{[e_1}
\dots\delta^{d_a}_{e_a]}+
ai\, {\mathbb E}_{[e_1}^{\;\;\;\;[d_1}\delta_{e_2}^{d_2}\dots \delta_{e_a]}^{d_a]}\)
\prod_{k=1}^{a-1}\(u+i\bs+i\frac{a-1}{2}-i k\)\;.
\eeqa
We see that the the last factor agrees with \eq{Wtau}.
Removing the scalar factor and performing the shift in accordance with 
\eq{Wtau} we see that $\tau_a(u)$ is built out of the following ``reduced"
$\cal L$-operators
\beq\la{Lred}
{\cal L}^{\rm reduced}_{\bar d,\bar e}(u)=
\(u- i (a-1)  \bs\) \delta^{d_1}_{[e_1}
\dots\delta^{d_a}_{e_a]}+
ai\, {\mathbb E}_{[e_1}^{\;\;\;\;[d_1}\delta_{e_2}^{d_2}\dots \delta_{e_a]}^{d_a]}\;.
\eeq

In section~\ref{sec:detloc} we are interested in the following combinations $\tau_a(u)+u \d_{\theta_\alpha}\tau_a(u)|_{u^{L-1}}=
I_{a,L-1}+\d_{\theta_\alpha} I_{a,L-2}$, which give a combination of generators acting on one site of the chain. From \eq{Lred}
we immediately obtain
\beqa
\la{Fex}&& \hat I_{a,L-1}+\d_{\theta_\alpha} \hat I_{a,L-2}=\\
&&\nn\[\(-\theta_\alpha- i (a-1)  \bs\) \delta^{d_1}_{[e_1}
\dots\delta^{d_a}_{e_a]}+
ai\, {\mathbb E}_{[e_1}^{\;\;\;\;[d_1}\delta_{e_2}^{d_2}\dots \delta_{e_a]}^{d_a]}\]{\Lambda}^{e_1}_{\;\;\;\;[d_1}
\dots
{\Lambda}^{e_a}_{\;\;\;\;d_a]}\\
\nn &&=
\(-\theta_\alpha- i (a-1)  \bs\)\chi_a
+i\sum_{j=1}^a
(-1)^{j-1}
{\rm tr}({\mathbb E}^t\Lambda^j)\chi_{a-j}
\;
\eeqa
where we use the superscript $t$ in $\mE^t$ to denote transposition in the auxiliary space and we used the identities
\beqa
ai\, {\mathbb E}_{[e_1}^{\;\;\;\;[d_1}\delta_{e_2}^{d_2}\dots \delta_{e_a]}^{d_a]}\Lambda^{e_1}_{d_1}\dots
\Lambda^{e_a}_{d_a}
&=&\sum_{j=1}^a
(-1)^{j-1}
{\rm tr}({\mathbb E}^t\Lambda^j)\chi_{a-j}\\
{\delta}_{[e_1}^{[d_1}\delta_{e_2}^{d_2}\dots \delta_{e_a]}^{d_a]}\Lambda^{e_1}_{d_1}\dots
\Lambda^{e_a}_{d_a}
&=&\chi_{a}\;.
\eeqa
Finally, we simplify \eq{Fex} to
\beq
\label{appcorr}
\sum_{j=1}^a
{\rm tr}\(({\mathbb E}^t-\bs)(-\Lambda)^j\)\;\chi_{a-j}
=
\(i\theta_\alpha+ \bs\)I_{a,L}+i I_{a,L-1}+i\d_{\theta_\alpha} I_{a,L-2}\;, 
\eeq
which is what we use in section~\ref{sec:detloc}.

\section{$\it Mathematica$ implementation of general measure elements}
\la{app:code}
In this appendix we give a simple implementation of our general formula \eq{Msl3resNfinal}. The code is purely for demonstration purposes and is not particularly optimised for long chains. 

First, we introduce some notations in {\it Mathematica}
\begin{mmaCell}[moredefined={sig0, xs, xsinv, Delta, r, t, SGN, MyPerm},morefunctionlocal={a, al, z, i, j, be},morepattern={z_, sig_, sig, lst_, lst, al_, n_, n, \#}]{Input}
  (*Trivial permutation*)
  sig0:=Table[Table[a,\{a,Nc-1\}],\{al,L\}]//Flatten;
  xs:=Flatten[Table[x[al,a],\{al,L\},\{a,Nc-1\}]];
  (*Subset of x's with given value of sigma*)
  xsinv[z_,sig_]:=xs[[(Position[sig,\mmaPat{z}]//Flatten)]];
  (*Vandermond determinant*)
  Delta[lst_]:=Product[lst[[i]]-lst[[j]],\{i,Length[lst]\}
                                        ,\{j,i+1,Length[lst]\}];                      
  (*Vandermond determinant*)                                        
  r[al_,n_]:=Product[Pochhammer[n+1-I (t[al]-t[be]),2s-1],\{be,L\}];
  (*Signature of a permutation*)
  SGN[sig_]:=Signature[Flatten[Table[xsinv[z,sig],\{z,Nc-1\}]]];
  (*Creates all permutations with fixed alpha's*)
  MyPerm:=Flatten[Table[Flatten@Table[p[al],\{al,L\}],
    Evaluate[Sequence@@Table[\{p[al],
    Permutations[Take[#,\{(Nc-1)(al-1)+1,(Nc-1)al\}]]\},\{al,L\}]]],L-1]&
\end{mmaCell}
With these helper definitions the implementation of the measure
$M_{\svy,\svx}$ is very simple and reads
\begin{mmaCell}[addtoindex=15,moredefined={M, sig0, SGN, Delta, xsinv, t, r, MyPerm},morepattern={ms_, ns_, ms, ns, al_, a_},morefunctionlocal={ks, al, a, z}]{Input}
  M[ms_,ns_]:=
  If[
  (*check SoV charges are the same*)
  Total[ms]==Total[ns],
    Sum[
      sig=ks-ms+sig0;
      If[
      (*check if sig is a valid permutation*)
      Union[Tally[sig]]==Union[Tally[sig0]],
        Do[k[al,a]=ks[[(al-1)(Nc-1)+a]],\{al,L\},\{a,Nc-1\}];
        (*the main formula*)
        SGN[sig]/SGN[sig0] *
          Product[Delta[xsinv[z,sig]]/Delta[t/@Range[L]],\{z,Nc-1\}] *
          Product[r[al,k[al,a]]/r[al,0],\{al,L\},\{a,Nc-1\}]
          /.x[al_,a_]->t[al]+I s+I k[al,a],0]
    ,\{ks,MyPerm[ns]\}]
  ,0]
\end{mmaCell}
The usage is the following, one has to initialise global variables $\verb"Nc"=N$ and $L$. Then one can call the function $M(m's,n's)$
to obtain the corresponding element of the measure $M_{\svx,\svy}$.
For example,
for $\sl(4)$ length $3$ spin chain
the measure element $M_{\svy,\svx}$ with $\svy=|1,1,0;3,1,1;1,1,0\rangle$
and $\svx=\langle 2,0,0;4,1,0;2,0,0|$ can be computed as follows
\begin{mmaCell}[addtoindex=26,moredefined={M, t}]{Input}
  Nc=4;
  L=3;
  M[\{1,1,0,3,1,1,1,1,0\}
   ,\{2,0,0,4,1,0,2,0,0\}]/.t[a_]->\mmaSub{t}{a}//FunctionExpand//Factor
\end{mmaCell}
\begin{mmaCell}[addtoindex=0]{Output}
  -((\mmaSup{s}{4} (s+1) \mmaSup{(2 s+1)}{3} (2 s+3) \mmaSup{(2 s+i \mmaSub{t}{1}-i \mmaSub{t}{2})}{2} (2 s+i \mmaSub{t}{1}-i \mmaSub{t}{2}+1)
  (2 s+i \mmaSub{t}{1}-i \mmaSub{t}{2}+2) (2 s+i \mmaSub{t}{1}-i \mmaSub{t}{2}+3) (2 s-i \mmaSub{t}{1}+i \mmaSub{t}{2}) (2 s-i \mmaSub{t}{1}+i \mmaSub{t}{2}+1)
  (2 s+i \mmaSub{t}{1}-i \mmaSub{t}{3}) (2 s+i \mmaSub{t}{1}-i \mmaSub{t}{3}+1) (2 s+i \mmaSub{t}{2}-i \mmaSub{t}{3}) (2 s+i \mmaSub{t}{2}-i \mmaSub{t}{3}+1) 
  (2 s-i \mmaSub{t}{1}+i \mmaSub{t}{3}) (2 s-i \mmaSub{t}{1}+i \mmaSub{t}{3}+1) \mmaSup{(2 s-i \mmaSub{t}{2}+i \mmaSub{t}{3})}{2} (2 s-i \mmaSub{t}{2}+i \mmaSub{t}{3}+1) 
  (2 s-i \mmaSub{t}{2}+i \mmaSub{t}{3}+2) (2 s-i \mmaSub{t}{2}+i \mmaSub{t}{3}+3))
  /(3 \mmaSup{(\mmaSub{t}{1}-\mmaSub{t}{2})}{2} (\mmaSub{t}{1}-\mmaSub{t}{2}-i) (\mmaSub{t}{1}-\mmaSub{t}{2}+i) (\mmaSub{t}{1}-\mmaSub{t}{2}+2 i) (\mmaSub{t}{1}-\mmaSub{t}{2}-3 i)
  (\mmaSub{t}{1}-\mmaSub{t}{2}-4 i) (\mmaSub{t}{1}-\mmaSub{t}{3}-i) (\mmaSub{t}{1}-\mmaSub{t}{3}+i) (\mmaSub{t}{1}-\mmaSub{t}{3}-2 i) (\mmaSub{t}{1}-\mmaSub{t}{3}+2 i)
  \mmaSup{(\mmaSub{t}{2}-\mmaSub{t}{3})}{2} (\mmaSub{t}{2}-\mmaSub{t}{3}-i) (\mmaSub{t}{2}-\mmaSub{t}{3}+i) (\mmaSub{t}{2}-\mmaSub{t}{3}-2 i) (\mmaSub{t}{2}-\mmaSub{t}{3}+3 i) (\mmaSub{t}{2}-\mmaSub{t}{3}+4 i)))
\end{mmaCell}
or one can reproduce $1,3$ element of the matrix in \eq{measuredata}
with this code
\begin{mmaCell}[addtoindex=5,moredefined={M, t}]{Input}
  Nc=3;
  L=2;
  M[\{0,0,1,1\},\{0,0,2,0\}]/.t[2]->0/.t[1]->\mmaSub{t}{1,2}//FullSimplify//Factor
\end{mmaCell}
\begin{mmaCell}{Output}
  \mmaFrac{s (2 s+1) (2 s+i \mmaSub{t}{1,2}) (i \mmaSub{t}{1,2}+2 s+1)}{\mmaSub{t}{1,2} (\mmaSub{t}{1,2}-i)}
\end{mmaCell}


\begin{thebibliography}{99}
  
  
  







\bibitem{LIJC}
 London Integrability Journal Club (LIJC) online seminar series, 
 www.integrability-london.weebly.com 






\bibitem{Beisert:2010jr}
  N.~Beisert {\it et al.},
  ``Review of AdS/CFT Integrability: An Overview,''
  Lett.\ Math.\ Phys.\  {\bf 99} (2012) 3
  [arXiv:1012.3982 [hep-th]].



\bibitem{Sklyanin:1984sb} 
  E.~K.~Sklyanin,
  ``The Quantum Toda Chain,''
  Lect.\ Notes Phys.\  {\bf 226}, 196 (1985).
  %%CITATION = LNPHA,226,196;


\bibitem{Sklyanin:1987ih}
E.~K.~Sklyanin,
``Separation of variables in the Gaudin model,''
J. Sov. Math. \textbf{47} (1989), 2473-2488
doi:10.1007/BF01840429

\bibitem{Sklyanin:1991ss}
  E.~K.~Sklyanin,
  ``Quantum inverse scattering method. Selected topics,''
  In: Quantum Group and Quantum Integrable Systems: Nankai Lectures
  on Mathematical Physics : Nankai Institute of Mathematics, China 2-18 April
  1991 (World Scientific 1992), pp 63-97
  [hep-th/9211111].




\bibitem{SklyaninFBA}
  E.~K.~Sklyanin,
  ``Functional Bethe Ansatz,'', in \textit{Integrable and Superintegrable
Systems}, ed. by B. A. Kupershmidt, pp. 8-33, Singapore: World Scientific.


\bibitem{Sklyanin:1995bm}
  E.~K.~Sklyanin,
  ``Separation of variables - new trends,''
  Prog.\ Theor.\ Phys.\ Suppl.\  {\bf 118} (1995) 35
  doi:10.1143/PTPS.118.35
  [solv-int/9504001].



\bibitem{Sklyanin:1992sm}
  E.~K.~Sklyanin,
  ``Separation of variables in the quantum integrable models related to the Yangian Y[sl(3)],''
  J.\ Math.\ Sci.\  {\bf 80} (1996) 1861
   [Zap.\ Nauchn.\ Semin.\  {\bf 205} (1993) 166]
  doi:10.1007/BF02362784
  [hep-th/9212076].





\bibitem{Smirnov2001}
F.~Smirnov, ``Separation of variables for quantum integrable
models related to $U_q(\widehat{sl}_N)$'', math-ph/0109013





\bibitem{Gromov:2016itr} 
  N.~Gromov, F.~Levkovich-Maslyuk and G.~Sizov,
  ``New Construction of Eigenstates and Separation of Variables for SU(N) Quantum Spin Chains,''
  JHEP {\bf 1709}, 111 (2017)
  doi:10.1007/JHEP09(2017)111
  [arXiv:1610.08032 [hep-th]].
  

\bibitem{Liashyk:2018qfc}
A.~Liashyk and N.~A.~Slavnov,
``On Bethe vectors in $\mathfrak{gl}_3$-invariant integrable models,''
JHEP \textbf{06} (2018), 018
doi:10.1007/JHEP06(2018)018
[arXiv:1803.07628 [math-ph]].






\bibitem{Maillet:2018bim} 
  J.~M.~Maillet and G.~Niccoli,
  ``On quantum separation of variables,''
  J.\ Math.\ Phys.\  {\bf 59}, no. 9, 091417 (2018)
  doi:10.1063/1.5050989
  [arXiv:1807.11572 [math-ph]].
 
  

\bibitem{Ryan:2018fyo}
P.~Ryan and D.~Volin,
``Separated variables and wave functions for rational gl(N) spin chains in the companion twist frame,''
J. Math. Phys. \textbf{60} (2019) no.3, 032701
doi:10.1063/1.5085387
[arXiv:1810.10996 [math-ph]].








  \bibitem{Maillet:2018czd}
  J.~M.~Maillet and G.~Niccoli,
  ``Complete spectrum of quantum integrable lattice models associated to Y(gl(n)) by separation of variables,''
  SciPost Phys.\  {\bf 6} (2019) 071
  doi:10.21468/SciPostPhys.6.6.071
  [arXiv:1810.11885 [math-ph]].
  

  \bibitem{Maillet:2018rto}
  J.~M.~Maillet and G.~Niccoli,
  ``Complete spectrum of quantum integrable lattice models associated to $\mathcal{U}_{q} (\widehat{gl_{n}})$ by separation of variables,''
  arXiv:1811.08405 [math-ph].
  
  
  

\bibitem{Maillet:2019nsy}
J.~M.~Maillet and G.~Niccoli,
``On quantum separation of variables beyond fundamental representations,''
[arXiv:1903.06618 [math-ph]].



\bibitem{Maillet:2019hdq}
J.~M.~Maillet and G.~Niccoli,
``On Separation of Variables for Reflection Algebras,''
J. Stat. Mech. \textbf{1909} (2019) no.9, 094020
doi:10.1088/1742-5468/ab357a
[arXiv:1904.00852 [math-ph]].



\bibitem{Maillet:2019ayx}
J.~M.~Maillet, G.~Niccoli and L.~Vignoli,
``Separation of variables bases for integrable $gl_{\mathcal{M}|\mathcal{N}}$ and Hubbard models,''
[arXiv:1907.08124 [math-ph]].



\bibitem{Ryan:2020rfk}
P.~Ryan and D.~Volin,
``Separation of variables for rational gl(n) spin chains in any compact representation, via fusion, embedding morphism and Backlund flow,''
[arXiv:2002.12341 [math-ph]].










\bibitem{Beisert:2005fw}
N.~Beisert and M.~Staudacher,
``Long-range psu(2,2|4) Bethe Ansatze for gauge theory and strings,''
Nucl. Phys. B \textbf{727} (2005), 1-62
doi:10.1016/j.nuclphysb.2005.06.038
[arXiv:hep-th/0504190 [hep-th]].




\bibitem{Cavaglia:2019pow}
A.~Cavaglia, N.~Gromov and F.~Levkovich-Maslyuk,
``Separation of variables and scalar products at any rank,''
JHEP \textbf{09} (2019), 052
doi:10.1007/JHEP09(2019)052
[arXiv:1907.03788 [hep-th]].



  \bibitem{Gromov:2019wmz}
N.~Gromov, F.~Levkovich-Maslyuk, P.~Ryan and D.~Volin,
``Dual Separated Variables and Scalar Products,''
Phys. Lett. B \textbf{806} (2020), 135494
doi:10.1016/j.physletb.2020.135494
[arXiv:1910.13442 [hep-th]].



\bibitem{SmirnovClassM}
F.~Smirnov, V.~Zeitlin,
``Affine Jacobians of spectral curves and integrable models'', arXiv:math-ph/0203037



\bibitem{SmirnovQuantM}
F.~Smirnov, V.~Zeitlin,
``On The Quantization of Affine Jacobi Varieties of Spectral Curves''. Statistical Field Theories. – Springer, Dordrecht, 2002, p. 79-89.



\bibitem{Maillet:2020ykb}
J.~M.~Maillet, G.~Niccoli and L.~Vignoli,
``On Scalar Products in Higher Rank Quantum Separation of Variables,''
[arXiv:2003.04281 [math-ph]].


\bibitem{Gromov:2013pga}
N.~Gromov, V.~Kazakov, S.~Leurent and D.~Volin,
``Quantum Spectral Curve for Planar $\mathcal{N} = 4$ Super-Yang-Mills Theory,''
Phys. Rev. Lett. \textbf{112} (2014) no.1, 011602
doi:10.1103/PhysRevLett.112.011602
[arXiv:1305.1939 [hep-th]].






%%QSC for correlators
  


\bibitem{Basso:2015zoa}
B.~Basso, S.~Komatsu and P.~Vieira,
``Structure Constants and Integrable Bootstrap in Planar N=4 SYM Theory,''
[arXiv:1505.06745 [hep-th]].
%156 citations counted in INSPIRE as of 06 Oct 2020


  \bibitem{Cavaglia:2018lxi} 
  A.~Cavagli\`a, N.~Gromov and F.~Levkovich-Maslyuk,
  ``Quantum spectral curve and structure constants in $ \mathcal{N}=4 $ SYM: cusps in the ladder limit,''
  JHEP {\bf 1810}, 060 (2018)
  doi:10.1007/JHEP10(2018)060
  [arXiv:1802.04237 [hep-th]].
  
  
  
  

  \bibitem{Giombi:2018qox} 
  S.~Giombi and S.~Komatsu,
  ``Exact Correlators on the Wilson Loop in $\mathcal{N}=4$ SYM: Localization, Defect CFT, and Integrability,''
  JHEP {\bf 1805}, 109 (2018)
  [arXiv:1802.05201 [hep-th]].



  \bibitem{Giombi:2018hsx} 
  S.~Giombi and S.~Komatsu,
  ``More Exact Results in the Wilson Loop Defect CFT: Bulk-Defect OPE, Nonplanar Corrections and Quantum Spectral Curve,''
  J.\ Phys.\ A {\bf 52}, no. 12, 125401 (2019)
  doi:10.1088/1751-8121/ab046c
  [arXiv:1811.02369 [hep-th]].
  
  
  
  
  

\bibitem{Cavaglia:2020hdb}
A.~Cavaglia, D.~Grabner, N.~Gromov and A.~Sever,
``Colour-twist operators. Part I. Spectrum and wave functions,''
JHEP \textbf{06} (2020), 092
doi:10.1007/JHEP06(2020)092
[arXiv:2001.07259 [hep-th]].


\bibitem{Belliard:2012pr}
S.~Belliard, S.~Pakuliak, E.~Ragoucy, N.~A.~Slavnov, S.~Pakuliak, E.~Ragoucy and N.~A.~Slavnov,
``Algebraic Bethe ansatz for scalar products in SU(3)-invariant integrable models,''
J. Stat. Mech. \textbf{1210} (2012), P10017
doi:10.1088/1742-5468/2012/10/P10017
[arXiv:1207.0956 [math-ph]].


\bibitem{Belliard:2012is}
S.~Belliard, S.~Pakuliak, E.~Ragoucy and N.~A.~Slavnov,
``Highest coefficient of scalar products in SU(3)-invariant integrable models,''
J. Stat. Mech. \textbf{1209} (2012), P09003
doi:10.1088/1742-5468/2012/09/P09003
[arXiv:1206.4931 [math-ph]].




\bibitem{Slavnov:2015qoa}
N.~A.~Slavnov,
``Scalar products in GL(3)-based models with trigonometric R-matrix. Determinant representation,''
J. Stat. Mech. \textbf{1503} (2015) no.3, P03019
doi:10.1088/1742-5468/2015/03/P03019
[arXiv:1501.06253 [math-ph]].



\bibitem{Hutsalyuk:2016yii}
A.~Hutsalyuk, A.~Liashyk, S.~Z.~Pakuliak, E.~Ragoucy and N.~A.~Slavnov,
``Scalar products of Bethe vectors in models with $\mathfrak{g}\mathfrak{l}(2|1)$ symmetry 2. Determinant representation,''
J. Phys. A \textbf{50} (2017) no.3, 034004
doi:10.1088/1751-8121/50/3/034004
[arXiv:1606.03573 [math-ph]].









\bibitem{Derkachov:2008aq}
S.~E.~Derkachov and A.~N.~Manashov,
``Factorization of R-matrix and Baxter Q-operators for generic sl(N) spin chains,''
J. Phys. A \textbf{42} (2009), 075204
doi:10.1088/1751-8113/42/7/075204
[arXiv:0809.2050 [nlin.SI]].


\bibitem{Derkachov:2001yn}
  S.~E.~Derkachov, G.~P.~Korchemsky and A.~N.~Manashov,
  ``Noncompact Heisenberg spin magnets from high-energy QCD: 1. Baxter Q operator and separation of variables,''
  Nucl.\ Phys.\ B {\bf 617} (2001) 375
  doi:10.1016/S0550-3213(01)00457-6
  [hep-th/0107193].




  \bibitem{Derkachov:2002tf} 
  S.~E.~Derkachov, G.~P.~Korchemsky and A.~N.~Manashov,
  ``Separation of variables for the quantum SL(2,R) spin chain,''
  JHEP {\bf 0307}, 047 (2003)
  doi:10.1088/1126-6708/2003/07/047
  [hep-th/0210216].
  
 
 
 

  
 

  
  

  


 \bibitem{Sklyanin:1989cg}
  E.~K.~Sklyanin,
  ``New Approach To The Quantum Nonlinear Schrodinger Equation,'' 
  J.\ Phys.\ A {\bf 22} (1989) 3551 (see also the Russian version)




\bibitem{Chernyak:2020lgw}
D.~Chernyak, S.~Leurent and D.~Volin,
``Completeness of Wronskian Bethe equations for rational $\mathfrak{gl}_{m|n}$ spin chain,''
[arXiv:2004.02865 [math-ph]].


\bibitem{Kulish:1983rd}
P.~P.~Kulish and N.~Y.~Reshetikhin,
``DIAGONALIZATION OF GL(N) INVARIANT TRANSFER MATRICES AND QUANTUM N WAVE SYSTEM (LEE MODEL),''
J. Phys. A \textbf{16} (1983), L591-L596
doi:10.1088/0305-4470/16/16/001
  
 

\bibitem{Cherednik}
I.V.~Cherednik,
``An analogue of the character formula for Hekke algebras,''
Funct Anal Its Appl \textbf{21} (1987), 172–174
doi:/10.1007/BF01078042


\bibitem{Bazhanov:1989yk}
V.~Bazhanov and N.~Reshetikhin,
``Restricted Solid on Solid Models Connected With Simply Based Algebras and Conformal Field Theory,''
J. Phys. A \textbf{23} (1990), 1477
doi:10.1088/0305-4470/23/9/012


\bibitem{Zabrodin:1996vm}
A.~Zabrodin,
``Discrete Hirota's equation in quantum integrable models,''
Int. J. Mod. Phys. B \textbf{11} (1997), 3125
doi:10.1142/S0217979297001520
[arXiv:hep-th/9610039 [hep-th]].


\bibitem{Sklyanin:1992eu}
  E.~K.~Sklyanin,
  ``Separation of variables in the classical integrable SL(3) magnetic chain,''
  Commun.\ Math.\ Phys.\  {\bf 150} (1992) 181
  doi:10.1007/BF02096572
  [hep-th/9211126].

\bibitem{Bazhanov:1996dr}
V.~V.~Bazhanov, S.~L.~Lukyanov and A.~B.~Zamolodchikov,
``Integrable structure of conformal field theory. 2. Q operator and DDV equation,''
Commun. Math. Phys. \textbf{190} (1997), 247-278
doi:10.1007/s002200050240
[arXiv:hep-th/9604044 [hep-th]].

  \bibitem{Krichever:1996qd} 
  I.~Krichever, O.~Lipan, P.~Wiegmann and A.~Zabrodin,
  ``Quantum integrable systems and elliptic solutions of classical discrete nonlinear equations,''
  Commun.\ Math.\ Phys.\  {\bf 188}, 267 (1997)
  [hep-th/9604080].

\bibitem{Kazakov:2010iu}
V.~Kazakov, S.~Leurent and Z.~Tsuboi,
``Baxter's Q-operators and operatorial Backlund flow for quantum (super)-spin chains,''
Commun. Math. Phys. \textbf{311} (2012), 787-814
doi:10.1007/s00220-012-1428-9
[arXiv:1010.4022 [math-ph]].

\bibitem{Bazhanov:2010jq}
V.~V.~Bazhanov, R.~Frassek, T.~Lukowski, C.~Meneghelli and M.~Staudacher,
``Baxter Q-Operators and Representations of Yangians,''
Nucl. Phys. B \textbf{850} (2011), 148-174
doi:10.1016/j.nuclphysb.2011.04.006
[arXiv:1010.3699 [math-ph]].

\bibitem{Frassek:2011aa}
R.~Frassek, T.~Lukowski, C.~Meneghelli and M.~Staudacher,
``Baxter Operators and Hamiltonians for 'nearly all' Integrable Closed $\mathfrak{gl}(n)$ Spin Chains,''
Nucl. Phys. B \textbf{874} (2013), 620-646
doi:10.1016/j.nuclphysb.2013.06.006
[arXiv:1112.3600 [math-ph]].





\bibitem{Costa:2010rz} M.~S.~Costa, R.~Monteiro, J.~E.~Santos and D.~Zoakos, ``On three-point correlation functions in the gauge/gravity duality,''
  JHEP {\bf 1011} (2010) 141
  doi:10.1007/JHEP11(2010)141
  [arXiv:1008.1070 [hep-th]].



  





\bibitem{Gerotto:2017sat}
L.~Gerotto and T.~McLoughlin,
``Diagonal Form Factors in Landau-Lifshitz Models,''
JHEP \textbf{03} (2019), 180
doi:10.1007/JHEP03(2019)180
[arXiv:1710.02138 [hep-th]].


\bibitem{Slavnov:2018kfx}
N.~A.~Slavnov,
``Algebraic Bethe ansatz,''
[arXiv:1804.07350 [math-ph]].



\bibitem{Slavnov:2019hdn}
N.~A.~Slavnov,
``Introduction to the nested algebraic Bethe ansatz,''
doi:10.21468/SciPostPhysLectNotes.19
[arXiv:1911.12811 [math-ph]].



\bibitem{Smirnov:1998kv}
F.~A.~Smirnov,
``Quasiclassical study of form-factors in finite volume,''
[arXiv:hep-th/9802132 [hep-th]].


\bibitem{Caetano:2020dyp}
J.~Caetano and S.~Komatsu,
``Functional equations and separation of variables for exact $g$-function,''
JHEP \textbf{09} (2020), 180
doi:10.1007/JHEP09(2020)180
[arXiv:2004.05071 [hep-th]].



\bibitem{Kitanine:2015jna}
N.~Kitanine, J.~M.~Maillet, G.~Niccoli and V.~Terras,
``On determinant representations of scalar products and form factors in the SoV approach: the XXX case,''
J. Phys. A \textbf{49} (2016) no.10, 104002
doi:10.1088/1751-8113/49/10/104002
[arXiv:1506.02630 [math-ph]].




\bibitem{Kitanine:2016pvg}
N.~Kitanine, J.~M.~Maillet, G.~Niccoli and V.~Terras,
``The open XXX spin chain in the SoV framework: scalar product of separate states,''
J. Phys. A \textbf{50} (2017) no.22, 224001
doi:10.1088/1751-8121/aa6cc9
[arXiv:1606.06917 [math-ph]].



\bibitem{Niccoli:2020zla}
G.~Niccoli, H.~Pei and V.~Terras,
``Correlation functions by Separation of Variables: the XXX spin chain,''
[arXiv:2005.01334 [math-ph]].



\bibitem{Slavnov:2020ngt}
N.~A.~Slavnov,
``Generating function for scalar products in the algebraic Bethe ansatz,''
Teor. Mat. Fiz. \textbf{204} (2020) no.3, 453-465
doi:10.1134/S004057792009010X





\bibitem{Belliard:2019bfz}
S.~Belliard and N.~A.~Slavnov,
``Why scalar products in the algebraic Bethe ansatz have determinant representation,''
JHEP \textbf{10} (2019), 103
doi:10.1007/JHEP10(2019)103
[arXiv:1908.00032 [math-ph]].





\bibitem{McGovern:2019sdd}
J.~McGovern,
``Scalar insertions in cusped Wilson loops in the ladders limit of planar $ \mathcal{N} $ = 4 SYM,''
JHEP \textbf{05} (2020), 062
doi:10.1007/JHEP05(2020)062
[arXiv:1912.00499 [hep-th]].



\bibitem{Gurdogan:2015csr} 
  Ö.~Gürdoğan and V.~Kazakov,
  ``New Integrable 4D Quantum Field Theories from Strongly Deformed Planar $\mathcal N = $ 4 Supersymmetric Yang-Mills Theory,''
  Phys.\ Rev.\ Lett.\  {\bf 117}, no. 20, 201602 (2016)
  Addendum: [Phys.\ Rev.\ Lett.\  {\bf 117}, no. 25, 259903 (2016)]
  doi:10.1103/PhysRevLett.117.201602, 10.1103/PhysRevLett.117.259903
  [arXiv:1512.06704 [hep-th]].



\bibitem{Gromov:2019aku} 
  N.~Gromov and A.~Sever,
  ``The Holographic Fishchain,''
  arXiv:1903.10508 [hep-th].


\bibitem{Derkachov:2018rot} 
  S.~Derkachov, V.~Kazakov and E.~Olivucci,
  ``Basso-Dixon Correlators in Two-Dimensional Fishnet CFT,''
  JHEP {\bf 1904}, 032 (2019)
  doi:10.1007/JHEP04(2019)032
  [arXiv:1811.10623 [hep-th]].

  


  


\bibitem{Derkachov:2019tzo}
S.~Derkachov and E.~Olivucci,
``Exactly solvable magnet of conformal spins in four dimensions,''
Phys. Rev. Lett. \textbf{125} (2020) no.3, 031603
doi:10.1103/PhysRevLett.125.031603
[arXiv:1912.07588 [hep-th]].
































































  

%%%%%%%%%%%%%%%%%%%%%%%%%%%%%%
%%%%%% OLD REFS
%%%%%%%%%%%%%%%%%%%%%%%%%%%%%%%%%%%%
%%%%%%%%%%%%%%%%%%%%%%%%




\bibitem{Basso:2019xay}
B.~Basso, G.~Ferrando, V.~Kazakov and D.~l.~Zhong,
``Thermodynamic Bethe Ansatz for Biscalar Conformal Field Theories in any Dimension,''
Phys. Rev. Lett. \textbf{125} (2020) no.9, 091601
doi:10.1103/PhysRevLett.125.091601
[arXiv:1911.10213 [hep-th]].


\bibitem{Derkachov:2020zvv}
S.~Derkachov and E.~Olivucci,
``Exactly solvable single-trace four point correlators in $\chi$CFT$_4$,''
[arXiv:2007.15049 [hep-th]].






\bibitem{Derkachov:2018ewi}
S.~E.~Derkachov and P.~A.~Valinevich,
``Separation of variables for the quantum $SL(3,\mathbb C)$ spin magnet: eigenfunctions of Sklyanin $B$-operator,''
J. Math. Sci. \textbf{242} (2019) no.5, 658-682
doi:10.1007/s10958-019-04505-5
[arXiv:1807.00302 [math-ph]].


\bibitem{Ferrando:2020vzk}
G.~Ferrando, R.~Frassek and V.~Kazakov,
``QQ-system and Weyl-type transfer matrices in integrable $SO(2r)$ spin chains,''
[arXiv:2008.04336 [hep-th]].

\bibitem{Ekhammar:2020enr}
S.~Ekhammar, H.~Shu and D.~Volin,
``Extended systems of Baxter Q-functions and fused flags I: simply-laced case,''
[arXiv:2008.10597 [math-ph]].


\bibitem{Gromov:2018cvh}
N.~Gromov and F.~Levkovich-Maslyuk,
``New Compact Construction of Eigenstates for Supersymmetric Spin Chains,''
JHEP \textbf{09} (2018), 085
doi:10.1007/JHEP09(2018)085
[arXiv:1805.03927 [hep-th]].

\bibitem{Maillet:2018rto}
J.~M.~Maillet and G.~Niccoli,
``Complete spectrum of quantum integrable lattice models associated to $\mathcal{U}_{q} (\widehat{gl_{n}})$ by separation of variables,''
J. Phys. A \textbf{52} (2019) no.31, 315203
doi:10.1088/1751-8121/ab2930
[arXiv:1811.08405 [math-ph]].

\bibitem{Pei:2020ljw}
H.~Pei and V.~Terras,
``On scalar products and form factors by Separation of Variables: the antiperiodic XXZ model,''
[arXiv:2011.06109 [math-ph]]


\bibitem{molev2007yangians}
A.~Molev, 
``Yangians and classical Lie algebras,''
American Mathematical Soc. 143 (2007).


\bibitem{Kuniba:1994na}
A.~Kuniba and J.~Suzuki,
``Analytic Bethe Ansatz for fundamental representations of Yangians,''
Commun. Math. Phys. \textbf{173} (1995), 225-264
doi:10.1007/BF02101234
[arXiv:hep-th/9406180 [hep-th]].


\bibitem{Kuniba:1995gi}
A.~Kuniba, Y.~Ohta and J.~Suzuki,
``Quantum Jacobi-Trudi and Giambelli formulae for U-q(B(r)(1)) from analytic Bethe ansatz,''
J. Phys. A \textbf{28} (1995), 6211-6226
doi:10.1088/0305-4470/28/21/024
[arXiv:hep-th/9506167 [hep-th]].


\bibitem{Tsuboi:1998ne}
Z.~Tsuboi,
``Analytic Bethe Ansatz And Functional Equations Associated With Any Simple Root Systems Of The Lie Superalgebra $sl(r+1|s+1)$,''
Physica A \textbf{252} (1998), 565-585
doi:10.1016/S0378-4371(97)00625-0
[arXiv:0911.5387 [math-ph]].


\bibitem{Tsuboi:1998sc}
Z.~Tsuboi,
``Analytic Bethe ansatz related to a one-parameter family of finite-dimensional representations of the Lie superalgebra $sl(r+1|s+1)$,''
J. Phys. A \textbf{31} (1998), 5485-5498
doi:10.1088/0305-4470/31/24/010
[arXiv:0911.5389 [math-ph]].



  


\bibitem{Gromov:2010kf} 
  N.~Gromov and V.~Kazakov,
  ``Review of AdS/CFT Integrability, Chapter III.7: Hirota Dynamics for Quantum Integrability,''
  Lett.\ Math.\ Phys.\  {\bf 99}, 321 (2012)
  doi:10.1007/s11005-011-0513-x
  [arXiv:1012.3996 [hep-th]].
  




\bibitem{Okuboosc}
S.~Okubo, ``Algebraic identities among U(n) infinitesimal generators'', J.\ Math.\ Phys.\ {\bf 16}, 528-535 (1975)



\bibitem{Palev:1996kh}
T.~D.~Palev,
``A Holstein-Primakoff and Dyson realizations for the Lie superalgebra gl(m/n+1),''
J. Phys. A \textbf{30} (1997), 8273-8278
doi:10.1088/0305-4470/30/23/023
[arXiv:hep-th/9607222 [hep-th]].



% \bibitem{Krichever:1996qd}
% I.~Krichever, O.~Lipan, P.~Wiegmann and A.~Zabrodin,
% ``Quantum integrable systems and elliptic solutions of classical discrete nonlinear equations,''
% Commun. Math. Phys. \textbf{188} (1997), 267-304
% doi:10.1007/s002200050165
% [arXiv:hep-th/9604080 [hep-th]].





\end{thebibliography}
\end{document}